\documentclass[useAMS,usenatbib]{mnras}

\usepackage{graphicx, bm, amssymb, xcolor}
\usepackage{verbatim}
\usepackage[all]{hypcap}
\usepackage{xspace}
\usepackage{multirow,bigdelim}
\usepackage[fleqn]{amsmath}

\usepackage{epsfig}
\usepackage{aas_macros}
\usepackage{natbib}
\usepackage{times,txfonts}
\usepackage{morefloats}
\usepackage{tabularx}
\usepackage{lscape}

\newcommand{\msun}{\mbox{M$_\odot$}}
\newcommand{\yr}{\mbox{${\rm yr}$}}
\newcommand{\myr}{\mbox{${\rm Myr}$}}
\newcommand{\gyr}{\mbox{${\rm Gyr}$}}
\newcommand{\pc}{\mbox{${\rm pc}$}}
\newcommand{\mpc}{\mbox{${\rm Mpc}$}}
\newcommand{\kpc}{\mbox{${\rm kpc}$}}
\newcommand{\kms}{\mbox{${\rm km}~{\rm s}^{-1}$}}
\newcommand{\cmc}{\mbox{${\rm cm}^{-3}$}}

\newcommand{\dex}{\mbox{${\rm dex}$}}
\newcommand{\feh}{\mbox{$[{\rm Fe}/{\rm H}]$}}
\newcommand{\iqr}{\mbox{${\rm IQR}$}}
\newcommand{\riqr}{\mbox{$r_{\rm IQR}$}}
\newcommand{\dfehdt}{\mbox{${\rm d}\feh/{\rm d}\log{t}$}}
\newcommand{\ngc}{\mbox{$N_{\rm GC}$}}
\newcommand{\ngcp}{\mbox{$N_{\rm GC}'$}}
\newcommand{\sfr}{\mbox{${\rm SFR}$}}
\newcommand{\ssfr}{\mbox{${\rm sSFR}$}}
\newcommand{\mvir}{\mbox{$M_{200}$}}
\newcommand{\rvir}{\mbox{$R_{200}$}}
\newcommand{\vmax}{\mbox{$V_{\rm max}$}}
\newcommand{\rvmax}{\mbox{$R_{V_{\rm max}}$}}
\newcommand{\cnfw}{\mbox{$c_{\rm NFW}$}}
\newcommand{\ttf}{\mbox{$\tau_{25}$}}
\newcommand{\tfz}{\mbox{$\tau_{50}$}}
\newcommand{\tsf}{\mbox{$\tau_{75}$}}
\newcommand{\tmax}{\mbox{$\tau_{\rm max}$}}
\newcommand{\ta}{\mbox{$\tau_{\rm a}$}}
\newcommand{\za}{\mbox{$z_{\rm a}$}}
\newcommand{\tf}{\mbox{$\tau_{\rm f}$}}
\newcommand{\zf}{\mbox{$z_{\rm f}$}}
\newcommand{\tmm}{\mbox{$\tau_{\rm mm}$}}
\newcommand{\zmm}{\mbox{$z_{\rm mm}$}}
\newcommand{\tam}{\mbox{$\tau_{\rm am}$}}
\newcommand{\zam}{\mbox{$z_{\rm am}$}}
\newcommand{\thub}{\mbox{$\tau_{\rm H}$}}
\newcommand{\nbrz}{\mbox{$N_{{\rm br},z>2}$}}
\newcommand{\nbr}{\mbox{$N_{\rm br}$}}
\newcommand{\rz}{\mbox{$r_{z>2}$}}
\newcommand{\nleaf}{\mbox{$N_{\rm leaf}$}}
\newcommand{\rbl}{\mbox{$r_{\rm bl}$}}
\newcommand{\fexs}{\mbox{$f_{\rm ex,*}$}}
\newcommand{\fexgc}{\mbox{$f_{\rm ex,GC}$}}

\newcommand{\ncorr}{\mbox{$N_{\rm corr}$}}
\newcommand{\pref}{\mbox{$p_{\rm ref}$}}
\newcommand{\peff}{\mbox{$p_{\rm eff}$}}

\newcommand{\fcorr}{\mbox{$f_{\rm corr}$}}
\newcommand{\mosaics}{MOSAICS\xspace}
\newcommand{\emosaics}{E-MOSAICS\xspace}
\newcommand{\eagle}{EAGLE\xspace}

\newcommand{\be}{\begin{equation}}
\newcommand{\ee}{\end{equation}}
\newcommand{\bea}{\begin{eqnarray}}
\newcommand{\eea}{\end{eqnarray}}

\newcommand{\appropto}{\mathrel{\vcenter{
  \offinterlineskip\halign{\hfil$##$\cr
    \propto\cr\noalign{\kern1pt}\sim\cr\noalign{\kern-2pt}}}}}

\defcitealias{pfeffer18}{Paper~I}

\title[The E-MOSAICS project]{\vspace{-6mm}The E-MOSAICS project: tracing galaxy formation and assembly with the age-metallicity distribution of globular clusters\vspace{-5mm}}

\author{J.~M.~Diederik Kruijssen,$^{1}$\thanks{E-mail: \href{kruijssen@uni-heidelberg.de}{kruijssen@uni-heidelberg.de}} Joel~L.~Pfeffer,$^2$ Robert~A.~Crain$^2$ and Nate Bastian$^2$\\
$^1$Astronomisches Rechen-Institut, Zentrum f\"{u}r Astronomie der Universit\"{a}t Heidelberg, M\"{o}nchhofstra\ss e 12-14, 69120 Heidelberg, Germany\\
$^2$Astrophysics Research Institute, Liverpool John Moores University, IC2, Liverpool Science Park, 146 Brownlow Hill,
Liverpool L3 5RF, United Kingdom\vspace{-4mm}}

\begin{document}

\date{Accepted 2019 April 2. Received 2019 April 2; in original form 2018 March 5\vspace{-2mm}}

\pagerange{\pageref{firstpage}--\pageref{lastpage}} \pubyear{2019}

\maketitle

\label{firstpage}

\begin{abstract}
We present 25 cosmological zoom-in simulations of Milky Way-mass galaxies in the `MOdelling Star cluster population Assembly In Cosmological Simulations within \eagle' (\emosaics) project. \emosaics couples a detailed physical model for the formation, evolution, and disruption of star clusters to the \eagle galaxy formation simulations. This enables following the co-formation and co-evolution of galaxies and their star cluster populations, thus realising the long-standing promise of using globular clusters (GCs) as tracers of galaxy formation and assembly. The simulations show that the age-metallicity distributions of GC populations exhibit strong galaxy-to-galaxy variations, resulting from differences in their evolutionary histories. We develop a formalism for systematically constraining the assembly histories of galaxies using GC age-metallicity distributions. These distributions are characterised through 13 metrics that we correlate with 30 quantities describing galaxy formation and assembly (e.g.~halo properties, formation/assembly redshifts, stellar mass assembly time-scales, galaxy merger statistics), resulting in 20 statistically (highly) significant correlations. The GC age-metallicity distribution is a sensitive probe of the mass growth, metal enrichment, and minor merger history of the host galaxy. No such relation is found between GCs and major mergers, which play a sub-dominant role in GC formation for Milky Way-mass galaxies. Finally, we show how the GC age-metallicity distribution enables the reconstruction of the host galaxy's merger tree, allowing us to identify all progenitors with masses $M_*\ga10^8~\msun$ for redshifts $1\leq z\leq2.5$. These results demonstrate that cosmological simulations of the co-formation and co-evolution of GCs and their host galaxies successfully unlock the potential of GCs as quantitative tracers of galaxy formation and assembly.
\end{abstract}

\begin{keywords}
galaxies: evolution --- galaxies: formation --- galaxies: haloes --- galaxies: star formation --- globular clusters: general\vspace{-4mm}
\end{keywords}

\section{Introduction} \label{sec:intro}
It is one of the major goals of modern astrophysics to reconstruct the formation and assembly histories of galaxies and their dark matter haloes. The identification of the physical mechanisms shaping the present-day galaxy population may enable bridging the gap between cosmological models and the observable baryonic mass in the Universe. The main difficulty in overcoming this problem is that observational galaxy formation studies must deal with instantaneous snapshots of the galaxy population -- it is not possible to follow the evolution of individual systems in time. These difficulties could be remedied by studying the evolution of large galaxy samples across different cosmic epochs \citep[e.g.][]{vandokkum10b,patel13,papovich15}, potentially in connection to empirical models \citep[e.g.][]{moster13,behroozi13}, but fundamentally this approach relies on statistical inference rather than probing the physical mechanisms driving galaxy formation and evolution directly. Such statistics of the galaxy population are unable to provide insight into the assembly histories of individual galaxies such as the Milky Way. Additional constraints on galaxy formation and assembly are especially desirable for the early phases of galaxy evolution at intermediate-to-high redshift ($z>1$), where the spatial resolution and sensitivity of observations are limited relative to the low-redshift ($z<1$) Universe.

All galaxies with stellar masses $M_\star>10^9~\msun$ host rich populations of massive, dense and (mostly) old star clusters. These globular clusters (GCs) have been an active field of study for many decades (if not centuries, see~\citealt{herschel89}), yet their origin remains debated. Once thought to have formed under conditions specific to the early Universe \citep[e.g.][]{peebles68,fall85}, the discovery of GC-like clusters forming today by the {\it Hubble Space Telescope} \citep{holtzman92,whitmore99} led to a surge of work proposing that GCs might be the relics of regular star and cluster formation during the epoch of peak star formation activity in the Universe \citep[e.g.][]{ashman92,elmegreen97,fall01,kravtsov05,kruijssen15b}. These (largely analytical) models have been very successful at reproducing several of the main properties of GC populations. Throughout these works, the awareness grew that if GCs do indeed originate from the height of cosmic star formation, then they may be excellent tracers of galaxy formation and assembly \citep[e.g.][]{harris91,forbes97,brodie06}.

Unfortunately, the promise of using GCs as tracers of galaxy formation and assembly has largely remained unfulfilled. Making progress has required finding answers to two key open questions in current GC research (see \citealt{kruijssen14c} and \citealt{forbes18} for recent reviews):
\begin{enumerate}
\item
What are the important physical mechanisms that shape the GC populations observed at $z=0$? 
\item
What do the properties of GCs populations reveal about the formation and assembly histories of their host galaxies?
\end{enumerate}
These questions can only be addressed in the context of a reasonable hypothesis for GC formation and evolution. In addition, they require the technological ability to construct a model for the co-formation and co-evolution of GCs and galaxies.

Motivated by the successes of previous works proposing that GCs are the relics of normal (but intense) star formation across cosmic time, as well as the recent successes of galaxy formation models to reproduce a broad range of properties of the galaxy population \citep[e.g.][]{vogelsberger14,schaye15,dave17,kaviraj17}, we have started the \emosaics\footnote{This is an acronym for `MOdelling Star cluster population Assembly In Cosmological Simulations within \eagle'.} project. The goal of this project is to address the above two questions by self-consistently modelling stellar cluster formation and evolution in galaxy formation simulations. Specifically, we couple the semi-analytic cluster formation and evolution model \mosaics \citep{kruijssen11,kruijssen12c} in a subgrid fashion to the \eagle simulations of galaxy formation \citep{schaye15,crain15}. In their respective fields, both of these models have been able to provide accurate representations of real-Universe systems (see Section~\ref{sec:emosaics}).

This paper is the second of a pair of reference papers describing the initial results of \emosaics. In the first paper \citep[hereafter \citetalias{pfeffer18}]{pfeffer18}, we address the first of the above two questions by describing the physical model in detail, validating it across a wide range of tests, and identifying the physics relevant to cluster formation and evolution during galaxy formation in a set of 10 cosmological zoom-in simulations of Milky Way-like galaxies. In brief, we find that modelling GC populations from the early Universe to $z=0$ requires models for the cluster formation efficiency (the fraction of star formation occurring in bound stellar clusters), the initial cluster mass function, cluster disruption by tidal perturbations and evaporation, cluster migration during galaxy assembly, and dynamical friction. While the combination of these elements had been proposed in analytical work \citep[e.g][]{kruijssen15b}, it is demonstrated in \citetalias{pfeffer18} that their self-consistent modelling in a galaxy formation context allows us to accurately follow their environmental dependence and reproduce a wide variety of resulting galaxies and GC populations. This environmental dependence is shown to be critical for reproducing the variety of galaxies observed in the local Universe, underlining the necessity of simulating a sample of galaxies. Finally, the results of \citetalias{pfeffer18} show that, at present and for the foreseeable future, sub-grid methods are the only numerically feasible way of studying the entire GC population over cosmological volumes and cosmic time.

In this paper, we aim to address the second of the above two questions and identify in what way GCs can be used to trace galaxy formation and assembly. Various observables describing GC populations may plausibly carry the imprints of the host galaxy formation history, such as the specific frequency (i.e.~the number of GCs per unit galaxy luminosity or mass), the spatial distribution and kinematics of the GC population, the GC metallicity distribution, and the GC age and mass distributions \citep[e.g.][]{brodie06,kruijssen14c,lamers17,forbes18}. Within the Milky Way out to $\sim50~\kpc$, it is possible to measure the ages of GCs to a precision of about $1~\gyr$ \citep{marinfranch09,dotter10,dotter11,vandenberg13}. The distribution of GCs in age-metallicity space has been proposed to be a powerful probe of galaxy assembly \citep{forbes10,leaman13}, potentially enabling the identification of dwarf galaxy accretion and episodes of active star formation. In view of our modelling of Milky Way-like galaxies, this first paper focuses on tracing galaxy formation with the age-metallicity distribution of GCs. A variety of future \emosaics papers will address other GC-related observables, such as GC formation histories, spatial distributions, metallicity distributions, kinematic distributions, specific frequencies, high-redshift luminosity functions, and the number of GCs per unit dark matter halo mass.

In this work, we expand the initial set of 10 zoom-in simulations of $L^\star$, Milky Way-like galaxies from \citetalias{pfeffer18} to a total of 25 simulations. By connecting several observables describing the age-metallicity distribution of GCs with quantitative metrics characterising galaxy formation and assembly histories (e.g.~through galaxy merger trees), we fulfil the potential of GCs as quantitative tracers of galaxy formation. Next to using the absolute ages of GCs, we also consider quantities based on their relative ages, which can be inferred to greater precision. This plausibly enables the application of the insights from this paper to galaxies and their GC populations beyond the Milky Way (Usher et al.~in prep.).

The structure of this paper is as follows. In Section~\ref{sec:emosaics}, we summarise the physical models for cluster formation and evolution and for galaxy formation and evolution, as well as introduce the 25 simulations used in this paper. The age-metallicity distributions of the resulting GC populations at $z=0$ are presented and characterised in Section~\ref{sec:agez}. In Section~\ref{sec:hist}, we quantify the galaxy formation and assembly histories and demonstrate how they are related to the properties of the GC age-metallicity distribution. This relation is expanded in Section~\ref{sec:recon}, where we show how the merger trees of galaxies can be reconstructed using the age-metallicity distribution of GCs. The paper is concluded with a discussion in Section~\ref{sec:disc} and a summary of our conclusions in Section~\ref{sec:concl}. In a follow-up paper \citep{kruijssen18c}, we apply the insights drawn from our analysis to the age-metallicity distribution of GCs in the Milky Way and place quantitative constraints on the Milky Way's assembly history.

\section{Cosmological zoom-in simulations of Milky Way-mass galaxies and their GC populations} \label{sec:emosaics}
Here, we briefly summarise how the \emosaics simulations model galaxy formation and evolution, describe the sub-grid model for stellar cluster formation and evolution, and describe the set of 25 cosmological zoom-in simulations of Milky Way-mass disc galaxies used in this work. In the subsequent sections, we analyse these simulations to determine the relation between the GC age-metallicity distribution at $z=0$ and the formation and assembly history of the host galaxy.

\subsection{Summary of the physical model}
We first describe the \eagle (Section \ref{sec:eagle}) and \mosaics (Section \ref{sec:mosaics}) components of the E-MOSAICS model. Since new simulations to the E-MOSAICS suite are introduced here, we retain a relatively detailed description of the two components, similar to that provided in Section~2 of \citetalias{pfeffer18}. All simulations examined in this study assume a $\Lambda$CDM cosmogony, described by the parameters advocated by the \citet{planck14}, namely $\Omega_0 = 0.307$, $\Omega_{\rm b} =
0.04825$, $\Omega_\Lambda= 0.693$, $\sigma_8 = 0.8288$, $n_{\rm s} = 0.9611$, $h = 0.6777$, and $Y = 0.248$.

\subsubsection{The \eagle galaxy formation model} \label{sec:eagle}
\eagle \citep{schaye15,crain15} is a campaign of cosmological, hydrodynamical simulations that model the formation and evolution of galaxies in a $\mathrm{\Lambda CDM}$ cosmogony. The simulations are evolved by a modified version of the smoothed particle hydrodynamics (SPH) and TreePM gravity solver \textsc{Gadget~3}, last described by \citet{springel05c}. Besides the inclusion of a series of subgrid routines governing key physical processes that govern galaxy formation, which are described in detail below, the modifications include the implementation of the pressure-entropy formulation of SPH presented by \citet{hopkins13c}, the time-step limiter of \citet{durier12}, and switches for artificial viscosity and artificial conduction of the forms proposed by, respectively, \citet{cullen10} and \citet{price08}.

The simulations implement the element-by-element radiative cooling and photoionization heating scheme of \citet{wiersma09}, which considers 11 species (H, He and 9 metal species). The (net) cooling rate is computed assuming the incidence of a spatially-uniform, temporally-evolving radiation field comprising the cosmic microwave background and the metagalactic ultraviolet/X-ray background produced by galaxies and quasars, as described by \citet{haardt01}. The gas is assumed to be optically thin and in ionization equilibrium. Gas with density greater than a metallicity-dependent threshold \citep{schaye04}, and which is within 0.5 decades of a Jeans-limiting temperature floor (see below), is eligible for stochastic conversion to a collisionless stellar particle. The probability of conversion is proportional to the particle's star formation rate (SFR), which is a function of its pressure \citep{schaye08}. By construction, this scheme reproduces the observed `star formation relation' between the gas mass (density) and the SFR (density) \citep{kennicutt98}. 

Each stellar particle is assumed to represent a simple stellar population (SSP) described by the \citet{chabrier03} initial mass function (IMF). The return of mass and metals from evolving stellar populations to the interstellar medium (ISM) is implemented with the scheme of \citet{wiersma09b}, which tracks the abundances of the same 11 elements considered when computing the radiative cooling and photoionization heating rates. Black holes (BHs) are seeded in dark matter haloes (identified using the friends-of-friends algorithm) that do not already have a BH when they reach a mass of $10^{10} \msun/h$, and they grow via gas accretion (at the minimum of the Bondi-Hoyle and Eddington rates) and by merging with other BHs \citep{springel05b,rosasguevara15,schaye15}. Feedback resulting from star formation \citep{dallavecchia12} and the accretion of mass onto BHs \citep{booth09,schaye15} is implemented as the stochastic heating of gas particles. AGN feedback is therefore implemented as a single heating mode, but mimics quiescent `radio-like' and vigorous `quasar-like' AGN modes when the BH accretion rate is a small ($\ll 1$) or large ($\sim 1$) fraction of the Eddington rate, respectively \citep[][]{mccarthy11}. 

Modelling the cold, dense phase of the ISM requires high resolution and treatments of the relevant physical processes, both of which are generally lacking from simulations of large cosmological volumes. To account for these omissions, gas in \eagle is subject to a polytropic temperature floor, $T_{\rm eos}(\rho_{\rm g})$, which corresponds to the equation of state $P_{\rm eos} \propto \rho_{\rm g}^{4/3}$. This relation is normalised to $T_{\rm eos} = 8000~{\rm K}$ at $n_{\rm H} \equiv X_{\rm H,0}\rho/m_{\rm H} = 10^{-1}~\cmc$, where $X_{\rm H,0}=0.752$ is the hydrogen mass fraction of gas with primordial composition. The exponent of $4/3$ is used as it ensures that the Jeans mass, and the ratio of the Jeans length to the SPH kernel support radius, are independent of the density \citep{schaye08}, thus limiting artificial fragmentation. Gas with $\log_{10} T > \log_{10} T_{\rm eos}(\rho_{\rm g}) + 0.5$ is ineligible for star formation, irrespective of its density.

As articulated by \citet[see their Section~2]{schaye15}, cosmological simulations also presently lack the resolution and physics necessary to compute, ab-initio, the efficiency of the feedback processes that regulate and quench galaxy growth. In \eagle, this problem is addressed by calibrating the subgrid efficiencies of feedback associated with star formation and gas accretion onto BHs to reproduce appropriate observables. The efficiency of the former is a smoothly-varying function of the metallicity and density of gas local to newly-formed stellar particles, and is calibrated to reproduce the present-day galaxy stellar mass function, and the size-mass relation of disc galaxies. The subgrid efficiency of AGN feedback is assumed to be constant, and is calibrated to reproduce the relation between the mass of central BHs and the stellar mass of their host galaxy at $z=0$ \citep[see also][]{booth09}. \citet{schaye15} argue that parameters may need to be recalibrated as the resolution of the simulation is changed; for this reason the parameters adopted for the Reference (`Ref') \eagle model are slightly different to those that yield the most accurate reproduction of the calibration diagnostics at a factor of 8 (2) better mass (spatial) resolution (the `Recal' model). 

The \eagle simulations have been shown to reproduce a broad range of observed galaxy properties and scaling relations, such as the evolution of the stellar masses \citep{furlong15} and sizes \citep{furlong17} of galaxies, their luminosities and colours \citep{trayford15}, their cold gas properties \citep{lagos15,lagos16,bahe16,marasco16,crain17}, and the properties of circumgalactic and intergalactic absorption systems \citep{rahmati15,rahmati16,turner16,turner17,oppenheimer16,oppenheimer18}.

\subsubsection{The \mosaics star cluster model} \label{sec:mosaics}
In current state-of-the-art simulations of galaxy formation, it is not possible to resolve the formation and evolution of the entire stellar cluster population from the Big Bang till the present day. To model the stellar cluster populations of the simulated galaxies, we combine the \eagle galaxy formation model with the semi-analytic star cluster formation and evolution model \mosaics (MOdelling Star cluster population Assembly In Cosmological Simulations, \citealt{kruijssen08,kruijssen09c,kruijssen11}), which was originally aimed at modelling the dynamical evolution of an initial cluster population due to tidally-limited evaporation and tidal shocks in the evolving potential of large-scale numerical simulations of galaxy formation and evolution. In \citetalias{pfeffer18}, we expanded \mosaics to include a physical model for the initial properties of the cluster population, capturing the environmental dependence of the cluster formation efficiency (CFE) and the maximum cluster mass, using the models of \citet{kruijssen12d} and \citet{reinacampos17}, respectively. We also expanded the model with a simple post-processing description of cluster destruction by inspiral due to dynamical friction.

In our simulations, \mosaics is called whenever a gas particle is converted into a stellar particle of mass $m_{\rm s}$. The particle mass is divided into a field star mass budget $m_{\rm field}=(1-\Gamma)m_{\rm s}$ and a cluster mass budget $m_{\rm clusters}=\Gamma m_{\rm s}$ using the CFE ($\Gamma$), which indicates the fraction of star formation occurring in gravitationally bound clusters \citep[introduced by][]{bastian08}. In \mosaics, the CFE is environmentally dependent according to the model of \citet[Section~7.3.3]{kruijssen12d} and depends on the local gas density, velocity dispersion, and temperature. This results in an effective increase with the local gas pressure, from $\Gamma\approx1$~per~cent at $P/k\approx10^{2.4}~{\rm K}~\cmc$ to $\Gamma\approx50$~per~cent at $P/k\approx10^{6.6}~{\rm K}~\cmc$, consistently with observations of the ISM and cluster populations in the local Universe \citep[e.g.][]{goddard10,adamo15b,johnson16,sun18}. The cluster mass budget of the particle $m_{\rm clusters}$ is then distributed over a stochastically-drawn, subgrid cluster population that follows a \citet{schechter76} initial cluster mass function (ICMF), i.e.~a power law with an exponential truncation at the high-mass end:
\be
\label{eq:icmf}
\frac{{\rm d}N}{{\rm d}M} \propto M^\alpha\exp{(-M/M_{\rm c,*})} ,
\ee
where the slope $\alpha=-2$ is chosen to be consistent with observations \citep[e.g.][]{portegieszwart10,longmore14}. We adopt a hard minimum mass limit of $10^2~\msun$, but clusters generated with masses $M<5\times10^3~\msun$ are discarded immediately after their formation to reduce the memory footprint of the calculation. The exponential truncation mass $M_{\rm c,*}$ is obtained using a slightly modified form of the \citet{reinacampos17} model (see \citetalias{pfeffer18}), which simultaneously reproduces the maximum cluster and cloud masses in local-Universe galaxies and at high redshift. In this model, cluster masses are limited by feedback in environments of low angular velocities ($\Omega\la0.6~\myr^{-1}$) and low surface densities ($\Sigma\la10^2~\msun~\pc^{-2}$), and by large-scale centrifugal forces in all other cases. As a result, the maximum cluster mass in our models increases with the gas pressure and decreases with the orbital frequency. At the particle resolution of our simulations (see Section~\ref{sec:sims} below), stellar particles rarely host more than one GC at $z=0$. Finally, the clusters are assigned a half-mass radius, which in the fiducial simulations is kept constant at $r_{\rm h}=4~\pc$ throughout their evolutionary histories. Clearly, this is a simplification -- in reality, we would expect some size evolution of the clusters \citep[e.g.][]{gieles11b}. We have tested different choices of the cluster radius and its time evolution in \citetalias{pfeffer18}, finding that within reasonable limits, it does not significantly affects the statistics of the resulting cluster populations. We refer the interested reader to \citetalias{pfeffer18} for the numerical details of how the cluster population is generated.

After the generation of the subgrid initial cluster population in a spawned stellar particle, the clusters undergo mass loss due to stellar evolution following the \eagle implementation of the \citet{wiersma09b} model, which uses stellar lifetimes from \citet{portinari98} and a \citet{chabrier03} IMF. In addition, the \mosaics model accounts for cluster mass loss through two dynamical mechanisms. The first of these mechanisms is gradual cluster evaporation due to two-body relaxation in the local tidal field, of which the mass loss rate is expressed as
\be
\label{eq:dmdtrlx}
\left(\frac{{\rm d}M}{{\rm d}t}\right)_{\rm rlx}=-\frac{\msun}{{\rm t}_{0,\odot}}\left(\frac{M}{\msun}\right)^{1-\gamma}\left(\frac{T}{{\rm T}_\odot}\right)^{1/2} ,
\ee
where ${\rm t}_{0,\odot}=21.3~\myr$ is a characteristic disruption time-scale parameter at the solar galactocentric radius \citep{lamers05,lamers06b,kruijssen09} with tidal field strength ${\rm T}_\odot=7.01\times10^2~\gyr^{-2}$ \citep{kruijssen11}, $T$ is the tidal field strength (i.e.~$T=\max{(\lambda)}+\Omega^2$, where $\max{(\lambda)}$ is the largest eigenvalue of the local tidal field tensor), and $\gamma$ is the mass dependence of the cluster disruption time \citep[$t_{\rm dis}\propto M^\gamma$,][]{lamers05}, which takes a value of $\gamma=0.62$ for a cluster with a \citet{king66} density profile with King parameter $W_0=5$. In \citetalias{pfeffer18}, we have tested choices appropriate for other King parameters but found negligible difference in the resulting mass loss rates.

The second dynamical mass loss mechanism considered in our simulations is cluster disruption by tidal shocks, i.e.~gravitational perturbations from ambient structure (e.g.~giant molecular clouds, spiral arms, the host galaxy disc) that induce variations in individual components of the tidal field tensor. The resulting mass loss rate is given by
\be
\label{eq:dmdtsh}
\left(\frac{{\rm d}M}{{\rm d}t}\right)_{\rm sh}=-\frac{20.4~\msun}{\myr}\left(\frac{r_{\rm h}}{4~\pc}\right)^{3}\left(\frac{I_{\rm tid}}{10^4~\gyr^{-2}}\right)\left(\frac{\Delta t}{10~\myr}\right)^{-1} ,
\ee
where the coefficient is derived from the full expression in \citet[eqs.~9, 17, and 23]{kruijssen11}, the tidal heating parameter $I_{\rm tid}$ is the square of the integral of the tidal tensor over the duration of the tidal shock \citep{gnedin99b,prieto08}, including a correction factor for the damping of the energy injection by adiabatic expansion \citep{weinberg94a,weinberg94b,weinberg94c}, and $\Delta t$ is the time since the previous shock. Several papers in the literature compare the mass loss rates due to evaporation and tidal shocks. They find that, as long as a model for the ISM is included, tidal shocks always dominate over evaporation (e.g.~\citealt{gieles06,lamers06b,kruijssen11,gieles16}; \citetalias{pfeffer18}).

Finally, we include a simple description for cluster destruction by dynamical friction in post-processing. Because the clusters exist as a subgrid component of the stellar particles, applying dynamical friction on the fly would lead to the unphysical result that the field star population within a stellar particle would experience the same drag as the clusters. We therefore discard clusters at the end of the simulation that at any point of their evolution had ages in excess of their dynamical friction time-scales for spiralling into the host galaxy's centre. Again, we refer the interested reader to \citetalias{pfeffer18} for the numerical details on the modelling of cluster mass loss and disruption by stellar evolution, two-body relaxation-driven evaporation, tidal shocks, and dynamical friction.

The \mosaics model (and elements thereof) has been applied to predict and explain a wide variety of observables describing $z=0$ cluster populations, such as their age and mass distributions \citep{kruijssen11c,adamo15,miholics17}, spatial distributions and kinematics \citep{kruijssen11,kruijssen12c}, cluster formation efficiencies, and maximum mass-scales (\citealt{adamo15b,johnson16}; \citetalias{pfeffer18}; \citealt{ward18}). In \emosaics, we extend these applications to the GC population, with the twofold goal of providing insight into the origin of GCs and of leveraging their unfulfilled potential as tracers of galaxy formation and assembly.

\subsection{Summary of the zoom-in simulations} \label{sec:sims}
In \citetalias{pfeffer18}, we present a set of 10 cosmological zoom-in simulations \citep[cf.][]{katz93} of disc-dominated, Milky Way-mass galaxies drawn from the \eagle Recal-L025N0752 volume with size $25~\mpc^3$. Here, we omit the limit on the disc fraction and extend the simulation suite to a volume-limited sample of 25 such galaxies, which represent all `Milky Way-mass' haloes in the Recal model, defined by the halo mass range $11.85<\log(\mvir/\msun)<12.48$. In these simulations, only the immediate environment of the target galaxy is modelled at high resolution. We generate the initial conditions of the 15 additional haloes fully analogously to those of \citetalias{pfeffer18} (see that paper for details), such that the radius of the high-resolution region at $z=0$ is at least 600~proper~kpc (pkpc). Beyond that radius, the typical particle mass increases with distance and all mass is modelled as a collisionless fluid. This rather large size of the high-resolution region ensures that none of the simulated galaxies are contaminated by low-resolution particles and also leads to the inclusion of several `bonus' galaxies that are not satellites of the target galaxies, but are still uncontaminated. With a couple of exceptions (see \citetalias{pfeffer18}), these galaxies are of lower mass than the target galaxies, with $M_*=10^8$--$10^9~\msun$. In this paper, we focus on the formation and assembly history of Milky Way-like galaxies, and thus we restrict our analysis to the target galaxies of the zoom-in simulations.

To within 4~per~cent, the particle masses used in the simulations are $m_{\rm g}=2.25\times10^5~\msun$ for the gas particles and $m_{\rm dm}=1.2\times10^6~\msun$ for the high-resolution dark matter particles. The gravitational softening length is 1.33~comoving~kpc for $z>2.8$, or 4~per cent of the mean particle separation length, and 0.35~pkpc for $z<2.8$. The simulations therefore marginally resolve the Jeans length at the star formation threshold. The SPH smoothing length decreases with the local density and has a lower limit of 10~per~cent of the gravitational softening length. For these choices of mass and spatial resolution, the simulations resolve (satellite) galaxies with stellar masses $M_*\geq2\times10^7~\msun$ (corresponding to $z=0$ halo masses $\mvir\ga2\times10^{10}~\msun$, see \citealt{moster13}) with at least $10^2$ baryonic particles and $10^4$ dark matter particles. These masses are similar to those of the lowest-mass dwarf galaxies in the Local Group hosting GCs \citep[e.g.][]{georgiev10,kruijssen12b,larsen12,larsen14,larsen18}, indicating that the simulations are able to capture the formation of even the most metal-poor GCs in their host galaxies. We save 29 simulation snapshots in the redshift range $z=0$--$20$, which is identical to the \eagle runs. Finally, the haloes (galaxies) are identified using a friends-of-friends algorithm \citep{davis85} and {\sc SUBFIND} \citep{springel01b,dolag09}, using the method described in \citet{schaye15}, with the generation of merger trees as described in \citetalias{pfeffer18}.

\begin{table}
  \caption{Properties of the 25~Milky Way-mass, $L^*$ galaxies at $z=0$ in the cosmological zoom-in simulations considered in this work. From left to right, the columns show: simulation ID; log halo mass; log stellar mass; log star-forming gas mass; log non-star-forming gas mass; SFR averaged over the last 300~Myr in $\msun~\yr^{-1}$. All masses are in units of $\msun$ and the baryonic galaxy properties are measured within 30~pkpc. To give an indication of the typical values and dynamic ranges, the final three rows list the median, the interquartile range, and the total range (i.e.~$\max-\min$) of each column.} 
\label{tab:sims}
  \begin{tabular}{lccccc}
   \hline
   Name & $\log \mvir$ & $\log M_*$ & $\log M_{\rm SF}$ & $\log M_{\rm NSF}$ & $\sfr$ \\ 
   \hline
   MW00 & $11.95$ & $10.28$ & $9.39$ & $10.34$ & $0.63$ \\ 
   MW01 & $12.12$ & $10.38$ & $9.55$ & $11.05$ & $0.93$ \\ 
   MW02 & $12.29$ & $10.56$ & $9.82$ & $11.19$ & $1.65$ \\ 
   MW03 & $12.17$ & $10.42$ & $9.82$ & $11.04$ & $1.72$ \\ 
   MW04 & $12.02$ & $10.11$ & $9.29$ & $10.84$ & $0.35$ \\ 
   MW05 & $12.07$ & $10.12$ & $8.51$ & $10.32$ & $0.08$ \\ 
   MW06 & $11.96$ & $10.31$ & $9.89$ & $10.86$ & $2.44$ \\ 
   MW07 & $11.86$ & $10.16$ & $9.81$ & $10.86$ & $1.52$ \\ 
   MW08 & $11.87$ & $10.12$ & $9.34$ & $10.78$ & $1.08$ \\ 
   MW09 & $11.87$ & $10.16$ & $9.62$ & $10.52$ & $1.36$ \\ 
   MW10 & $12.36$ & $10.48$ & $9.47$ & $11.38$ & $0.93$ \\ 
   MW11 & $12.15$ & $10.06$ & $9.26$ & $11.04$ & $0.63$ \\ 
   MW12 & $12.34$ & $10.44$ & $9.69$ & $11.36$ & $1.13$ \\ 
   MW13 & $12.38$ & $10.37$ & $8.81$ & $11.22$ & $0.49$ \\ 
   MW14 & $12.34$ & $10.59$ & $9.70$ & $11.45$ & $1.91$ \\ 
   MW15 & $12.16$ & $10.15$ & $9.78$ & $10.99$ & $2.15$ \\ 
   MW16 & $12.32$ & $10.54$ & $7.16$ & $10.40$ & $0.00$ \\ 
   MW17 & $12.29$ & $10.49$ & $9.42$ & $10.97$ & $0.57$ \\ 
   MW18 & $12.25$ & $10.00$ & $9.40$ & $10.75$ & $1.31$ \\ 
   MW19 & $12.20$ & $9.93$ & $9.64$ & $11.20$ & $1.13$ \\ 
   MW20 & $11.97$ & $10.10$ & $9.60$ & $11.04$ & $0.71$ \\ 
   MW21 & $12.12$ & $10.03$ & $9.25$ & $10.60$ & $0.50$ \\ 
   MW22 & $12.15$ & $10.43$ & $7.55$ & $10.33$ & $0.00$ \\ 
   MW23 & $12.19$ & $10.53$ & $9.97$ & $11.24$ & $3.30$ \\ 
   MW24 & $12.06$ & $10.29$ & $9.40$ & $10.74$ & $0.65$ \\ 
   \hline
   Median & $12.15$ & $10.29$ & $9.47$ & $10.97$ & $0.93$ \\ 
   IQR & $0.27$ & $0.32$ & $0.41$ & $0.45$ & $0.95$ \\ 
   Range & $0.52$ & $0.66$ & $2.81$ & $1.13$ & $3.30$ \\ 
   \hline
  \end{tabular} 
\end{table}
We summarise the properties of all 25 simulated haloes in \autoref{tab:sims}. The first 10 of these (MW00--MW09) are the simulations presented in \citetalias{pfeffer18}. The halo masses of the full set span a factor of 3.3 (which is 2.7 for MW00--MW09), whereas the stellar masses span a factor of 4.6 (which is 3.2 for MW00--MW09). In terms of total masses, the addition of 15 further haloes thus does not significantly increase the variety of systems covered by the \emosaics MW suite. However, we will show in Section~\ref{sec:hist} that the variety of galaxy formation and assembly histories is increased significantly, with a wider range of $z=0$ SFRs (see the final column of \autoref{tab:sims}) and merger tree topologies. Most importantly for the goal of this paper, the expansion of the sample to 25 haloes improves the statistics of the sample sufficiently to correlate GC-related and galaxy formation-related quantities, thus tracing galaxy formation and assembly histories using the GC age-metallicity distribution.

\begin{figure*}
\includegraphics[width=\hsize]{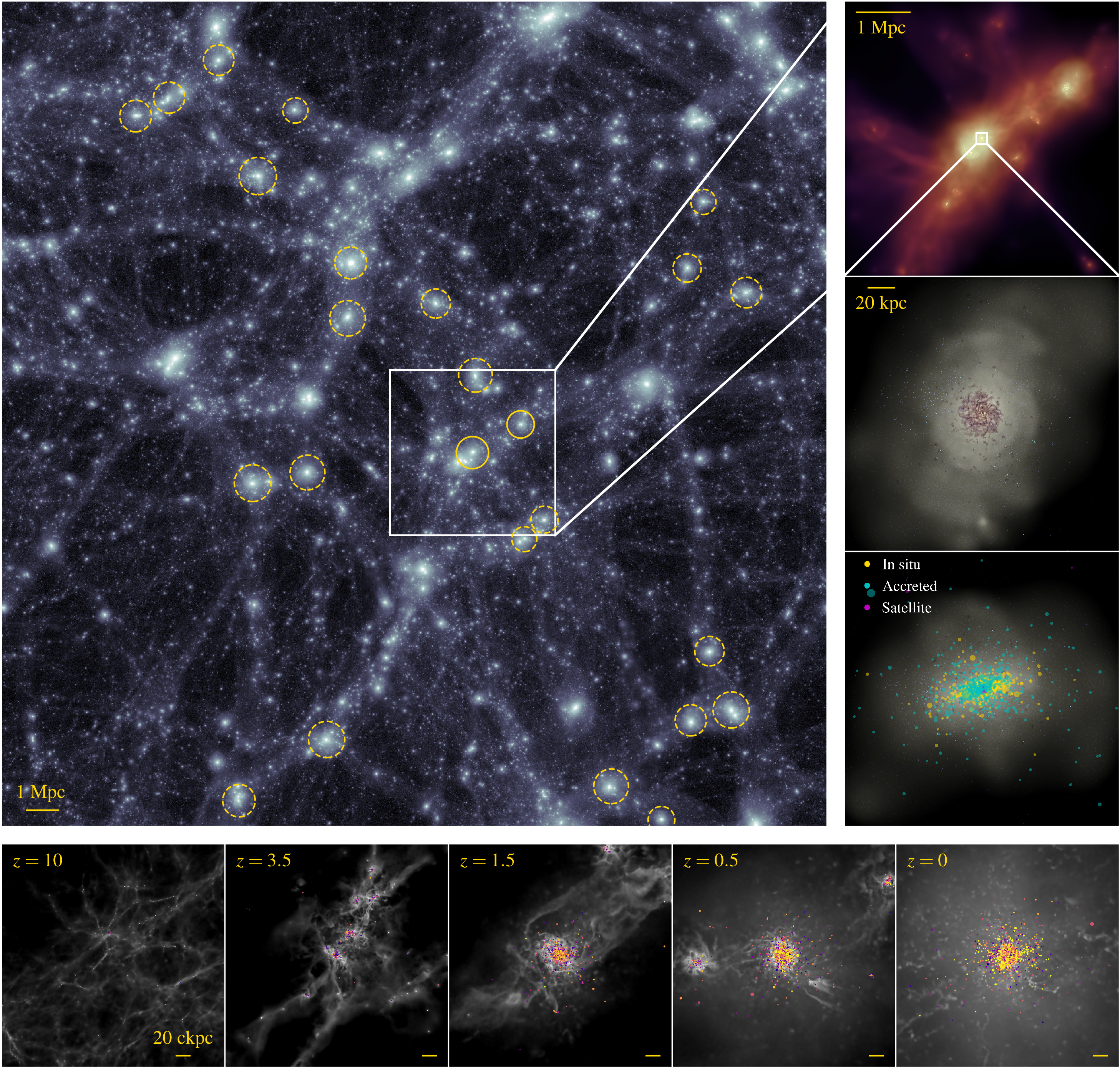}%
\caption{
\label{fig:sims}
Visualisation of the E-MOSAICS simulations as in Figure~1 of \citetalias{pfeffer18}, expanded to highlight all 25 Milky Way-mass ($L^\star$) galaxies from \autoref{tab:sims}. The main panel shows the dark matter distribution of the \eagle Recal-L025N0752 simulation at $z=0$, with the resimulated haloes marked with yellow circles, of which the sizes indicate the virial radii. Solid circles denote the two galaxies featured in the top right panel, which shows gas density coloured by temperature, with red and white hues corresponding to $T=10^5~{\rm K}$ and $T=10^6~{\rm K}$, respectively. The two middle right-hand panels zoom in on one galaxy (MW23) and show mock optical images, with the bottom panel including massive ($M\geq5\times10^4~\msun$) stellar clusters as dots, coloured by their origin -- those that currently reside in a satellite are shown in magenta, those that formed in a satellite galaxy and were subsequently accreted are shown in cyan, whereas those formed in the main progenitor are shown in yellow. The bottom row visualises the assembly of the same galaxy and its cluster population, from $z=10$ to $z=0$, with gas shown in grey scale and coloured dots again representing the massive star clusters, this time coloured by their metallicities ($-2.5\leq\feh\leq0.5$).
}
\end{figure*}
\autoref{fig:sims} shows the locations (yellow circles, with radii indicating the virial radii) of the 25~haloes in their parent \eagle Recal-L025N0752 volume. This extends Figure~1 of \citetalias{pfeffer18} and visualises that the grown sample draws haloes from a broader range of cosmic environments than our previous set of 10 haloes, and is therefore likely to span to a wider variety of assembly histories. This is an important improvement when aiming to identify systematic trends between the properties of the GC population and the host galaxy assembly history, as we will turn to in Sections~\ref{sec:hist} and~\ref{sec:recon}. In the smaller panels, \autoref{fig:sims} also illustrates that the GC populations at $z=0$ originate from a varied range of formation environments. Some of the GCs formed in-situ, during the build-up of the main progenitor, whereas others formed ex-situ, in lower-mass satellites that have since been accreted. The resulting variety of assembly histories of the GC population is expected to manifest itself in age-metallicity space, because the GCs' chemical composition and the timing of their formation must depend on the formation, enrichment, and assembly histories of their natal galaxies.

\subsection{Validation by comparison to the Galactic GC population} \label{sec:valid}
\begin{figure*}
\includegraphics[width=0.92\hsize]{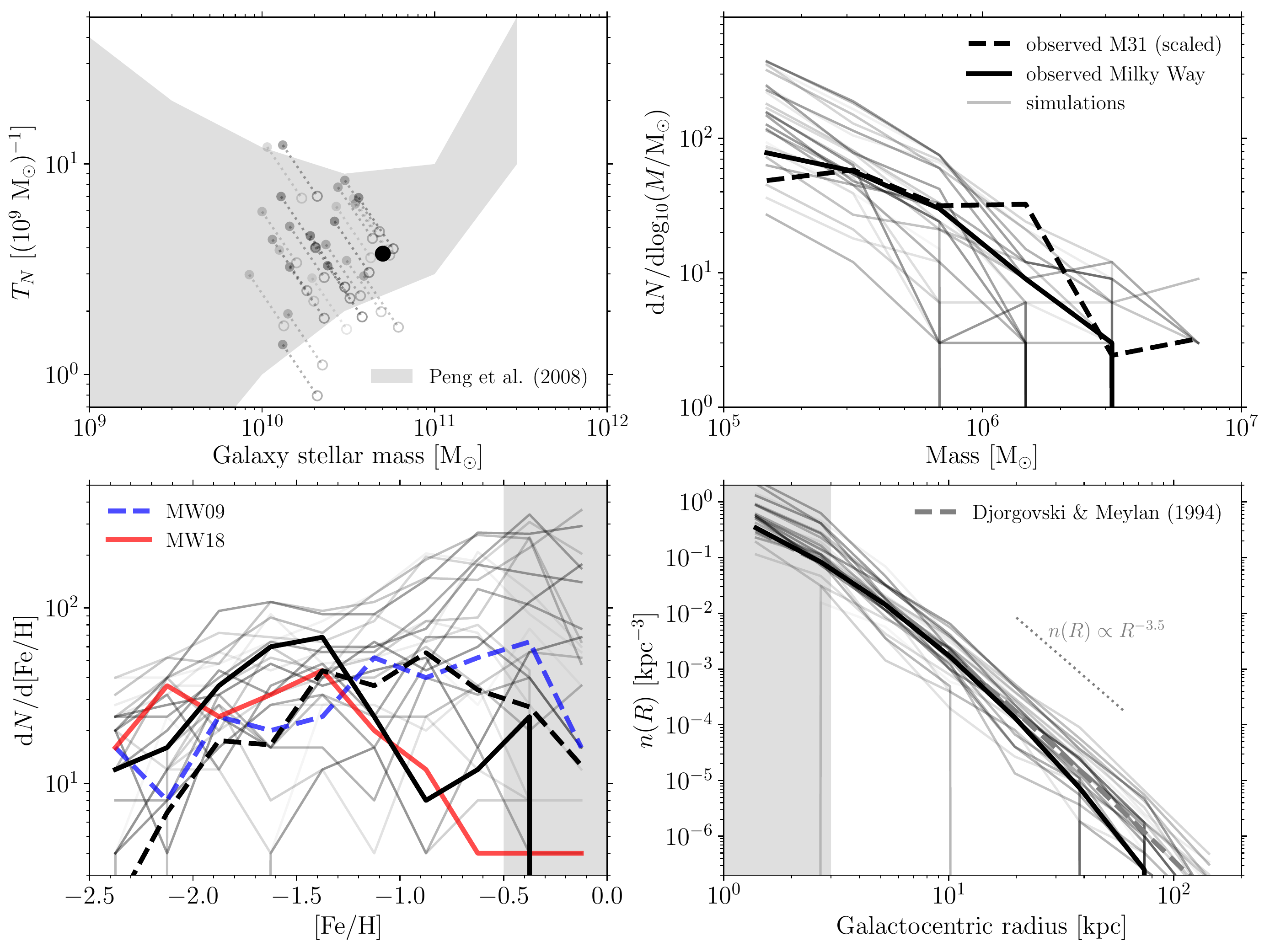}%
\caption{
\label{fig:compobs}
Comparison of the properties of the simulated GC populations to the Galactic GC system. The panels show the specific frequency ($T_N=\ngc/M_*$) as a function of galaxy stellar mass (top left), the upper ($M>10^5~\msun$) end of the $z=0$ GC mass function (top right), the GC metallicity distribution (bottom left), and the GC radial density profile (bottom right). In all panels, the solid black line or symbol represents the observed GC population of the Milky Way (taken from \citealt[2010 edition]{harris96}, assuming a constant $M/L_V=2~\msun~{\rm L}_\odot^{-1}$), the dashed black line indicates the observed GC population of M31 (taken from \citealt{caldwell11}, scaled to the same number of GCs as the in the Milky Way sample), and the grey lines show the simulations, with higher simulation IDs having lighter shades. In the top left panel, the grey-shaded region shows the range of specific frequencies spanned by galaxies in the Virgo Cluster \citep{peng08}. The dotted lines connect the true modelled values of $T_N$ to ones that are corrected for the underproduction of stars in \eagle and the under-destruction of GCs at the considered masses ($M>10^5~\msun$) in \emosaics, which move the points to the right by a factor of 2 and downward by a factor of 1.75 (see Section~\ref{sec:gcmetrics}), respectively. In the bottom-left panel, coloured lines highlight two simulations of which the metallicity distributions resembles those of GCs in M31 (blue dashed) and the Milky Way (red solid), and the grey area indicates the metallicity range for which GCs are excluded in the other panels -- the analysis of this work is restricted to $-2.5<\feh<-0.5$. In the bottom right panel, the grey dashed line shows the density profile derived for the Galactic GC system by \citet{djorgovski94}, and the grey area indicates the radius range for which GCs are excluded in the other panels (see the text).
}
\end{figure*}
In several published (\citetalias{pfeffer18}; \citealt{kruijssen18c,usher18,reinacampos18,hughes19}), submitted \citep[e.g.][]{reinacampos19,pfeffer19}, and upcoming papers, we are carrying out a detailed comparison of the modelled GC populations at $z=0$ to observed GC populations in the local Universe. Here, we carry out a brief comparison for a number of key observables to demonstrate that the \emosaics simulations successfully reproduce a variety of properties of GC populations, whereas disagreement remains for some observables. Broadly speaking, the results can be understood in terms of two common results. Firstly, the properties of GC populations are sensitive to differences in galaxy assembly histories. This implies a significant variation of the observables between galaxies and also means that reproducing the Milky Way is not necessarily a goal in itself, because it represents just a single galaxy with a single assembly history. Secondly, the \emosaics simulations underestimate GC disruption, because the cold ISM is not resolved (see fig.~17 of \citetalias{pfeffer18} and the accompanying discussion). This results in an under-destruction of GCs, which most strongly affects metal-rich ones born at late cosmic times (typically $z<1$), because these spend their entire lives in their natal, disruptive environments (see Appendix~\ref{sec:appfeh}). By contrast, metal-poor, old GCs are less affected by the under-destruction, because their natal galaxies are generally being tidally stripped on a short time-scale.

We limit the influence of GC under-destruction throughout this paper by only considering massive ($M>10^5~\msun$) GCs, which undergo little dynamical mass loss anyway, and by excluding GCs with high metallicities ($\feh>-0.5$), which matches the Galactic GC population for which ages have been measured (see Section~\ref{sec:agez}). In the the comparison of this section, we additionally exclude GCs associated with recent star formation in the disc, i.e.~at small galactocentric radii ($R<3~\kpc$, which are generally hard to detect in observations due to being projected onto the centres of galaxies) and young ages ($\tau<8~\gyr$, below which the Milky Way hosts few massive GCs). These cuts remove the clusters from the simulations that are likely to have been disrupted by a cold ISM had it been included. In the longer term, this shortcoming of the simulations will be addressed by increasing the numerical resolution and improving the ISM model (also see \citetalias{pfeffer18} for a discussion). Appendix~\ref{sec:appfeh} demonstrates that any remaining under-destruction of (mostly metal-rich) GCs in the final sample has a small effect on the results presented in this work.

\autoref{fig:compobs} shows a comparison between the properties of the GC populations simulated in \emosaics and the observed GC population of the Milky Way. In the top-left panel, the specific frequency (here expressed as the number of GCs per unit stellar mass $T_N$, see e.g.~\citealt{harris91} and \citealt{peng08}) of the Galactic GC population falls in the range of specific frequencies observed at the same galaxy mass in the Virgo Cluster \citep{peng08}. The \emosaics galaxies span the same range and on average fall within a factor of 2 of the Galactic value of $T_N$. The specific frequencies in the simulations are obtained by counting the number of GCs with masses $M>10^5~\msun$ and doubling that number to account for lower-mass GCs. This mirrors the common practice in observational studies \citep[e.g.][]{peng08}. In \autoref{fig:compobs}, the dotted lines and open symbols indicate the effect of correcting for the under-destruction of GCs in \emosaics (see Section~\ref{sec:gcmetrics}), as well as the slight underproduction of stars in Milky Way-mass galaxies in the \eagle model (see figs.~4 and~8 of \citealt{schaye15}). Making these corrections changes the agreement with the Galactic GC population quantitatively, but not qualitatively, due to the considerable spread in specific frequencies found in \emosaics. This spread results from differences in the formation and assembly histories of the host galaxies, suggesting that the observed spread in $T_N$ has the same origin.

The top-right panel of \autoref{fig:compobs} shows the GC mass function for the mass range ($M>10^5~\msun$) considered in this work. The mass functions produced by the simulations follow the same shape as that of the Galactic GC population, with similar slopes, curvature, and maximum mass scales. The GC mass function of M31 also falls within the range spanned by the simulations. As shown by fig.~16 of \citetalias{pfeffer18}, this agreement is driven by a combination of our environmentally-dependent model for the maximum cluster mass at formation \citep{reinacampos17} and cluster destruction by dynamical friction. Which of these two mechanisms dominates the maximum GC mass varies from galaxy to galaxy. The vertical scatter between the simulated GC mass functions is a direct result of the range of specific frequencies in the top-left panel and thus reflects differences in host galaxy assembly history. Finally, the shown GC mass functions only deviate from that of Galactic GCs at masses $M\la10^5~\msun$, due to the underestimated disruption rate in \emosaics. This is visible already in the lowest-mass bin shown here, where the black line in \autoref{fig:compobs} curves down more strongly than the grey lines. This trend continues at lower masses (see fig.~17 of \citetalias{pfeffer18}) and motivates our choice of GC mass range. For the GC mass range considered, which matches the masses typically accessible in extragalactic studies \citep[e.g.][]{jordan07}, the observations and simulations agree.

The bottom-left panel of \autoref{fig:compobs} shows the metallicity ($\feh$) distribution of the simulated GC populations, as well as those of the Milky Way and M31. Again, GC under-disruption is responsible for the relative excess of GCs at high metallicities in the simulations (see Appendix~\ref{sec:appfeh}). However, the metallicity distributions at $\feh<-1.0$ turn out to be reasonably consistent with observations. While it is often assumed that GC metallicity distributions are bimodal (often based on optical colour bimodallity), direct spectroscopic metallicity measurements have not corroborated this universal picture, with many early type galaxies hosting unimodal GC metallicity distributions \citep[see e.g.][as well as the difference between the Milky Way and M31]{usher12}. This large variety of metallicity distributions is an important prediction of \emosaics and reaffirms that exactly reproducing the metallicity distribution of Galactic GCs is not a goal in itself. Future observations will be able to explicitly test this prediction.

Specifically, we find that $\sim55\%$ of the GC populations exhibit bimodal metallicity distributions (Pfeffer et al.~in prep.), whereas \citet{usher12} find this for 7 out of 11 GC populations in early-type galaxies (or $64\%$). \autoref{fig:compobs} contains both examples, as Galactic GCs follow a bimodal distribution, whereas those in M31 follow a unimodal one. In the \emosaics simulations, the variation in metallicity distributions again arises due to differences in host galaxy assembly history, which is an interesting feature in the context of this work. Note that our further analysis is restricted to $-2.5<\feh<-0.5$, omitting the grey-shaded area in the bottom-left panel of \autoref{fig:compobs}. This minimises the effects of the under-disruption of metal-rich GCs.

Across the full metallicity range, the $\feh$ distributions of GCs observed in the Milky Way and M31 fall largely within the range spanned by the simulations. The metallicity distribution of GCs in M31 agrees well with the simulated ones from \emosaics and closely follows that of MW09, which is highlighted in \autoref{fig:compobs}. The metallicity distribution of Galactic GCs is similar to that of MW18, but overall it is more metal-poor than most of the simulated ones. In part, this may result from the \eagle galaxy formation model, which overestimates the metallicities of galaxies with masses $M_*<10^9~\msun$, where metal-poor GCs are expected to have formed, by $0.1{-}0.5~\dex$ \citep[fig.~13]{schaye15}. Another possible explanation would be that the initial GC mass function in low-mass galaxies (forming low-metallicity stars and GCs) differs from that in more massive galaxies, such that more stars are born in massive clusters \citep[cf.][]{larsen12,larsen14}. There are dynamical reasons why this may happen \citep{trujillogomez19}, which presents a promising avenue for future work. Most importantly in the context of this work, the results of Section~\ref{sec:hist} show that the metrics used to characterise the GC metallicity distribution are poor tracers of the host galaxy formation and assembly history. The slope of the GC age-metallicity distribution is a considerably better probe, and is also less sensitive to which GCs are (or are not) disrupted than the absolute metallicity distribution itself. The presented results are therefore unlikely to be strongly affected by any biases in the GC metallicity distribution.

Finally, the bottom-right panel of \autoref{fig:compobs} shows the distribution of simulated and Galactic GCs as a function of galactocentric radius, together with the distribution observed in the Milky Way. For reference, the figure includes lines indicating the canonical slope of $-3.5$ and the density profile $n(R)\propto(1+R/R_{\rm c})^{-4}$ with $R_{\rm c}=2~\kpc$ that have been derived for the Galactic GC system \citep{djorgovski94}. As for the GC mass functions, the simulations and observations are in good agreement, showing the same global shape and slope. This agreement is reflected also by the half-number (median) radii of the GC populations. The observed median radius of Galactic GCs (black line) is $5.9~\kpc$, whereas across the 25 simulated galaxies (grey lines) we obtain median radii of $7.8\pm5.4~\kpc$ (mean $\pm$ standard deviation), which are in excellent agreement.\footnote{When calculating these numbers, we have used identical selection functions, considering GCs with $M>10^5~\msun$, $-2.5<\feh<-0.5$, and an extended galactocentric radius range of $1<R/\kpc<200$.}

As before, the vertical scatter of the radial density profiles is driven by the variations in specific frequency shown in the top-left panel of \autoref{fig:compobs} and thus traces differences in host galaxy assembly history. While reproducing the spatial distribution of Galactic GCs is thus not difficult (all that is required is a simulated galaxy with the right assembly history), it is interesting to note that \emosaics includes a number of galaxies with a spatial distribution of GCs nearly identical to that of the Milky Way. This detailed agreement is not necessarily meaningful. For instance, the steepening of the observed distribution at $R>20~\kpc$ is thought to have been shaped by the accretion of a massive satellite \citep{deason13}, which may correspond to the Kraken \citep{kruijssen18c} or Sausage/Gaia-Enceladus \citep{myeong18,helmi18} accretion events. Importantly, it shows that detailed features in the shape of the radial number profile of the GC population are driven by stochastic events such as satellite accretion -- the relevant point of comparison here is the global shape and slope of the distributions. As \autoref{fig:compobs} shows, \emosaics reproduces the global properties of the spatial distribution of Galactic GCs.

In addition to the four observables shown in \autoref{fig:compobs}, the \emosaics simulations quantitatively reproduce the `blue tilt' of metal-poor GCs \citep{usher18}, the existence of GCs associated with fossil stellar streams \citep{hughes19}, and the age distributions of GCs \citep{reinacampos19}. In the remainder of this work, we focus on the distribution of GCs in age-metallicity space, because of its great diagnostic power for constraining the formation and assembly history of the host galaxy. As demonstrated in \citet{kruijssen18c}, the variety of GC age-metallicity distributions produced by the \emosaics simulations encompass the observed distribution of GCs in the Milky Way. Detailed comparisons to additional observables will be the topic of future work. In summary, the populations of massive ($M>10^5~\msun$) GCs of low-to-intermediate metallicity ($-2.5<\feh<-0.5$) produced in the \emosaics simulations have the right numbers, masses, spatial distributions, ages, and relations between these in comparison to observations.

\section{Characterising a great variety of GC age-metallicity distributions} \label{sec:agez}
In this section, we present the variety of modelled GC age-metallicity distributions obtained from the simulations, which are then quantitatively characterised through 13 different parameters. In Section~\ref{sec:hist}, these parameters will be correlated with a different set of quantities describing the galaxy formation and assembly histories, with the goal of showing how the GC age-metallicity distribution traces the galaxy formation process.

\subsection{Variety of GC age-metallicity distributions} \label{sec:gcvariety}
\begin{figure*}
\includegraphics[width=\hsize]{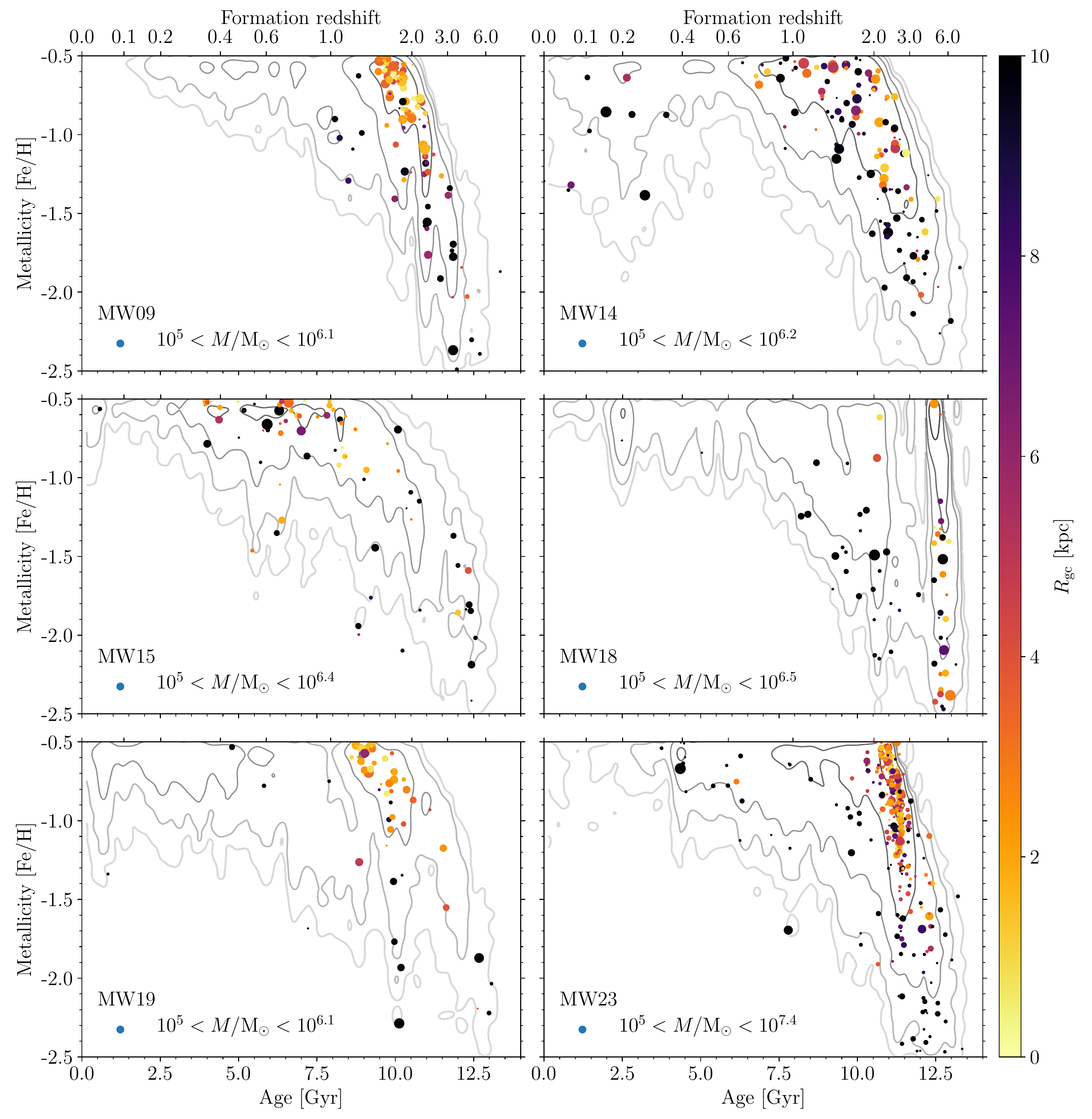}%
\caption{
\label{fig:agez}
Age-metallicity distributions of GCs (coloured dots) for six of our Milky Way-mass simulations at $z=0$ (MW09, MW14, MW15, MW18, MW19, MW23; indicated in the bottom-left corner of each panel). The symbol colour indicates the galactocentric radius according to the colour bar and the symbol size reflects the logarithm of the GC mass across the range indicated in each panel. For reference, the distribution of field stars is shown by the grey contours, indicating mass densities of $\{1,1.5,2,2.5,3\}\times10^7~\msun~\gyr^{-1}~{\rm dex}^{-1}$ from light to dark. This small subset of 6 out of the 25 simulations presented in this paper already shows a remarkable variety of age-metallicity distributions, which arises due to differences in the formation and assembly histories of the host galaxies.
}
\end{figure*}
\autoref{fig:agez} shows the GC age-metallicity distributions for a subset of six galaxies from our suite of Milky Way-mass simulations. These were chosen to illustrate the great variety of age-metallicity distributions and largely encompass the variation seen among the full sample of 25 simulations. The GC samples are limited to masses of $M>10^5~\msun$ at $z=0$, so that they are unlikely to have been strongly affected by cluster disruption (see Section~\ref{sec:valid} and \citetalias{pfeffer18}), and to metallicities $-2.5<\feh<-0.5$, to mimic the range of metallicities for which fairly comprehensive observational measurements exist of the ages of Milky Way GCs \citep[e.g.][]{forbes10,dotter10,dotter11,vandenberg13}. The simulated GC populations have barely any GCs with $\feh<-2.5$. We colour the GCs by galactocentric radius to give an indication of their in-situ or ex-situ origin. For reference, \autoref{fig:agez} also includes the complete age-metallicity distribution of the field stars constituting the host galaxy as grey contours.

Inspection of \autoref{fig:agez} reveals a wide range of features immediately relevant to the link between GC formation and the formation and assembly history of the host galaxy.
\begin{enumerate}
\item
GCs trace the density peaks in the field star age-metallicity distribution. Without exception, the contours enclosing the highest densities of field stars in age-metallicity space are also associated with GCs. This is not necessarily surprising, because the GC populations simulated in \emosaics are a natural byproduct of the star formation process. The only case where one would plausibly expect a larger overdensity of GCs is MW18 (middle-right panel of \autoref{fig:agez}), which has a very high field star density at $\{\tau/\gyr,\feh\}=\{12.5,-0.75\}$ with only one associated GC. This overdensity of field stars marks a nuclear starburst that is accompanied by the formation of more than 100~GCs, all of which have been destroyed by tidal shocks and dynamical friction. It thus provides an example of how cluster disruption can erase part of the correlation between GCs and the field star population. None the less, \autoref{fig:agez} clearly shows that the distribution of GCs in age-metallicity space may be used to probe the formation and enrichment history of its host galaxy.
\item
The GCs span the full range of metallicities shown here ($-2.5<\feh<-0.5$), but generally gravitate towards old ($\tau>10~\gyr$) ages, even though nearly every galaxy has a population of younger GCs (with minimum GC ages ranging from $\tau_{\rm min}=1$--$8~\gyr$). In a few extreme cases, the GC population can be considerably younger. For instance, MW15 (middle-left panel of \autoref{fig:agez}) has a median age of $\widetilde{\tau}\approx8~\gyr$.
\item
Following on the previous two statements, the GC age-metallicity distributions show that most of the modelled galaxies undergo an initial, rapid phase of star formation and metal enrichment, which is usually accompanied by the formation of most of the GC population. This initial phase of star and cluster formation generally takes place during the first few $\gyr$, elevating the metallicity from $\feh\approx-2.5$ to nearly solar. Such episodes of intense star formation are accompanied by efficient cluster formation up to high maximum mass-scales (see Figures~5--8 of \citetalias{pfeffer18}), implying that this is the dominant epoch of in-situ GC formation within our sample of Milky Way-like galaxies.
\item
Galaxy mass and metallicity are observed to follow a positive correlation, of which the normalisation slowly increases with cosmic time \citep[e.g.][]{tremonti04,erb06,mannucci09}. This relation is mirrored by the GCs. At fixed age, GCs at the low end of the metallicity range formed in low-mass galaxies and are thus typically accreted (signified by large galactocentric radii in \autoref{fig:agez}), whereas those at the high end of the metallicity range formed in-situ (evidenced by their small galactocentric radii). This results in `satellite branches' of lower-metallicity GCs (dark colours in \autoref{fig:agez}) emerging from the `main branch' of the rapidly-enriched, in-situ GC population (dominated by light colours in \autoref{fig:agez}). These branches are the trails of disrupted satellites and are most evident for MW09, MW18, MW19, and MW23 (top-left, middle-right, bottom-left, and bottom-right panels of \autoref{fig:agez}, respectively).
\item
The wide range of progenitor masses generates a wide range of GC metallicities at any given age. As a result, even the metal-rich host galaxies considered here can host significant populations of low-metallicity GCs with young ages of a few $\gyr$ (see e.g.~MW14 in the top-right panel of \autoref{fig:agez}). Most often, this is caused by accretion events of lower-mass satellite galaxies, but a population of GCs with unusually-low metallicities can also signify the gradual tidal stripping of surviving satellites or the accretion of low-metallicity gas. In fact, the formation of low-metallicity GCs in MW14 at $\tau\approx2.5~\gyr$ and $\feh<-1$ is enabled by two major mergers between $z=0.3$ and $z=0$, which drive low-metallicity gas accretion and trigger a young generation of stars and clusters. This is illustrated by the fact that the field star distribution also shows a clear density enhancement at the same ages and metallicities as the GCs. MW14 illustrates that the GC age-metallicity distribution can be reduced to a combination of simple building blocks in the form of accreted galaxies, but at the same time can exhibit significant deviations from the enrichment histories of these progenitors due to the complexity of the accretion processes.
\end{enumerate}

Together, the above set of features gives rise to a wide variety of GC age-metallicity distributions with different morphologies. Some follow a wide band of metallicities that gradually increase with time (MW14 and MW15), whereas others experienced rapid star formation and metal enrichment, leading to a steep and narrow sequence of GCs in the age metallicity plane. If these galaxies accrete satellites with GC populations, the age metallicity distributions become forked, with the ex-situ low-metallicity satellite branch being shallower than the in-situ high-metallicity GC main branch (MW09, MW18, and MW23). Depending on the satellites' enrichment histories and the number of GCs brought in, the low-metallicity branch may be so prominent that an inverse fork appears, in which a satellite branch and the main branch are initially independent and merge towards lower redshifts (MW19, where the satellite branch merges into the main branch at $z<0.5$).

Observationally, the Milky Way is the only massive ($L^*$) galaxy with a GC population for which the age-metallicity distribution of GCs has been measured \citep[see e.g.][]{marinfranch09,forbes10,dotter10,dotter11,vandenberg13}. Having such a limited sample size has led to the suggestion in the literature that reproducing the Galactic GC age-metallicity distribution should be a goal for numerical models of GC populations, because it may benchmark the GC formation and disruption models used \citep[see e.g.][]{renaud17,choksi18}. The great variety of GC age-metallicity distributions obtained in our sample of Milky Way-mass simulations using a single GC formation and disruption model demonstrates that the contrary may be true -- given a suite of simulations with a sufficiently wide range of formation and assembly histories, there should exist a simulation that reproduces the GC age-metallicity distribution of the Milky Way. This is unlikely to be a suitable probe of the mechanisms governing GC formation and evolution. Instead, we find that the variety of GC age-metallicity distributions predominantly mirrors the underlying variety of host galaxy formation and assembly histories, suggesting that the distribution may be used to constrain the formation and assembly history of the host galaxy. 

These results further imply that simulations of single galaxies and their GC populations have limited predictive power. During the analysis of the \emosaics simulations, it became evident that the expansion of the suite from 10 galaxies in \citetalias{pfeffer18} to 25 galaxies in the present work is necessary to span a sufficient variety of host galaxy formation histories. Even this volume-limited sample of galaxies from the $25~\mpc$ \eagle Recal-L025N0752 volume is still somewhat restricted, but we demonstrate in Section~\ref{sec:hist} that the galaxy properties have a sufficiently large dynamic range to obtain statistically significant and physically meaningful correlations with the properties of the GC population.

Comparing the age-metallicity distributions obtained in \emosaics to that of Galactic GCs \citep{forbes10,dotter10,dotter11,vandenberg13}, we recognise certain qualitative similarities. The observed distribution of Galactic GCs is bifurcated like MW09, MW18, and MW23. In the context of the \emosaics models, its main branch is largely attributed to in-situ GC formation, whereas its satellite branch is almost exclusively populated by GCs from accreted (dwarf) galaxies. This corroborates the previous interpretations by \citet{forbes10} and \citet{leaman13} (albeit only partially, see \citealt{kruijssen18c}) with a cosmologically-motivated galaxy formation model. In the next sections, we further quantify the link between the GC age-metallicity distribution and the host galaxy's formation, enrichment, and assembly history, with the goal of enabling a deeper interpretation of the observed properties of (Galactic) GCs.

The close correspondence between the GC age-metallicity distribution and the formation, enrichment, and assembly history of the host galaxy's progenitors is highly promising, especially in combination with the predicted wide variety of distributions, because it implies a relevant diagnostic power. In particular, the spread and slope of the age-metallicity distribution seem to be indicators of the richness of the host galaxy's formation and assembly history, as well as of how quickly the host galaxy formed and enriched. We quantify these ideas further in Section~\ref{sec:hist}. Simply by eye, it is also possible to identify features in age-metallicity space that are likely to trace accreted galaxies. This may be used to reconstruct a galaxy's assembly history through its merger tree, which we demonstrate in Section~\ref{sec:recon}.

\subsection{Quantitative characterisation of GC age-metallicity distributions} \label{sec:gcmetrics}
\begin{figure*}
\includegraphics[width=0.88\hsize]{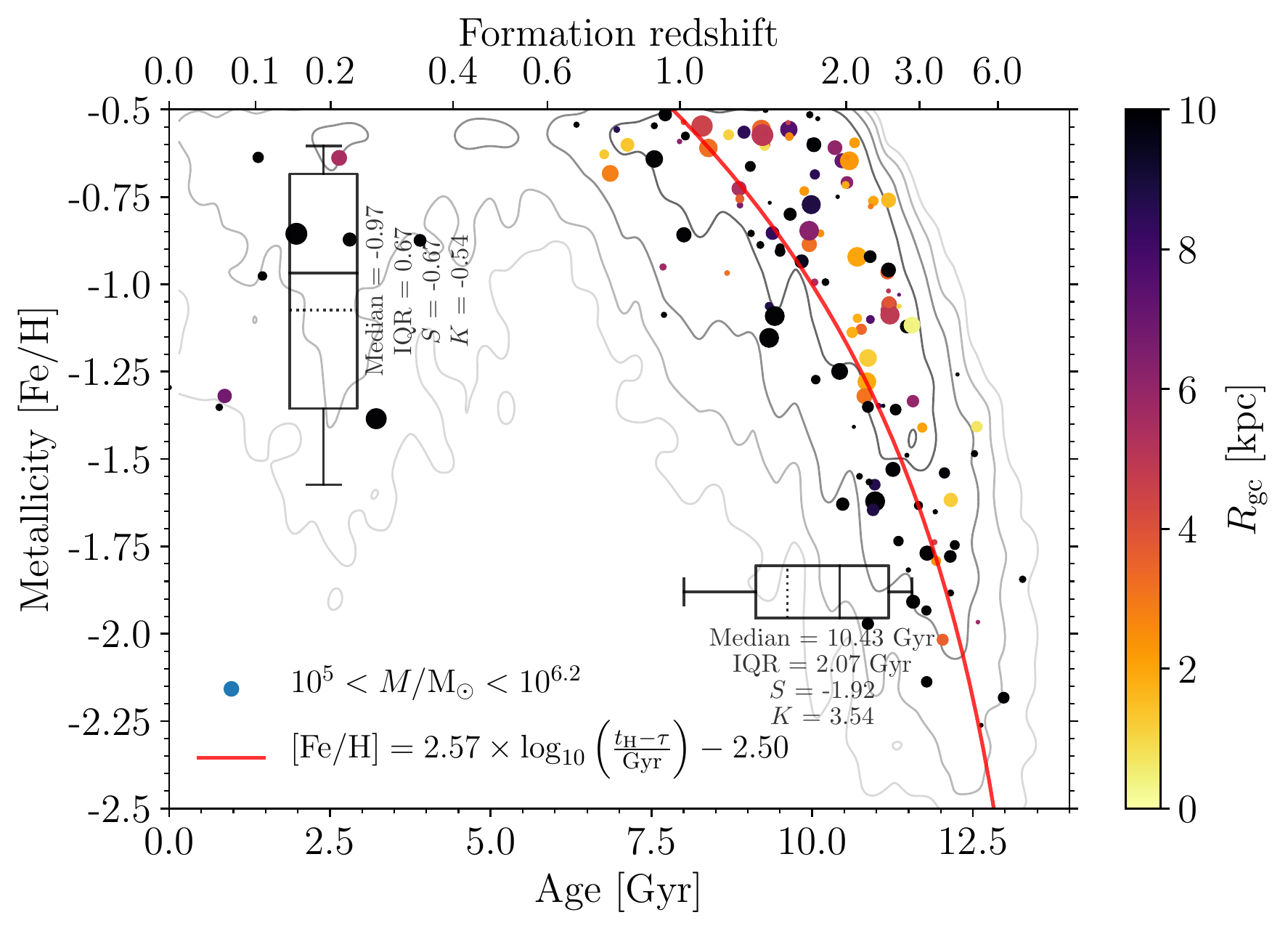}%
\caption{
\label{fig:agezmetrics}
Age-metallicity distribution of GCs (coloured dots) for simulation MW14 at $z=0$, with symbol colours and sizes as in \autoref{fig:agez}. As before, the distribution of field stars is shown by the grey contours. We now include several metrics used to characterise the age-metallicity distribution. The distributions in both the age and metallicity dimensions are represented by two box plots, with quoted median, interquartile range, skewness, and kurtosis. In the box plots, the solid line indicates the median, and the dotted line marks the mean. The interquartile ranges can be combined to describe the two-dimensional distribution of GCs in the age-metallicity plane -- their product represents the overall spread, whereas their ratio reflects the aspect ratio of the distribution. In addition, we fit the function of equations~(\ref{eq:fit}) and~(\ref{eq:fit2}) to the data points (red line), with the best-fitting form indicated in the legend. The slope of this function ($\dfehdt$) provides the nominal metal enrichment rate traced by the GCs, whereas its intercept ($\feh_0$) represents the `initial' GC metallicity at $1~\gyr$ after the Big Bang. These quantitative metrics are listed for all 25 simulations in Appendix~\ref{sec:appmetrics_gc}, together with the number of GCs shown, i.e.~those with present-day masses $M>10^5~\msun$ and metallicities $-2.5<\feh<-0.5$.
}
\end{figure*}

We expand the above qualitative interpretation by quantitatively characterising the GC age-metallicity distributions of the 25 simulations using a variety of relevant quantities. In Section~\ref{sec:hist}, these will be correlated with a second set of quantities describing the galaxy formation and assembly histories. The quantities are illustrated in \autoref{fig:agezmetrics}, which repeats the age-metallicity distribution of MW14 from the top-right panel of \autoref{fig:agez} with the inclusion of our quantitative metrics, visualised by box plots to represent the one-dimensional distributions of the data on each axis and a red line indicating the best-fitting form of equation~(\ref{eq:fit}) below. The full set of metrics is listed in Appendix~\ref{sec:appmetrics_gc} for all simulations.

Firstly, the variety of GC age-metallicity distributions in \autoref{fig:agez} can be captured by considering the various moments of the distributions along each of the axes. These are the median age $\widetilde{\tau}$ and median metallicity $\widetilde{\feh}$, as well as the interquartile range (i.e.~the difference between the 75th and 25th percentiles) of the age $\iqr(\tau)$ and of the metallicity $\iqr(\feh)$. We prefer using the interquartile range over the standard deviation, because it better represents the (sometimes strongly) non-Gaussian and asymmetric nature of the age-metallicity distributions. The median GC age traces when most of the GC population formed, which likely correlates with the main episodes of progenitor galaxy growth. Similarly, the median GC metallicity probes the host galaxy mass at the time of formation through the galaxy mass-metallicity relation. The spreads around both of these quantities show how quickly the galaxy and its GC population formed and how broad the mass range of progenitor galaxies might have been, potentially providing a means of distinguishing between in-situ growth by star formation and ex-situ growth by satellite accretion. We note that absolute GC ages have $\sim\gyr$ uncertainties \citep[e.g.][]{marinfranch09}, implying that the median may represent an inaccurate metric when applied to observations. The interquartile range is not deleteriously affected, because the spread of GC ages is (almost) insensitive to their normalisation.

We also consider the higher-order moments of the one-dimensional distributions. The skewness $S$ of the GC ages $S(\tau)$ and metallicities $S(\feh)$ indicates the degree of asymmetry around the median, with negative (positive) values of $S$ representing extended tails towards small (large) values such that the highest point density of the distribution resides at large (small) values. As such, a negatively (positively) skewed distribution `leans' towards the right (left). The skewness may indicate whether the peak of GC formation or metal enrichment either commenced suddenly (e.g.~MW18 and MW23, with low $S$) or ended abruptly (e.g.~MW09, with high $S$), thus probing whether the host galaxy underwent sudden major star formation episodes (e.g.~due to major mergers) or (intermittent) quenching.

The kurtosis $K$ of the GC ages $K(\tau)$ and metallicities $K(\feh)$ indicates the relative importance of outliers on either side of the median. Because the kurtosis reflects the fourth moment of the distribution, the contributions from data points within a standard deviation of the median are negligible. High values of $K$ reflect numerous outliers relative to the width of the central peak (i.e.~enhanced tails), whereas low values indicate a rarity of outliers (i.e.~underrepresented tails). We specifically consider the {\it excess} kurtosis, which is obtained by subtracting 3 from the formal kurtosis, such that a Gaussian distribution has an excess kurtosis of $K=0$ and the sign of $K$ measures the presence of outliers relative to that of a Gaussian. \autoref{tab:gcmetrics} shows that a high kurtosis is often accompanied by a small interquartile range, indicating a strongly peaked distribution (low $\iqr$) with broad wings (high $K$). Conversely, a negative kurtosis typically signifies a broad distribution (high $\iqr$) without significant wings (low $K$). The kurtosis thus quantifies whether GC formation is greatly enhanced at certain ages or metallicities (resulting in a dispersion much smaller than the total range, hence $K>0$) or proceeds continuously over a broader range of ages and metallicities (resulting in a dispersion close to the total range, hence $K<0$). The kurtosis indicates the presence of single star (and GC) formation episodes ($K(\tau)\gg0$ as in e.g.~MW18, MW19, and MW23) or gradual accretion over the full epoch of GC formation ($K(\tau)<0$ as in e.g.~MW15), as well as gradual ($K(\feh)<0$ as in e.g.~MW14) or rapid ($K(\feh)\gg0$ as in e.g.~MW23) metal enrichment.

An obvious problem with the higher-order metrics $S$ and $K$ is that they are dominated by outliers and thus depend sensitively on the completeness of the GC sample. When a complete census of the GC population is ensured as in our simulations, these quantities may provide useful insight into the formation and assembly history of the host galaxy. However, incomplete (observed) GC samples are unlikely to have the same skewness or kurtosis as the complete parent GC population. We therefore caution against the observational application of any insights gleaned from these two metrics in Section~\ref{sec:hist} and provide alternative, more reliable metrics below.

Each of the quantities discussed so far strictly considers either the GC age or the metallicity, but the GC age-metallicity distribution allows both to be used in conjunction. In addition, we now derive a number of quantities that reflect the two-dimensional nature of the age-metallicity distribution of GCs. The most straightforward of these is the combined interquartile range $\iqr^2=\iqr(\tau)\times\iqr(\feh)$, which represents the total spread in GC age-metallicity space. This combined spread effectively probes the number of star formation events through which the GC population was formed. As such, $\iqr^2$ can be useful for identifying extended assembly histories (through high values of $\iqr^2$ as in e.g.~MW14 and MW15) or single (starburst) events (through low values of $\iqr^2$ as in e.g.~MW23 -- although note that the starburst in MW18 revealed by the contours in \autoref{fig:agez} does not translate into a low $\iqr^2$, due to the destruction of the associated clusters by tidal shocks and dynamical friction). However, Section~\ref{sec:hist} shows that $\iqr(\tau)$ is often a better tracer of the galaxy assembly history than $\iqr^2$, likely due to the lack of variation of $\iqr(\feh)$ among the simulated galaxies.

A more useful combination of $\iqr(\tau)$ and $\iqr(\feh)$ is their ratio, i.e.~$\riqr=\iqr(\feh)/\iqr(\tau)$, which represents the aspect ratio of the age-metallicity distribution and thus acts as a proxy for the metal enrichment rate. Indeed, we see that the galaxy with the steepest GC age-metallicity distribution in \autoref{fig:agez} (MW23) also has the highest value of $\riqr$. This diagnostic is expected to trace the growth rate of the host galaxy and may therefore be positively correlated with the concentration parameter of the dark matter halo \citep[cf.][]{bullock01} or any other galaxy-related quantities that reflect the rapidity of galaxy assembly.

The most direct and powerful set of metrics considered here is obtained by fitting the GC age-metallicity distribution with a two-parameter function. The fit is performed by carrying out an orthogonal distance regression \citep{boggs90}, in which both axes represent free (unfixed) variables and the distance orthogonal to the best-fitting function is minimised. Given a sample of GCs with ages $\tau$ and metallicities $\feh$, we fit the GC age-metallicity distributions with the function
\be
\label{eq:fit}
\feh=\frac{{\rm d[Fe/H]}}{{\rm d}\log t}\times\log \left(\frac{\thub-\tau}{\gyr}\right)+\feh_0 ,
\ee
which is equivalent to the power law relation
\be
\label{eq:fit2}
10^{\rm [Fe/H]} = 10^{{\rm [Fe/H]}_0} \left(\frac{t}{\gyr}\right)^{{\rm d[Fe/H]}/{\rm d}\log t} .
\ee
Equation~(\ref{eq:fit}) effectively expresses the Fe abundance of GCs as a power law function of the time since the Big Bang $t=\thub-\tau$, with $\thub=13.82~\gyr$ the age of the Universe. The expression depends on two free parameters. The first of these is $\dfehdt$, which indicates the rapidity of metal enrichment in the progenitor galaxies as traced by GCs, and the second is $\feh_0$, which is the typical `initial' GC metallicity at $1~\gyr$ after the Big Bang. As shown by~\autoref{fig:agez}, our simulations predict a wide variety of age-metallicity distributions, with many of them exhibiting substantial scatter. Despite the fact that these distributions thus do not strictly follow a single function such as that described by equation~(\ref{eq:fit}), we find that this expression performs well in characterising the distributions in terms of the typical metallicity after the initial collapse of the host haloes $\feh_0$ and the subsequent metal enrichment rate $\dfehdt$. For instance, the galaxy that visually undergoes the most rapid enrichment (MW23) has a high value of $\dfehdt$, whereas those galaxies with slow enrichment and shallow distributions in the age-metallicity plane (e.g.~MW14 and MW15) are characterised by low values of $\dfehdt$. Unsurprisingly, $\dfehdt$ and $\riqr$ are positively correlated, with a Spearman rank correlation coefficient of $r=0.80$.

Finally, we include the total number of GCs ($N_{\rm GC}$) with masses $M>10^5~\msun$ and metallicities $-2.5<\feh<-0.5$ as a measure of the richness of the GC population. This number is expected to correlate with basic galaxy properties (such as its virial mass, radius, and velocity, cf.~\citealt{spitler09,durrell14,harris17}), and possibly the number of accreted dwarf galaxies, because these are characterised by a high number of GCs per unit galaxy mass \citep[e.g.][]{peng08}. When correlating $N_{\rm GC}$ to infer the formation and assembly history of observed galaxies, we caution that \emosaics somewhat overpredicts the number of GCs, because the cluster disruption rate is underestimated (see Section~\ref{sec:valid} and \citetalias{pfeffer18}). As a result, the number of GCs with masses $M>10^5~\msun$ surviving to $z=0$ is about a factor of $1.75$ too high (see Section~\ref{sec:phys}). We recommend dividing $N_{\rm GC}$ by $1.75$ when comparing the numbers in \autoref{tab:gcmetrics} to observations.

Of course, the above metrics describing the GC age-metallicity distribution can also be used in conjunction. For instance, the metal enrichment rates $\dfehdt$ and aspect ratios $\riqr$ of MW14 and MW15 are very similar, indicating the gradual growth and enrichment of their GC populations. However, the difference in age kurtosis $K(\tau)$ suggests that the GC formation history of MW14 was characterised by a single episode of rapid growth, indicative of a more rapid assembly history than MW15. This suggests that, while the growth of both galaxies has been gradual throughout cosmic time, it proceeded mostly in the form of gas accretion and in-situ star formation in MW15, whereas MW14 must have experienced significant growth through satellite accretion and galaxy mergers. In the next section, we take this type of inference a step further by discussing the formation and assembly histories of the modelled galaxies and demonstrating how these are correlated with the quantities describing the GC age-metallicity distribution discussed in this section.

\section{The relation to the formation and assembly history of the host galaxy} \label{sec:hist}
In this section, we present the variety of formation and assembly histories of the 25 simulated Milky Way-mass galaxies, which are quantitatively characterised through a wide range of physically relevant parameters. These are then correlated with the parameters describing the GC age-metallicity distributions from Section~\ref{sec:agez}, revealing how the age-metallicity distribution traces the formation and assembly history of the host galaxy.

\subsection{Variety of galaxy formation and assembly histories} \label{sec:galform}
In order to obtain meaningful relations between the properties of the GC population and a set of metrics characterising galaxy formation and assembly histories, both elements of the correlation must span a sufficiently large dynamic range. Section~\ref{sec:agez} shows that the age-metallicity distributions of the simulated GC populations differ greatly, but this should also hold for the formation and assembly histories of the 25 simulated galaxies. As indicated several times in this paper, the 25 simulated galaxies satisfy this requirement -- close inspection of their growth histories and merger trees shows that none of them are alike. We now quantify this statement by discussing the star formation histories of the entire sample and the merger trees of the six example galaxies from \autoref{fig:agez}.

\begin{figure}
\includegraphics[width=\hsize]{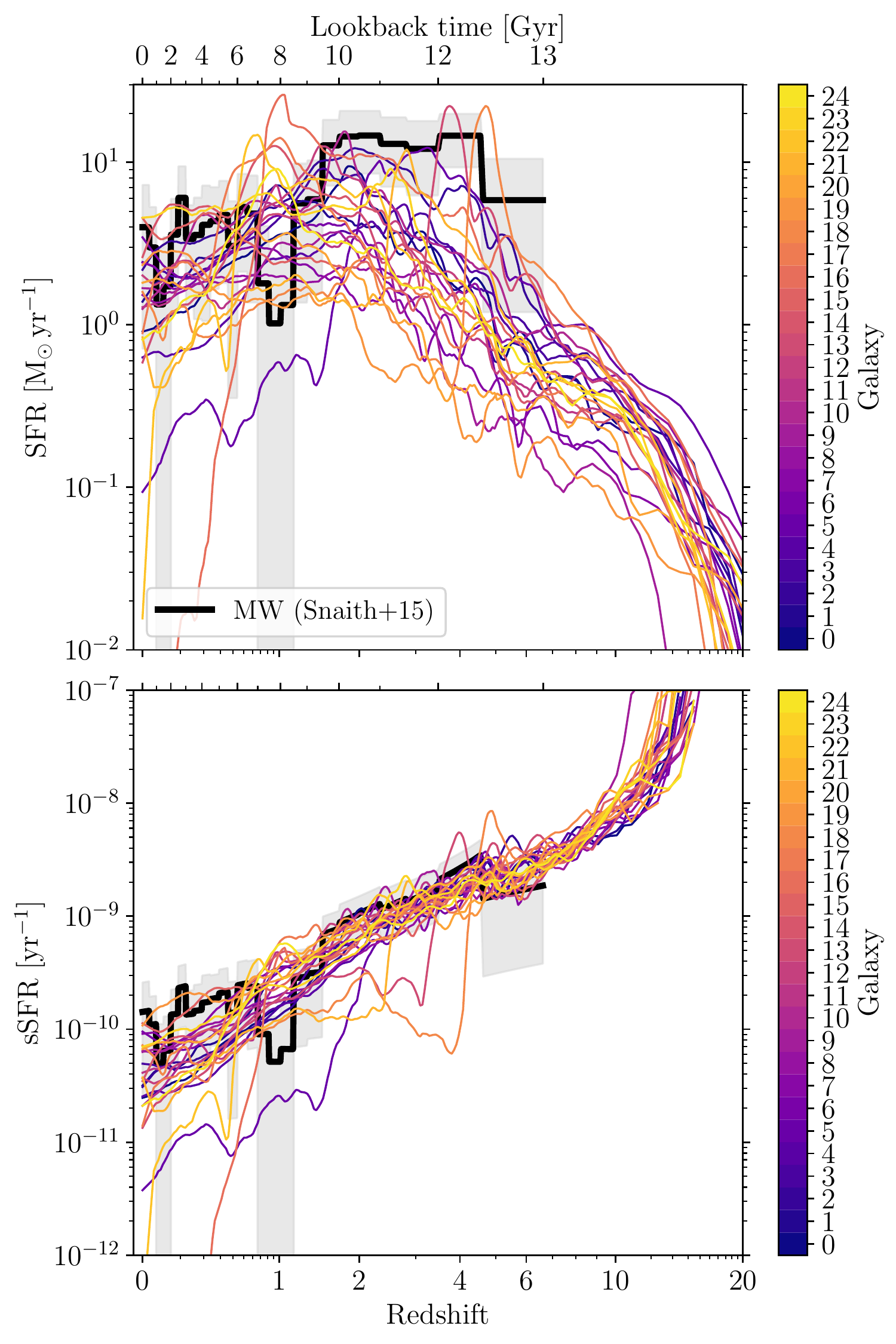}%
\caption{
\label{fig:sfh}
Star formation histories of the 25 simulated galaxies as a function of redshift and lookback time. Top panel: instantaneous star formation rate. Bottom panel: specific star formation rate, i.e.~normalised to the stellar mass of the galaxy. The SFR at each time is obtained in a $\Delta(1+z)$ redshift bin with a minimum width of 200~Myr. The black line represents the observed star formation history of the Milky Way from \citet{snaith14,snaith15}, with the grey-shaded area indicating the $1\sigma$ uncertainty. Relative to fig.~2 of \citetalias{pfeffer18}, we see that the expansion of the simulation suite to 25 galaxies yields a broader variety of absolute star formation histories in the top panel. However, the sample of specific star formation rates remains narrow.
}
\end{figure}
\autoref{fig:sfh} shows the SFR and specific SFR ($\ssfr\equiv\sfr/M_*$) as a function of redshift and lookback time for all 25 galaxies in our suite of simulations. It thus provides an update of fig.~2 in \citetalias{pfeffer18}, which showed the same for the first 10 simulations (MW00--MW09). At face value, the SFRs vary greatly between the different simulations -- at any given redshift, the range of instantaneous SFRs is at least an order of magnitude (and often more). This heterogeneity is even larger than for the 10 simulations shown in \citetalias{pfeffer18}. The increased variety of SFRs is unlikely to be related to differences in ISM properties or the star formation efficiency. As in \citetalias{pfeffer18}, the bottom panel of \autoref{fig:sfh} shows that the sSFR evolves relatively universally with redshift, demonstrating that the majority of the simulated galaxies are star-forming `main sequence' galaxies \citep[e.g.][]{brinchmann04,daddi07}. The only exceptions to this trend are MW05, MW16, and MW22, in which star formation is quenched between redshifts $z=0.5$--$2$ and the SFR subsequently drops precipitously (compare to \autoref{tab:sims} for the $z=0$ SFRs). The contrast between the spreads of the SFR and sSFR in \autoref{fig:sfh} means that variations in the SFR must be caused by variations in the galaxy mass growth history. The large dynamic range of SFRs thus translates to a wide range of mass accretion histories and, hence, a great variety of merger trees.

\begin{figure*}
\includegraphics[width=0.97\hsize]{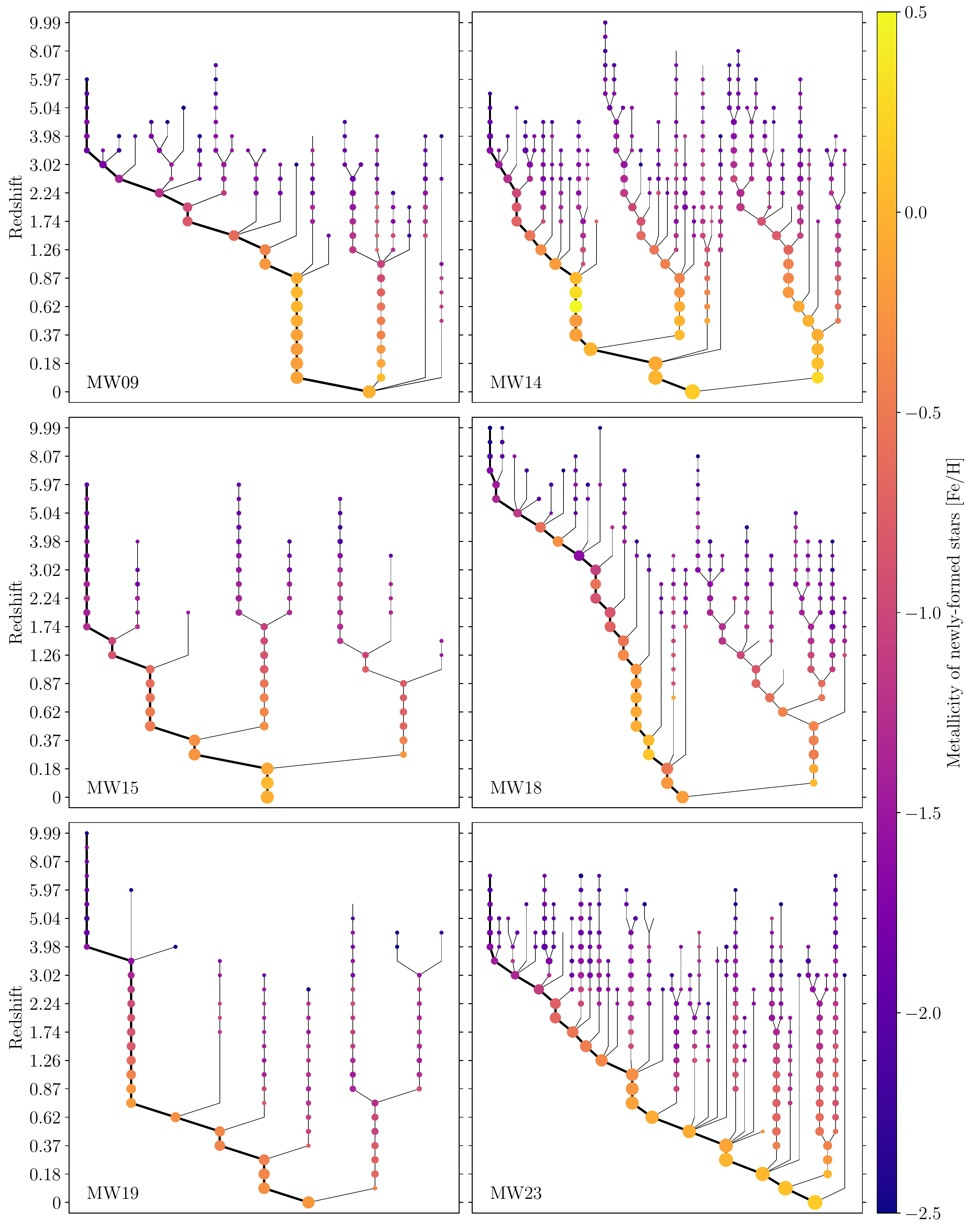}%
\caption{
\label{fig:trees}
Galaxy merger trees for six of our Milky Way-mass galaxies at $z=0$ (MW09, MW14, MW15, MW18, MW19, MW23; indicated in the bottom-left corner of each panel), excluding their satellite systems. The symbol colour indicates the mean metallicity of the stars in each galaxy that formed since the preceding snapshot and the symbol area scales with the cube root of the host galaxy stellar mass. Only galaxies with stellar mass $M_*>4.5\times10^6~\msun$ (corresponding to $20m_{\rm g}$) are included. The main branch is indicated by a thick black line on the left of each merger tree and follows the most massive progenitor in each merger. It thus represents growth by `in-situ' star formation, whereas the other galaxies provide `ex-situ' stars and clusters. This small subset of 6 out of the 25 simulations presented in this paper already shows a great variety of galaxy assembly histories.
}
\end{figure*}
The variety of galaxy assembly histories and merger trees suggested by these star formation histories is illustrated in \autoref{fig:trees}, which shows the merger trees of the six example galaxies from \autoref{fig:agez}. The dots in the merger tree represent the progenitors of the $z=0$ galaxy, with sizes indicating their stellar masses and colours showing the mean metallicity $\feh$ of the stars in each galaxy that formed since the previous snapshot. The minimum stellar mass for including galaxies in the merger tree is $M_*>4.5\times10^6~\msun$, which means that the vast majority of galaxies hosting GCs are accounted for -- the lowest-mass galaxy in which GCs are found in our simulations has a stellar mass of a few $10^6~\msun$. We omit galaxies with $M_*<4.5\times10^6~\msun$, because they are resolved with $<20$ particles.

In each of the merger trees, the main branch (thick line) indicates the mass growth and enrichment history of the main progenitor (following the most massive branch at each node while moving up). The mass evolution along the main branch is generally monotonic, because the stellar mass of the main progenitor only grows, except in a few minor exceptions where the stellar mass of the main progenitor decreases slightly due to stellar evolutionary mass loss. For the accreted satellites, the stellar mass does not evolve monotonically, because the galaxies are stripped prior to their merging into the main branch (illustrated by decreasing symbol sizes). The masses of the satellite branches thus represent the ex-situ mass growth of the main branch, whereas the mass evolution along the main branch after subtraction of the accreted satellite masses visualises the in-situ mass growth.

The metallicity at which stars are born evolves largely monotonically, but significant deviations exist. For instance, the main branch of MW18 undergoes an initial phase of rapid enrichment, after which it accretes a substantial reservoir of low-metallicity gas at $z\approx4$ and the metallicity of newborn stars drops by more than an order of magnitude, from $\feh\approx-0.3$ to $\feh\approx-1.6$. Similar (but less extreme) fluctuations exist in the other galaxies as well (e.g.~the main branches of MW09, MW14, and MW19). The main physical reason for the difference relative to the monotonic stellar mass evolution is that the metallicity evolution is considerably more complex, due to the combination of enrichment, gas accretion, and mergers, each of which can affect the metallicity in a variety of ways. This is one of the reasons why the galaxy mass-metallicity relation \citep[e.g.][]{tremonti04,erb06,mannucci09} exhibits more scatter than the sSFR `main sequence' \citep[e.g.][]{brinchmann04,daddi07}. In the GC age-metallicity plane, these $\feh$ fluctuations manifest themselves both as scatter and systematic offsets in metallicity. At any redshift, the satellites generally have lower masses and metallicities than the main branch. This again reflects the galaxy mass-metallicity relation and shows that the interpretation of sequences of low-$\feh$ GCs in \autoref{fig:agez} as `satellite branches' is accurate.

Comparing the six different merger trees in \autoref{fig:trees}, we see a great variety of assembly histories. Some galaxies (e.g.~MW14, MW18, and MW23) have a large number of well-resolved progenitors (i.e.~with $M_*>4.5\times10^6~\msun$), while others (e.g.~MW15 and MW19) have quiescent merger histories. Defining a major merger as having a satellite-to-main progenitor stellar mass ratio in excess of $1/4$ (corresponding to a symbol area difference of less than a factor $4^{1/3}\approx1.6$), only MW19 did not experience a major merger -- the other galaxies experienced one or multiple major mergers, with the most recent one having taken place anywhere between $z=0$ (MW14 and MW18) and $z=2.5$ (MW23). As a result, MW19 almost exclusively grew through in-situ star formation, whereas MW14 has a large fraction of stars that formed ex-situ.

Qualitatively, the merger trees largely confirm the insight gleaned from a qualitative analysis of the GC age-metallicity distributions in Section~\ref{sec:agez}. As expected from their satellite branches of low-$\feh$ GCs in the age-metallicity plane, MW09, MW18, and MW23 experienced extensive accretion of satellites with mass $M_*>4.5\times10^6~\msun$. A naive classification of the GC age-metallicity distribution of MW14 did not immediately hint at its rich merger history, but this is caused by the fact that its assembly is dominated by a series of {\it major} mergers -- due to having masses similar to the main progenitor, these satellite galaxies have similar metallicities at any given age, implying that their GCs overlap with the main branch in the age-metallicity plane. This degeneracy between the main branch and the satellite branch may be lifted statistically by considering how many GCs a galaxy that is forming stars at a given age-metallicity coordinate is expected to contribute (see Section~\ref{sec:recon}). Indeed, the fact that MW14 contains considerably more GCs than the other galaxies is interpreted in Section~\ref{sec:gcmetrics} to indicate the accretion of multiple satellites with high specific frequencies. Analogously, the low number of GCs in the age-metallicity plane of MW15 and MW19 suggest these galaxies had few mergers. Again, the merger trees of \autoref{fig:trees} confirm this impression.\footnote{Among the six example galaxies, MW18 represents the exception to this apparent relation between the number of GCs and the number of mergers. However, as stated before, this is the result of a nuclear starburst, during which the vast majority of the clusters form at such small radii and high ISM densities that they are all destroyed by tidal shocks and dynamical friction. Very few of the other galaxies experience such an unusual star formation episode.}

Even the timing of mergers is reasonably well-constrained by reading the GC age-metallicity distribution. For MW19, the presence of two pronounced branches with different metallicities at old ages in \autoref{fig:agez} implies the existence of a prominent satellite at $z\sim1.5$ with (at that time) a mass comparable to the main branch, which merges into the main branch at lower redshifts ($z<0.5$). The merger tree of MW19 shows that this is indeed the case. Conversely, the high age skewness $S(\tau)$ of MW09 discussed in Section~\ref{sec:gcmetrics} suggests a rapid truncation of the main GC formation episode, which based on the GC age-metallicity distribution should be at $z\sim1$. The merger tree in \autoref{fig:trees} now shows that MW09 experiences an episode without any merger activity at all from $0<z<0.9$. This period of quiescent star formation is not accompanied by any GC formation.

We also used the GC age-metallicity distributions to determine the burstiness of the star formation history. Specifically, the high age kurtosis $K(\tau)$ of MW18, MW19, and MW23, as well as the low $\iqr(\tau)$ of MW23, both suggest bursty star formation episodes that were accompanied by pronounced GC formation. This is confirmed in \autoref{fig:sfh} for MW18, which exhibits a strong SFR peak at $z=5$. No such extreme features can be identified for MW19 and MW23, but the latter does have a generally high SFR between $z=2.5$--$4.5$, which is the likely cause for its elevated age kurtosis and low $\iqr(\tau)$. Likewise, the age-metallicity distribution constrains the metal enrichment history -- the intermediate-to-high fitted GC age-metallicity slope $\dfehdt$ of MW18 and MW23 (directly tracing the combined metal enrichment rate of all progenitors hosting GCs) suggests rapid initial enrichment, which is mirrored by the colours in the merger trees of \autoref{fig:trees}, showing that these galaxies indeed attain $\feh=-0.5$ already at $z\ga1.5$.

Even the higher-order combinations of the GC age-metallicity diagnostics seem to be quite accurate. For instance, the similar enrichment rates $\dfehdt$ of MW14 and MW15 indicate that both systems gradually grew their GC populations. At the same time, the higher age kurtosis of MW14 suggests that its growth proceeded through a (small number of) dominant burst(s) (indicative of satellite accretion), whereas the low age kurtosis of MW15 suggests that its growth took place through gas accretion and in-situ star formation. Again, this distinction based on the GC age-metallicity distribution is confirmed by MW14's rich merger tree and MW15's low number of mergers, with the first merger taking place as late as $z\approx1.5$.

The above links between the (quantitative) properties of the GC age-metallicity distribution and the qualitative formation and assembly histories of the host galaxy warrant a systematic evaluation of their correlation. Therefore, we now quantitative characterise the galaxy formation and assembly histories, so that these metrics can be contrasted with the diagnostics describing the GC age-metallicity distribution.

\subsection{Quantitative characterisation of galaxy formation and assembly histories} \label{sec:galmetrics}
We expand the interpretation of galaxy formation and assembly through the star formation histories (\autoref{fig:sfh}) and merger trees (\autoref{fig:trees}) from Section~\ref{sec:galform} with a set of 30 quantitative metrics characterising the 25 simulated galaxies. In Section~\ref{sec:correlations}, we correlate these with the metrics describing the GC age-metallicity distribution from Section~\ref{sec:agez}. To describe the galaxies, we consider two sets of metrics, listed for all galaxies in Appendix~\ref{sec:appmetrics_gal}. The first set describes the global galaxy (mass growth) properties (\autoref{tab:galmetric1}), whereas the second set describes its merger tree (\autoref{tab:galmetric2}).

\autoref{tab:galmetric1} shows 14 quantities for the fiducial simulations discussed throughout this work. Some of the listed quantities specifically describe the properties of the host dark matter halo. Because the halo properties can be affected by baryonic physics \citep[e.g.][]{governato12,schaller15}, we have also run an analogous set of dark matter-only simulations of the same galaxies and verified that the obtained metrics are similar. Given the small differences ($\la10$~per~cent), we limit the following discussion to the fiducial baryonic simulations.

The first set of five quantities in \autoref{tab:galmetric1} are all measured instantaneously at $z=0$ and form the standard set of diagnostics to describe the haloes of galaxies in a $\Lambda$CDM cosmogony. These are the virial mass $\mvir$, the virial radius $\rvir$, the maximum circular velocity $\vmax$, the galactocentric radius at which $\vmax$ is reached $\rvmax$, and the concentration parameter of the dark matter halo $\cnfw$ (parameterised with a \citealt{navarro97} profile). The first four of these are correlated, as they all increase with the galaxy mass, whereas the halo concentration parameter traces the condensation redshift of the halo \citep{bullock01,correa15} and is weakly anti-correlated with the halo mass in our simulations. These quantities are all obtained by fitting the density profiles of the dark matter haloes at $z=0$.

The next four quantities in \autoref{tab:galmetric1} describe the mass growth history of the galaxy's halo mass. Because the mass growth proceeds continuously throughout cosmic time, we describe it in terms of four characteristic time-scales $\{\ttf, \tfz, \tsf, \tmax\}$, representing the lookback times at which $\{25, 50, 75, 100\}$~per~cent of the maximum halo mass is first attained. A few galaxies have $\tmax>0~\gyr$, which means that the halo mass was somewhat higher in the past than at $z=0$. These differences result from a temporary overestimation of the halo mass during mergers, caused by an ill-defined virial radius due to deviations from spherical symmetry. These differences are small -- the $z=0$ halo mass always exceeds two thirds of the maximum halo mass and usually falls within a few per~cent.

Next, we use five quantities to describe the balance between in-situ and ex-situ galaxy growth. We follow the approach of \citet{qu17}, who formulate the assembly (lookback) time $\ta$ and redshift $\za$ as the moment when the main progenitor (thick lines in the merger trees of \autoref{fig:trees}) has attained half of the $z=0$ stellar mass.\footnote{Contrary to \citet{qu17}, we use the actual stellar mass rather than the initial stellar mass. Fig.~B1 of \citet{qu17} shows  that the influence of this difference is minor for Milky Way-mass galaxies.} These are combined with the formation (lookback) time $\tf$ and redshift $\zf$, which marks the moment when all progenitors together have attained half of the $z=0$ mass (this is effectively the median age or median formation redshift of all stars in the galaxy at $z=0$). By definition, $\ta\leq\tf$ and $\za\leq\zf$, because once a certain mass is attained by the main branch, it must also have been attained by all progenitors combined. When $\ta$ and $\tf$ (or $\za$ and $\zf$) are similar, the galaxy largely formed in-situ, because the main branch dominates the population of progenitors. However, if $\ta\ll\tf$ (or $\za\ll\zf$), then the galaxy experienced considerable ex-situ growth, because all progenitors together attained $0.5M_*(z=0)$ well before the main branch did. We quantify the balance between in-situ and ex-situ galaxy growth through the dimensionless quantity
\be
\label{eq:deltat}
\delta_t=1-\frac{\ta}{\tf} ,
\ee
which measures the fraction of time between the formation time $\tf$ ($\approx$ median stellar age) and the present day that has elapsed when the main progenitor has attained half its final mass, i.e.~at the assembly time $\ta$.\footnote{This expression differs from the one in \citet[page~1665]{qu17}, because their equation contains a small typographical error. The results presented by \citet{qu17} are unaffected by this issue.} If this expression is close to zero, a galaxy formed mostly in-situ (e.g.~MW09 and MW15, which have $\delta_t\approx0.05$), whereas a high value indicates prominent ex-situ growth with a rich merger history (e.g.~MW14 and MW18, which have $\delta_t>0.4$). In the galaxy mass range covered by out 25 simulations, \citet{qu17} obtain typical values of $\delta_t=0.05$--$0.10$, very close to our median. Following \citet{qu17}, we somewhat arbitrarily define $\delta_t=0.1$ as the threshold separating negligible ($\delta_t<0.1$) and significant ($\delta_t>0.1$) ex-situ galaxy growth.\footnote{Only six of the simulations have $\delta_t>0.1$. Importantly, this concerns the stars -- for GCs, $\delta_t$ may be higher, because low-mass galaxies have higher specific frequencies (i.e.~number of GCs per unit stellar mass), and thus accreted satellites are expected to be contribute a larger fraction of the GC population than of the stellar mass. We will demonstrate this below using a pair of quantities that explicitly measure the ex-situ fractions.}

In \autoref{tab:galmetric2}, we list an additional set of 16 quantities that describe the merger trees of the simulated galaxies, quantifying their assembly histories. We reiterate that the resolution limit of the simulations imposes a minimum galaxy mass of $M_*=4.5\times10^6~\msun$. Unresolved, lower-mass galaxies are counted as smooth accretion.

The first set of five quantities describes the (relative) timing of the most recent major and minor galaxy mergers. These are the lookback time $\tmm$ and redshift $\zmm$ of the most recent major merger, the lookback time $\tam$ and redshift $\zam$ of the most recent merger in general (`any merger'), and the time-scale ratio
\be
\label{eq:rt}
r_t=\frac{\thub-\tmm}{\thub-\tam} ,
\ee
which indicates the fraction of time between the Big Bang (at a lookback time $\thub=13.82~\gyr$) and the time of the last merger (at $\tam$) that has elapsed at the time of the last major merger $\tmm$. It is a dimensionless measure of how far into the epoch of active galaxy merging major mergers still took place. This quantity distinguishes between galaxies that have their last major merger at similar redshifts, but have different levels of minor merger activity thereafter. Galaxies with recent major mergers have a value of $r_t$ close to unity, whereas galaxies without major mergers after the initial collapse of the main progenitor have $r_t$ close to zero.

The next set of five diagnostics describes the topology of the merger trees. The quantities $\nbrz$ and $\nbr$ represent the number of branches emerging from the main branch (thick line in \autoref{fig:trees}) at $z>2$, i.e.~at the peak epoch of GC formation, and at all redshifts, respectively. These count the number of mergers experienced by the main progenitor with galaxies of stellar mass $M_*\geq4.5\times10^6~\msun$. The ratio between these, i.e.
\be
\label{eq:rz}
\rz=\frac{\nbrz}{\nbr} ,
\ee
traces the rapidity of the galaxy assembly process. A high value ($\rz>0.5$) indicates that most of the merger activity took place at $z>2$, whereas a low value ($\rz<0.5$) indicates that most mergers occurred in the last $10.5~\gyr$. Across the 25 simulated galaxies, the median value is $\rz=0.40$ and only seven galaxies experience most of their mergers before $z=2$. From the set of six example galaxies in \autoref{fig:agez} and \autoref{fig:trees}, only MW18 is characterised by such an extremely rapid assembly.

The next two quantities in \autoref{tab:galmetric2} capture the (relative) importance of major and minor mergers. The first of these, $\nleaf$, represents the total number of progenitors with $M_*\geq4.5\times10^6~\msun$ as obtained by counting the leaves of the merger tree, which corresponds to the number of progenitors prior to any merger activity. Because each leaf eventually merges into the main branch (though possibly after a sequence of intermediate mergers), the number of leaves sets a maximum on the number of branches, i.e.~$\nbr\leq\nleaf$. If the number of leaves exceeds the number of branches, then multiple leaves must have merged before being accreted onto the main branch. This is quantified through the branch-to-leaf ratio
\be
\label{eq:rbl}
\rbl=\frac{\nbr}{\nleaf} ,
\ee
which reflects the balance between hierarchical and monolithic galaxy growth. If $\nbr\ll\nleaf$, the main branch undergoes few mergers, but each of these does contribute a large number of progenitors. This is indicative of major mergers, because massive galaxies have multiple progenitor galaxies, as well as hierarchical growth, as the leaves merged into an intermediate system before merging into the main branch. Conversely, if $\nbr\sim\nleaf$, the leaves typically merge directly into the main branch without experiencing any mergers along the way. This monolithic behaviour is indicative of minor mergers, because a low-mass galaxy is more likely to have only a single progenitor and to have formed entirely in-situ. The most illustrative cases of the branch-to-leaf ratio $\rbl$ among our six example galaxies are MW14 (major merger dominated, with $\rbl=0.34$) and MW23 (minor merger dominated, with $\rbl=0.74$). We invite the reader to compare these to the merger trees in \autoref{fig:trees}. The comparison shows that $\rbl$ is a highly useful diagnostic for characterising the topology of merger trees.

The next set of four quantities in \autoref{tab:galmetric2} address the numbers of mergers of various mass ratios. We reiterate that the merger statistics only consider progenitors with stellar masses $M_*\geq4.5\times10^6~\msun$. We list the number of `tiny mergers' $N_{<1:100}$, for which the satellite has a mass $1/100$ times that of the main progenitor at the time of the merger, as well as the number of minor mergers $N_{1:100-1:4}$, for which the satellite has a mass between $1/100$ and $1/4$ times that of the main progenitor, and the number of major mergers $N_{>1:4}$, for which the merging galaxies have stellar masses within a factor of 4 of each other. These three merger types together constitute all mergers ever experienced by the simulated galaxy with progenitors of mass $M_*\geq4.5\times10^6~\msun$.\footnote{Note that this mass limit implies that for main progenitor masses $M_*<4.5\times10^8~\msun$, we have $N_{<1:100}=0$ by definition. Across our 25 simulations, this typically affects galaxies at $z\ga2.5$.} As can be verified using \autoref{tab:galmetric2}, the total number of mergers $\nbr=N_{<1:100}+N_{1:100-1:4}+N_{>1:4}$. The fourth and final variable of this set represents the ratio between major and non-major mergers, defined as
\be
\label{eq:rmm}
r_{\rm mm}=\frac{N_{>1:4}}{N_{<1:100}+N_{1:100-1:4}} .
\ee
As in the discussion of $\nbr$ and $\rbl$ above, we see that both the richness of the merger tree and the relative importance of major mergers are reflected by $N_{<1:100}$, $N_{1:100-1:4}$, $N_{>1:4}$, and $r_{\rm mm}$. Specifically, the branch-to-leaf ratio $\rbl$ and the major merger ratio $r_{\rm mm}$ trace each other. MW14 (major merger dominated) has a high $r_{\rm mm}=0.27$, whereas the minor merger dominated MW23 has $r_{\rm mm}=0.04$, which confirm the earlier suggestion based on the branch-to-leaf ratios that MW14 and MW23 predominantly accreted massive and low-mass galaxies, respectively.

The final two quantities in \autoref{tab:galmetric2} are the fraction of all stars ($\fexs$) and all stars in GCs ($\fexgc$) with an ex-situ origin. These are obtained by summation of the stellar or GC mass formed over all branches other than the main branch in \autoref{fig:trees} and dividing by the total stellar mass at $z=0$ or the final mass of the GC population, respectively. Both quantities cover a wide dynamic range, from $\fexs=0.03$--$0.62$ and $\fexgc=0.09$--$0.60$. Among the six example galaxies, those with the most prominent major mergers (MW14 and MW18) have the highest ex-situ fractions, as expected.

The above discussion covers a total of 30 quantities describing galaxy formation and assembly. Quite clearly, not all of them are independent. The redshifts and lookback times correlate perfectly by definition, and the virial masses, radii, and velocities also show a high degree of correlation. Above all, any relations between the 30 metrics are non-linear and thus each of them provides a different functional form that can be used when determining the best-fitting relations with the GC-related diagnostics in Section~\ref{sec:correlations}. Even though the effective number of independent metrics is smaller, we include all quantities in the following discussion, so that we can select the simplest correlations when possible.

\subsection{Correlation between the GC age-metallicity distribution and the host galaxy formation and assembly history} \label{sec:correlations}
We have correlated all metrics describing the GC age-metallicity distribution (\autoref{tab:gcmetrics}) with all metrics describing the galaxy formation and assembly histories (\autoref{tab:galmetric1} and \autoref{tab:galmetric2}). In Appendix~\ref{sec:appspear}, we list the resulting Spearman rank correlation coefficients (\autoref{tab:correlations_sr}) and the associated $p$-values representing the probability that the correlation is achieved by random chance (\autoref{tab:correlations_sp}). This allows us to evaluate which correlations are statistically significant (we define this significance in Appendix~\ref{sec:appspear}). While each of these correlations can be directly reproduced from the numbers listed in \autoref{tab:gcmetrics}--\ref{tab:galmetric2}, we visualise a small subset of them below for illustration. The complementary Pearson correlation coefficients and $p$-values are provided in Appendix~\ref{sec:appcorr}, both for the original correlated quantities (evaluation a linear relation) and their logarithms (evaluating a power law relation).

\autoref{tab:correlations_sr} and \autoref{tab:correlations_sp} show that most of the expected correlations discussed in Section~\ref{sec:agez} between the metrics describing the GC age-metallicity distribution and those describing the host galaxy's formation and assembly history indeed exist in some form, but many of them only as a rough trend with relatively low Spearman correlation coefficients ($|r|<0.5$) and low levels of significance (with probabilities $p>0.05$ and hence $\log{p}>-1.3$ that the correlation arises due to random chance). Such trends lack predictive power. In the following discussion, we therefore focus on the correlations that have a high statistical significance (see Appendix~\ref{sec:appspear}) and enable the use of the GC age-metallicity distribution to constrain the formation and assembly history of the host galaxy.

\subsubsection{Statistically (in)significant correlations} \label{sec:significant}
\begin{figure*}
\center
\includegraphics[width=\hsize]{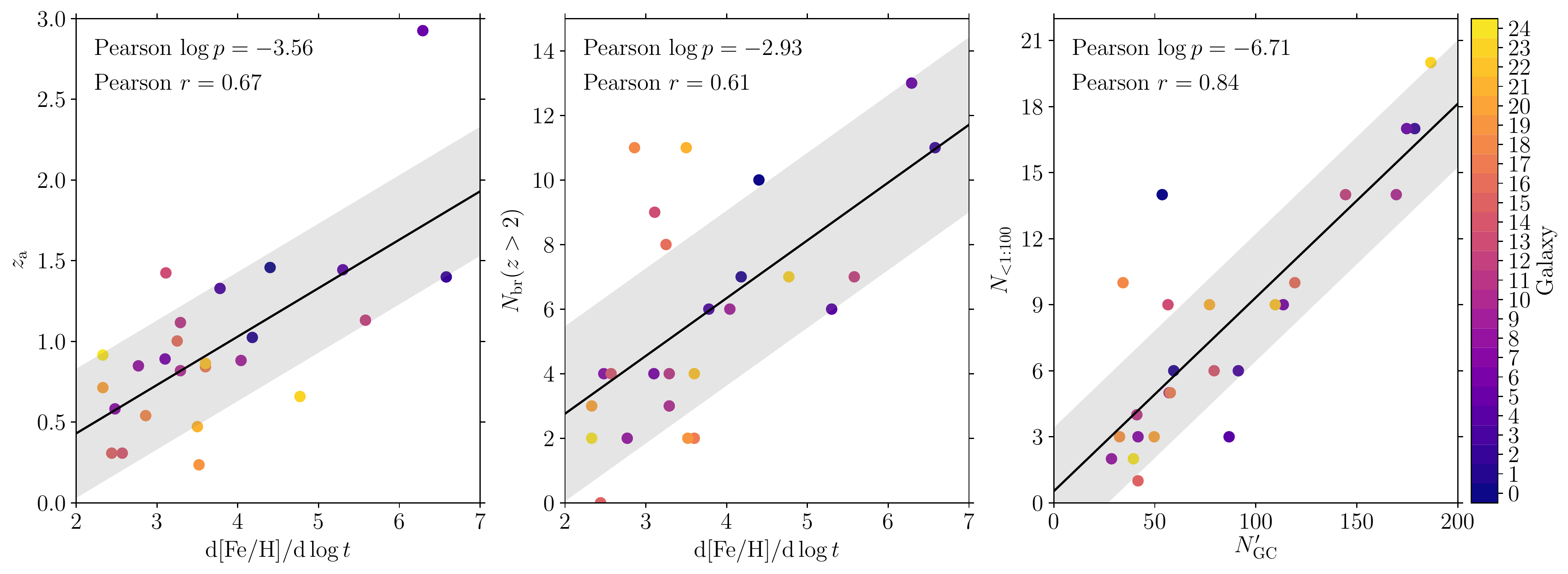}%
\caption{
\label{fig:correlations}
Examples of three statistically significant correlations between galaxy formation and assembly-related quantities from Section~\ref{sec:galmetrics} ($y$-axes) and the quantities describing the GC age-metallicity distribution from Section~\ref{sec:gcmetrics} ($x$-axes). Left panel: galaxy assembly redshift $\za$, which indicates when the main progenitor first attains 50~per~cent of the stellar mass at $z=0$, as a function of the best-fitting slope of the GC age-metallicity distribution $\dfehdt$. Middle panel: number of high-redshift ($z>2$) mergers $\nbrz$ as a function of $\dfehdt$. Right panel: number of tiny mergers $N_{<1:100}$ (recall that galaxies with stellar masses $M<4.5\times10^6~\msun$ are not counted as discrete mergers, but as smooth accretion), as a function of the corrected number of GCs $\ngcp$ (see the text) in the mass and metallicity range $M>10^5~\msun$ and $-2.5<\feh<-0.5$. Coloured symbols denote the 25 simulations as indicated by the colour bar on the right. In each panel, the solid line indicates the best-fitting linear regression, with the grey-shaded band marking the $1\sigma$ scatter of the data around the fit, and the Pearson $p$-values and correlation coefficients in the top-left corner indicating its statistical significance. This figures illustrates that the properties of the GC population quantitatively trace the host galaxy's formation and assembly history.
}
\end{figure*}
The Spearman correlation coefficients and (log) $p$-values of the statistically significant correlations are marked in red in \autoref{tab:correlations_sr} and \autoref{tab:correlations_sp}. Across the entire set of 338 correlations shown (omitting the redshifts from \autoref{tab:galmetric1} and \autoref{tab:galmetric1}), a number of interesting results immediately stand out.
\begin{enumerate}
\item
The most powerful GC-based predictors of the host galaxy formation and assembly history are related to the GC age distribution ($\widetilde{\tau}$, $\iqr(\tau)$), the metal enrichment rate ($\riqr$, $\dfehdt$), or the total number of GCs ($\ngc$). These mainly trace when and over which timespan the host galaxy grew (with prominent episodes of growth being positively correlated with GC formation), at which rate it grew and enriched (with the galaxy's enrichment being mirrored by the GC age-metallicity distribution), and how many (minor) mergers it experienced (with prevalent galaxy merging being positively correlated with rich GC populations).
\item
Metrics based on the GC metallicity distribution alone are weak quantitative probes, which is likely caused by the similarity of metallicity distributions across our sample of Milky Way-like galaxies (see \autoref{tab:gcmetrics}). The higher-order diagnostics of the age-metallicity skewness and kurtosis are also poor tracers of galaxy assembly, due to their high sensitivity to outliers. The only exception is the metallicity kurtosis $K(\feh)$, which indicates how strongly peaked the metallicity distribution is and carries a marginally significant correlation with the number of major mergers $N_{>1:4}$.
\item
The galaxy virial mass and radius correlate poorly with the GC age-metallicity distribution ($\mvir$, $\rvir$). These correlations are insignificant due to the relatively narrow range of halo masses considered here (see \autoref{tab:correlations_sp}), despite the strong correlation between $\mvir$ and $\ngc$ observed in galaxies over $\sim5$ orders of magnitude in $\mvir$ \citep[e.g.][]{spitler09,durrell14,harris17}. We quantify this interpretation by carrying out 10,000 Monte Carlo experiments in which we generate 25 values of $\ngc$ from the $\mvir$-$\ngc$ relation in \citet[eq.~4]{harris17} including its uncertainties, for the range of halo masses spanned by our simulations. Across these experiments, we obtain Spearman $r=0.27\pm0.19$ and $\log{p}=-0.94\pm0.79$, with the uncertainties indicating the standard deviations across the 10,000 experiments. These values are both fully consistent with the result obtained for the \emosaics galaxies ($r=0.35$ and $\log{p}=-1.07$). This demonstrates that the weak correlation of $\ngc$ with $\mvir$ and $\rvir$ is caused by spanning a range in $\mvir$ that is similar to the scatter on $\ngc$ in the observed $\mvir$-$\ngc$ relation.
\item
Surprisingly in view of the weak correlation between $\mvir$ and $\ngc$, we do find that $\vmax$ correlates well with $\ngc$ (yet $\rvmax$ does not). The correlation between $\vmax$ and $\ngc$ is not fundamental, because $\vmax$ increases both with $\mvir$ and $\cnfw$ \citep[e.g.][]{bullock01,mo10}, the second of which is correlated with the halo assembly history and is therefore strongly positively correlated with $\ngc$ (see below). By contrast, $\rvmax\propto\rvir/\cnfw\propto\mvir/\cnfw$ carries opposite dependences on $\mvir$ and $\cnfw$ that cancel (although it is also affected by baryonic physics). This implies that $\ngc$ is a poor proxy for $\rvmax$.
\item
For the Milky Way-mass galaxies simulated here, GCs trace the quantities describing the initial build-up of the host galaxy ($\cnfw$, $\ttf$, $\tfz$, $\ta$, $\tf$, $\rz$), whereas late-time galaxy growth proceeds independently of the GC properties ($\tsf$, $\tmax$, $\tmm$, $\tam$). This is not necessarily surprising, because GCs are generally older than the field star population \citep[e.g.][]{harris02,peng08,reinacampos19}. Galaxies growing a larger fraction of their mass at early times (characterised by large values of $\cnfw$, $\ttf$, $\tfz$, $\ta$, $\tf$, $\rz$), are thus expected to host older ($\widetilde{\tau}$), more rapidly-formed ($\iqr(\tau)$, $\riqr$, and $\dfehdt$), and richer ($\ngc$) GC populations (also see \citealt{mistani16}, who only consider low-mass GCs, i.e.~$M<10^5~\msun$). This quantitative connection between early galaxy formation and the GC population identified here opens up a new and potentially powerful way of constraining the host galaxy growth history. As a first example of this potential, we note the strong correlation between $\cnfw$ and $\ngc$ and propose that the halo concentration may be an important driver of the scatter around the observed $\mvir$-$\ngc$ relation.
\item
Major merger statistics do not strongly correlate with GC properties ($\tmm$, $r_t$, $\rbl$, $N_{>1:4}$, $r_{\rm mm}$), but the total merger history does, which is dominated by minor mergers ($\nbr$, $\nbrz$, $\nleaf$, $N_{<1:100}$, $N_{1:100-1:4}$).\footnote{Like the previous statement that GCs trace early galaxy growth, the finding that GCs properties mainly correlate with minor mergers may depend on the galaxy mass. We reiterate that our results apply to the halo mass range $11.85<\log{\mvir/\msun}<12.48$.} The majority of (minor) merger-related quantities is correlated with the median age, metal enrichment rate, or total number of the GC population. Clearly, (minor) mergers are associated with galaxy growth, which is reflected by the quantities describing the growth of the GC population and the steepness of the GC age-metallicity distribution. This provides a new way of characterising galaxy merger trees through their GC populations.
\item
Perhaps surprisingly, the balance between in-situ and ex-situ star and GC formation is not probed by any of the GC-related metrics ($\delta_t$, $\fexs$, $\fexgc$). In Section~\ref{sec:gcmetrics} and Section~\ref{sec:galform}, it was possible to connect some GC-related properties such as the age range and kurtosis to the relative importance of ex-situ galaxy growth on a case-by-case basis. However, the statistical tests of \autoref{tab:correlations_sr} and \autoref{tab:correlations_sp} demonstrate that the statistical basis for these connections is weak. While some imprints may be found in individual cases, the full galaxy sample shows that the GC age-metallicity distribution is shaped more by the assembly history than whether the stars or GCs formed in-situ or ex-situ. This is likely caused by the enormous variety of ex-situ environments, with some representing tiny dwarf galaxies and others closely resembling the main branch, thus prohibiting a distinction between in-situ and ex-situ formation. Conceptually, the idea of tracing ex-situ growth with GCs relies on satellites being considerably less metal-rich than the main progenitor, which is not always satisfied (see \autoref{fig:trees} and Section~\ref{sec:ageztrees}).
\end{enumerate}

Across the entire set of 338 correlations evaluated, we quantify 20 statistically significant relations between the metrics characterising the GC age-metallicity distribution and those describing the host galaxy formation and assembly history by fitting linear or power law relations. This provides a quantitative framework for reconstructing the assembly histories of a galaxy using its observed GC age-metallicity distribution. For all statistically significant correlations we fit functions of the form
\be
\label{eq:fit3}
y=\frac{{\rm d}y}{{\rm d}x}x+y_0 ,
\ee
where $x$ represents a GC-related quantity and $y$ a galaxy-related quantity. We fit equation~(\ref{eq:fit3}) both to the original quantities and to their logarithms. In the latter case, the fit represents a power law. We avoid taking the logarithm of quantities that can take a zero value by an appropriate transformation. For instance, logarithms of the lookback time are converted into logarithms of the time since the Big Bang by writing $\log{\tau}\rightarrow\log{(\thub-\tau)}$.

\begin{table*}
\caption{Summary of the 20 statistically significant correlations (i.e.~with Spearman and Pearson $p$-values of $p<\peff$) between quantities characterising galaxy formation and assembly (first and second columns) and those describing the GC age-metallicity distribution (third and fourth columns). For each correlation, the table lists the Spearman and Pearson correlation coefficient $r$ (fifth and seventh columns), the associated $p$-values that the correlation arises by random chance (sixth and eighth columns), the slope of the best-fitting linear regression ${\rm d}y/{\rm d}x$, the intercept of the best fit $y_0$, and the scatter of the simulations around the best-fitting relation.}
\label{tab:fit}
\begin{tabular} {@{}ccccccccccc@{}}
  \hline
 Quantity ($y$) & [units] & Correlates with ($x$) & [units] & Spearman $r$ & $\log$~Spearman $p$ & Pearson $r$ & $\log$~Pearson $p$ & ${\rm d}y/{\rm d}x$ & $y_0$ & Scatter  \\ 
  \hline
  $\log{\vmax}$ & $[\kms]$ & $\log{\ngcp}$ & [--] & $0.68$ & $-3.73$ & $0.66$ & $-3.54$ & $0.14$ & $1.99$ & $0.04$ \\
  $\cnfw$ & [--] & $\iqr(\tau)$ & $[\gyr]$ & $-0.63$ & $-3.15$ & $-0.63$ & $-3.17$ & $-1.79$ & $10.63$ & $1.58$ \\
  $\cnfw$ & [--] & $\riqr$ & $[\gyr^{-1}]$ & $0.67$ & $-3.57$ & $0.64$ & $-3.27$ & $5.95$ & $4.72$ & $1.56$ \\
  $\log{\cnfw}$ & [--] & $\log{\ngcp}$ & [--] & $0.60$ & $-2.78$ & $0.60$ & $-2.78$ & $0.31$ & $0.31$ & $0.10$ \\
  $\ttf$ & $[\gyr]$ & $\widetilde{\tau}$ & $[\gyr]$ & $0.61$ & $-2.96$ & $0.61$ & $-2.96$ & $0.97$ & $0.00$ & $1.21$ \\
  $\log{(\thub-\ttf)}$ & $[\gyr]$ & $\log{(\dfehdt)}$ & [--] & $-0.54$ & $-2.26$ & $-0.55$ & $-2.34$ & $-0.71$ & $0.89$ & $0.14$ \\
  $\log{(\thub-\tfz)}$ & $[\gyr]$ & $\log{(\dfehdt)}$ & [--] & $-0.59$ & $-2.69$ & $-0.58$ & $-2.65$ & $-0.65$ & $1.08$ & $0.12$ \\
  $\za$ & [--] & $\dfehdt$ & [--] & $0.56$ & $-2.48$ & $0.67$ & $-3.56$ & $0.30$ & $-0.17$ & $0.40$ \\
  $\tf$ & $[\gyr]$ & $\widetilde{\tau}$ & $[\gyr]$ & $0.60$ & $-2.80$ & $0.68$ & $-3.72$ & $1.30$ & $-5.76$ & $1.36$ \\
  $\nbrz$ & [--] & $\widetilde{\tau}$ & $[\gyr]$ & $0.75$ & $-4.74$ & $0.75$ & $-4.83$ & $2.67$ & $-22.67$ & $2.26$ \\
  $\nbrz$ & [--] & $\dfehdt$ & [--]  & $0.56$ & $-2.47$ & $0.61$ & $-2.93$ & $1.79$ & $-0.82$ & $2.71$ \\
  $\nbr$ & [--] & $\riqr$ & $[\gyr^{-1}]$ & $0.62$ & $-3.05$ & $0.62$ & $-3.05$ & $20.73$ & $4.21$ & $5.74$ \\
  $\nbr$ & [--] & $\dfehdt$ & [--] & $0.64$ & $-3.30$ & $0.65$ & $-3.33$ & $4.06$ & $0.21$ & $5.59$ \\
  $\nbr$ & [--] & $\ngcp$ & [--] & $0.76$ & $-4.92$ & $0.78$ & $-5.37$ & $0.11$ & $5.57$ & $4.59$ \\
  $\rz$ & [--] & $\widetilde{\tau}$ & $[\gyr]$ & $0.78$ & $-5.43$ & $0.81$ & $-6.15$ & $0.14$ & $-1.10$ & $0.10$ \\
  $\log{\nleaf}$ & [--] & $\log{\ngcp}$ & [--] & $0.72$ & $-4.24$ & $0.70$ & $-4.03$ & $0.73$ & $0.00$ & $0.18$ \\
  $N_{<1:100}$ & [--] & $\riqr$ & $[\gyr^{-1}]$ & $0.57$ & $-2.57$ & $0.64$ & $-3.22$ & $15.27$ & $-0.14$ & $4.05$ \\
  $N_{<1:100}$ & [--] & $\dfehdt$ & [--] & $0.59$ & $-2.77$ & $0.66$ & $-3.49$ & $2.98$ & $-3.03$ & $3.95$ \\
  $N_{<1:100}$ & [--] & $\ngcp$ & [--] & $0.75$ & $-4.73$ & $0.84$ & $-6.71$ & $0.088$ & $0.53$ & $2.89$ \\
  $\log{N_{1:100-1:4}}$ & [--] & $\log{\ngcp}$ & [--] & $0.55$ & $-2.36$ & $0.56$ & $-2.45$ & $0.62$ & $-0.53$ & $0.23$ \\
  \hline
\end{tabular}
\end{table*}
For illustration, \autoref{fig:correlations} shows three of the statistically significant correlations from \autoref{tab:correlations_sr} and \autoref{tab:correlations_sp}, together with the best-fitting relations according to equation~(\ref{eq:fit3}). These three examples are all interesting for different reasons.
\begin{enumerate}
\item
The first panel correlates the assembly redshift $\za$, when 50~per cent of the galaxy's final stellar mass first resides in the main progenitor, with the slope of the GC age-metallicity distribution $\dfehdt$. Galaxies with steeper GC age-metallicity distributions assembled earlier. Qualitatively, this is not surprising at all, because a steep age-metallicity distribution indicates rapid metal enrichment, which itself suggests rapid galaxy growth. However, the scatter around the relation is only $\sigma(\za)=0.40$, which means that $\dfehdt$ provides a useful quantitative constraint on the assembly redshift. The outlier with $\za\sim3$ is MW05, which quenches at $z\sim2$ and therefore has an elevated assembly redshift relative to galaxies that do sustain in-situ star formation until $z=0$.
\item
The second panel of \autoref{fig:correlations} shows one of the weakest correlations among the subset of statistically significant ones, i.e.~between the number of mergers onto the main progenitor prior to $z=2$ ($\nbrz$) and $\dfehdt$. Galaxies with steeper GC age-metallicity distributions experienced a larger number of high-redshift mergers. Again, this is not necessarily surprising, but with a scatter of just $\sigma(\nbrz)=2.71$, this panel shows the interesting result that a GC population provides quantitative insight into the merger history of its host galaxy. Given that $\dfehdt$ also correlates with $\za$, it is interesting to point out that $\za$ and $\nbrz$ are weakly correlated too, with a Spearman rank coefficient of $r=0.48$. This system of correlated quantities illustrates that it can be difficult to identify the most fundamental correlations among the set of statistically significant ones.
\item
The third and final panel of \autoref{fig:correlations} presents a more surprising result. It visualises the strong correlation between the richness of the GC population $\ngcp\equiv\ngc/\fcorr$ (i.e.~the number of GCs with masses $M>10^5~\msun$ and $-2.5<\feh<-0.5$, with a correction factor that is discussed below) and the number of tiny mergers $N_{<1:100}$ (defined with mass ratios $<1:100$ and a minimum satellite mass $M_*>4.5\times10^6~\msun$). This correlation is much stronger than either of both quantities correlates with the host galaxy halo mass, indicating that the accretion of low-mass satellites stimulates the formation, assembly, and survival of GC populations. Interestingly, no such correlation exists between $\ngcp$ and the number of major mergers (see \autoref{tab:correlations_sr} and \autoref{tab:correlations_sp}). We discuss the influence of (minor) galaxy merging on GC population assembly in more detail below in the context of the complete set of statistically significant correlations. For now, \autoref{fig:correlations} already illustrates that satellite accretion influences the properties of GC populations more strongly than major mergers.
\end{enumerate}

\autoref{tab:fit} summarises the best-fitting relations for all 20 significant correlations. For each correlation, the linear function or power law with the lowest Pearson $p$-value is listed, retaining only those correlations with Spearman and Pearson $p$-values $p<\peff$ (see Appendix~\ref{sec:appcorr} for the Pearson correlation coefficients and $p$-values). Some relations are redundant, because they connect equivalent lookback times or redshifts (cf.~$\{\ta,\tf,\tmm,\tam\}$ and $\{\za,\zf,\zmm,\zam\}$ in \autoref{tab:galmetric1} and \autoref{tab:galmetric2}). For these, we only list the relation with the lowest Pearson $p$-value. This results in a total of 20 relations, constraining 12 quantities describing the formation and assembly history of the host galaxy. The table lists the Spearman and Pearson correlation coefficients and $p$-values, as well as the slope and intercept of each best-fitting relation. The final column lists the scatter of the data around this relation, which is defined as the standard deviation of the data (with values $y_{\rm data}$) around the best-fitting relation, i.e.~as $\sigma(y-y_{\rm data})$. Comparing to the median values of the galaxy-related quantities in \autoref{tab:galmetric1} and \autoref{tab:galmetric2}, the relative uncertainties range from 9-70~per~cent. This means that the listed relations all have relatively high precision, providing useful constraints on galaxy formation and assembly histories.

In addition to the high precision of individual relations listed in \autoref{tab:fit}, several quantities can be determined using multiple independent GC-related quantities. Specifically, the dark matter halo concentration parameter $\cnfw$, the total number of mergers $\nbr$, and the number of tiny mergers $N_{<1:100}$ appear in three relations each. In addition, sets of two independent relations can be used to determine the lookback time at which 25~per~cent of the host halo mass has assembled $\ttf$ and the number of high-redshift (i.e.~$z>2$) mergers at $\nbrz$. In observational applications of these correlations, these quantities provide a helpful opportunity to evaluate the accuracy of the proposed relations between the GC population and their host galaxy's formation and assembly history.

\subsubsection{Physical interpretation of significant correlations} \label{sec:phys}
Qualitatively, the relations of \autoref{tab:fit} can be understood in terms of reasonable physical trends.
\begin{enumerate}
\item
The maximum circular velocity $\vmax\appropto c_{\rm NFW}^{0.22}M_{200}^{1/3}$ (where the scaling is appropriate for the range of concentration parameters listed in \autoref{tab:galmetric1}, cf.~\citealt{mo10}, chapter~11.1.2) and the number of GCs, $\ngc$, are both positively correlated with the concentration parameter (see below) and with the halo mass \citep[e.g.][]{durrell14,harris17}. These two correlations result in richer GC populations in higher-mass haloes with higher concentration parameters (and thus overdensities, see below). Together, they thus yield a strong correlation between $\vmax$ and $\ngc$.
\item
At fixed halo mass, the collapse redshift of galaxy haloes increases with the halo concentration parameter ($\cnfw$) \citep{navarro97,bullock01,correa15}, because the concentration parameter is related to the characteristic (over)density of a dark matter halo as $\rho_{\rm char}\propto\delta_{\rm char}\appropto c_{\rm NFW}^{2.60}$ (where the scaling is appropriate for the range of concentration parameters listed in \autoref{tab:galmetric1}, cf.~\citealt{mo10}, chapter~7.5.1), which in turn depends on the mean cosmic density at some characteristic epoch during the assembly history of the halo \citep{mo10}. In terms of the correlations discussed here, this results in more rapid galaxy assembly and metal enrichment at high $\cnfw$, which naturally implies a shorter time-scale for the assembly of the GC population. This explains the increase of $\cnfw$ with the interquartile range aspect ratio ($\riqr=\iqr(\feh)/\iqr(\tau)$) and its decrease with $\iqr(\tau)$. These two correlations are not independent and result from the same physics. After all, even if the GC population forms rapidly, i.e.~over a small $\iqr(\tau)$, the monotonicity of metal enrichment implies that the range of $\iqr(\feh)$ is unchanged. This means that $\riqr$ and $\iqr(\tau)$ are strongly covariant, as shown by their Spearman $r=-0.79$. Finally, the positive correlation between $\cnfw$ and $\ngc$ can be understood in terms of the conditions of star formation at high redshift -- haloes that collapse early (resulting in high $\cnfw$) spend a larger fraction of their growth under the high-pressure conditions that are conducive to massive cluster formation.
\item
The next set of five correlations reflect similar physical processes. These relate the lookback times or redshifts at which a certain percentile of the halo mass has been assembled ($\ttf$ and $\tfz$) or half of the stellar mass has either formed ($\tf$) or assembled into the main progenitor galaxy ($\za$) to the median GC age $\widetilde{\tau}$ and the metal enrichment rate traced by the GC population $\dfehdt$. More rapid early galaxy growth leads to larger characteristic lookback times, which are naturally traced by higher median GC ages and metal enrichment rates. Each of these correlations is physically reasonable, as is underlined by the fact that these four lookback times and redshifts all positively correlate with the median age and enrichment rate at some level in \autoref{tab:correlations_sr} and \autoref{tab:correlations_sp}. \autoref{tab:fit} lists the five strongest of these correlations.
\item
Similar trends are found among the quantities representing the numbers of (different kinds of) mergers. The number of mergers prior to $z=2$ ($\nbrz$) increases with the GC median age and metal enrichment rate, because $\nbrz$ reflects the rapidity of early galaxy assembly. A similar correlation applies to the total number of mergers ($\nbr$) and the interquartile range aspect ratio, the GC metal enrichment rate, and the number of GCs. Each of these correlations reflects the well-known result that a high merger rate is associated with rapid galaxy growth \citep[e.g.][]{qu17} and enrichment (cf.~\autoref{fig:trees}). This produces old GC populations with a narrow range of ages and steep age-metallicity relations, thus explaining the first four of these relations. The number of GCs increases with the number of mergers for two reasons. Firstly, a high merger rate is largely driven by minor mergers, because they are the most numerous. The accreted, sub-$L^*$ galaxies typically have a higher number of GCs per unit stellar mass than Milky Way-mass galaxies \citep[e.g.][]{peng08,larsen14,lamers17}, implying that such minor mergers must increase the number of GCs at a fixed $z=0$ galaxy mass range. Secondly, the accelerated galaxy growth that characterises galaxies with a large number of mergers is accompanied by elevated gas pressures (\citetalias{pfeffer18}). The fraction of star formation that results in GCs increases with the gas pressure too \citep[e.g.][]{elmegreen97,kruijssen15b}, which contributes to the relation between $\nbr$ and $\ngc$. The relation between merger-driven galaxy growth and GC formation also explains why the median age of the GC population is positively correlated with the fraction of mergers taking place at $z>2$ ($\rz$).
\item
The final five correlations again highlight a strong relation between the rapidity of galaxy assembly traced by GCs (through their number, $\ngc$, interquartile range aspect ratio $\riqr$, and metal enrichment rate, $\dfehdt$) and the accreted satellite population (through the total number of progenitors, $\nleaf$, the number of tiny mergers, $N_{<1:100}$, or the number of minor mergers, $N_{1:100-1:4}$). These correlations are again driven by a combination of processes. As before, the minor merger rate itself reflects the rapidity of galaxy growth (explaining the correlations with $\riqr$ and $\dfehdt$ and partially those with $\ngc$), but low-mass galaxies also have a higher number of GCs per unit stellar mass, implying that they contribute more GCs than stars to the central galaxy (partially explaining the correlations with $\ngc$). The final 11 relations listed in \autoref{tab:fit} thus reflect similar underlying physical processes.
\end{enumerate}
In summary, we see that the distribution of GCs in age-metallicity space traces the early phase of host galaxy formation and assembly. Interestingly, both the metal enrichment rate traced by the GC population and the number of GCs are good predictors of the number of minor mergers experienced by the host galaxy. This is unlikely to be caused by a hidden dependence of multiple quantities on the halo mass, because it correlates only weakly with most other galaxy-related quantities due to the relatively small range of halo masses spanned by our simulations. Instead, the reason for this correlation between GC properties and the number of minor mergers may be threefold. Firstly, minor mergers bring in GCs due to their high specific frequencies. Secondly, such accretion events may induce star formation and the associated formation of massive clusters. Thirdly, the accretion and disruption of satellite galaxies separates GCs from the disruptive ISM in their host galaxy, by facilitating their migration into a galaxy halo. This enhances GC survival, especially at early cosmic times \citep[e.g.][]{kravtsov05,kruijssen15b}. We note that no correlation exists between GC-related quantities and major merger statistics (see \autoref{tab:correlations_sr} and \autoref{tab:correlations_sp}), highlighting that major mergers are not the main drivers of GC formation \citep[cf.][]{ashman92,forbes97}, at least not in Milky Way-mass galaxies. Only at low redshift ($z\sim0$), when the conditions of star formation do not regularly lead to massive cluster formation, are major mergers expected to dominate the formation of `young GCs' \citep[e.g.][]{schweizer98,bastian06,kruijssen14c}.

In addition to the galaxy-related metrics listed in \autoref{tab:galmetric1} and \autoref{tab:galmetric2}, we have also correlated the GC-related quantities with the cosmic dark matter density within a $2~\mpc$ radius surrounding each halo, to probe any links between GC formation and large-scale cosmic structure. \autoref{fig:sims} shows that the 25 simulated galaxies reside in a variety of cosmic environments -- their ambient cosmic density spans a factor of 12.5. Despite this dynamic range, we find no correlation of the $2~\mpc$-averaged cosmic density with any of the GC-related quantities. This confirms the picture sketched above that the formation and assembly of GC populations is dominated by smaller-scale structure and correlates specifically with the richness of the full progenitor galaxy population, with an emphasis on the more numerous, minor mergers, as well as the collapse redshift and metal enrichment rate of the main progenitor.

\autoref{tab:fit} lists a large number of correlations involving the GC median age $\widetilde{\tau}$ (4 correlations) or the number of GCs $\ngc$ (6 correlations). Before proceeding, two points of discussion are in order concerning these quantities.
\begin{enumerate}
\item
As stated in Section~\ref{sec:gcmetrics}, relative GC ages can be measured with a reasonable accuracy, but absolute GC ages are notoriously difficult to obtain, with a typical uncertainty of $1$--$2~\gyr$ \citep[e.g.][]{marinfranch09,dotter10,dotter11,vandenberg13}. However, we do not expect this uncertainty to undermine the diagnostic power of the relations involving the median age in \autoref{tab:fit}, because they exhibit a scatter (final column) that is similar in magnitude to the slope of the relation (${\rm d}y/{\rm d}x$) in units of $\gyr^{-1}$. This means that a systematic age bias in excess of $1~\gyr$ is needed to significantly change the results \citep[also see][]{kruijssen18c}. The redundancy presented by the pairs of relations providing $\ttf$ and $\nbrz$ from the median age ($\widetilde{\tau}$) and the metal enrichment rate ($\dfehdt$) may provide an additional way of evaluating whether the absolute age calibration of an observed GC population is accurate.
\item
We discussed in \citetalias{pfeffer18} that \emosaics underestimates the cluster disruption rate, because the resolution of the simulations is insufficient to completely resolve the ISM and its ability to tidally perturb and heat the clusters. As a result, the number of GCs is slightly overestimated, even at the high ($M>10^5~\msun$) cluster masses to which we have restricted the age-metallicity distributions. To remedy this, we estimate the factor by which \emosaics overpredicts $\ngc$ by comparing the total GC mass function across all 25 simulations to the best-fitting evolved Schechter function to the observed GC mass function from \citet{jordan07} and integrating both over the interval $M=[10^5~\msun,\rightarrow\rangle$. Doing so shows that we overpredict $\ngc$ by a factor of $1.75$ in this GC mass range, resulting in a correction factor of $\fcorr=1/1.75\approx0.57$ with $\ngcp=\fcorr\ngc$. We have used this expression to correct all best-fitting relations in \autoref{tab:fit}, which indeed lists the dependences on the corrected $\ngcp$ rather than $\ngc$. This enables the relations listed in \autoref{tab:fit} to be applied directly to observed GC populations. By contrast, when applying the correlations from \autoref{tab:fit} to any of the \emosaics simulations, the simulated number of GCs $\ngc$ should be multiplied by $\fcorr$ before substituting it into the listed relations.
\end{enumerate}
Keeping these considerations in mind, the framework presented in this section quantitatively connects the GC population of the simulated galaxies to their formation and assembly histories. Specifically, we find that GCs are strong probes of the host galaxy's early formation and assembly history (through $\cnfw$, $\ttf$, $\tfz$, $\za$, $\tf$, $\nbrz$, and $\rz$), as well as of its merger tree, predominantly in connection to the more numerous minor mergers that the galaxy experienced (through $\nbr$, $\nleaf$, $N_{<1:100}$, and $N_{1:100-1:4}$). This demonstrates that GC age-metallicity distributions are strongly correlated with the growth of the host galaxy and explicitly realises the largely unfulfilled potential of tracing galaxy formation and assembly using the properties of observed GC populations.

\section{Reconstructing galaxy merger trees using the GC age-metallicity distribution} \label{sec:recon}
Having identified the key properties of the GC age-metallicity distribution that trace galaxy formation, we now consider the detailed structure of these distributions and show how their individual features can be used to reconstruct (part of) the merger tree of the host galaxy. This provides a unique opportunity to derive how real-Universe galaxies like the Milky Way assembled over cosmic history, which we address specifically in \citet{kruijssen18c}.

\subsection{GCs and galaxy merger trees in age-metallicity space} \label{sec:ageztrees}
\begin{figure*}
\includegraphics[width=0.96\hsize]{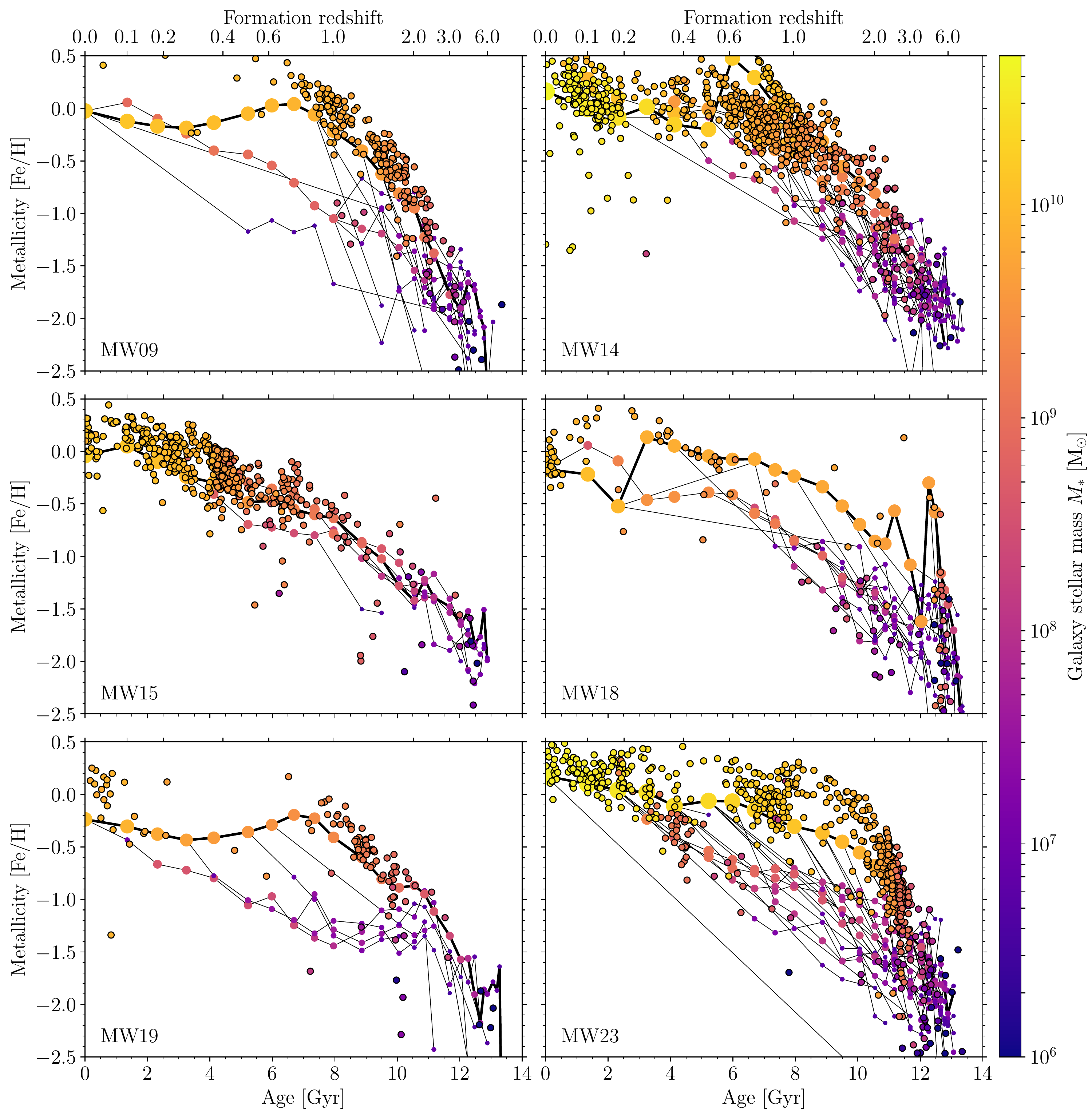}%
\caption{
\label{fig:ageztrees}
Galaxy merger trees for six of our Milky Way-mass galaxies at $z=0$ from \autoref{fig:trees} (MW09, MW14, MW15, MW18, MW19, MW23; indicated in the bottom-left corner of each panel), projected onto the age-metallicity plane from \autoref{fig:agez}, showing the median metallicities of the newly formed stellar particles in each progenitor as a function of age and formation redshift. As before, the main branch is indicated by a thick line, whereas accreted satellites are shown as thin lines. The symbols are coloured according to the instantaneous (host) galaxy stellar mass as indicated by the colour bar, both for the GCs (circles with solid edges) and for the merger tree (borderless circles). The symbols on the merger tree are only shown for galaxies with stellar masses $M_*\geq4.5\times10^6~\msun$. Relative to \autoref{fig:agez}, we have extended the metallicity range to be able to follow the main branch down to $z=0$. The overdensities of GCs along the branches of the merger trees show that (substructure in) the GC age-metallicity distribution traces the merger tree of the host galaxy.}
\end{figure*}
In order to connect directly the distribution of GCs in age-metallicity space to the host galaxy's merger tree, we can overlay the merger trees of \autoref{fig:trees} on the GC age-metallicity distributions of \autoref{fig:agez}. This is shown in \autoref{fig:ageztrees}, which considers the same six example galaxies as before. The correspondence between the merger trees and the GC age-metallicity distributions immediately confirms the interpretation made in the discussion of \autoref{fig:agez} that the low-metallicity `branches' towards younger ages are the result of satellite accretion. Indeed, the four galaxies in which we identified such a clear bifurcation into a `main branch' and a `satellite branch' (MW09, MW18, MW19, and MW23) all have merger trees with a pronounced population of lower-metallicity satellites that do not merge into the main branch until $z<0.5$. These GC populations are generally coloured similarly to the nearest branches of the satellites in the merger trees, indicating that the GCs formed in galaxies with masses matching those of the satellites.

The similarity in age-metallicity space of the GCs and the satellites shows that the presence of satellite branches in observed GC age-metallicity distributions provides an unambiguous way of identifying satellite accretion. However, the inverse does not hold -- MW14 undergoes two late ($z<0.5$) major mergers, but these are not easily distinguished when projecting its merger tree onto the age-metallicity plane. This is caused by the fact that the three galaxies involved have similar metallicities, thus obstructing the identification of separate branches. \autoref{fig:ageztrees} clearly shows that the age-metallicity plane is most useful for identifying accretion events with large mass ratios.

One way of possibly inferring the existence of multiple progenitor galaxies with similar metallicities or enrichment histories is to consider the typical number of GCs brought in by galaxies forming stars at a given age-metallicity coordinate. If there exists a significant overabundance of GCs at a given age and metallicity relative to the expected number of GCs, this may in principle be used to infer the number of progenitors occupying that part of age-metallicity space. We explore this idea in Section~\ref{sec:galrecon} and~\ref{sec:gcrecon} below, but emphasise that an elaborate statistical approach may be necessary to identify such age-metallicity degeneracies between progenitor galaxies with greater confidence.

Another important result from \autoref{fig:ageztrees} is that the upper envelope of the GC population traces the enrichment history of the main progenitor, even if the most metal-rich GCs are offset to higher metallicities by $\Delta\feh=0.0$--$0.6~\dex$. This offset is not necessarily surprising, because the high gas pressures leading to GC formation are preferentially found near galactic centres (see e.g.~Figure~4 of \citetalias{pfeffer18}), where the metallicity is greatest. As a result, the most metal-rich GCs form near the galactic centre at metallicities higher than the median metallicity of star formation in the entire galaxy. Another way in which the GC metallicity may be elevated above the main branch is in the case of a major merger, in which the lower-mass galaxy has experienced a more rapid enrichment history than the main branch and is thus forming its GCs at higher metallicities. Across our sample of 25 galaxies, we find that this is relatively rare and only occurs for any significant fraction of the main galaxy's history in MW16 (not shown in \autoref{fig:ageztrees}). The formation of GCs above the main branch in age-metallicity space is thus dominated by central star formation in the main progenitor. Correcting for this small bias through a nominal downward shift of $\Delta\feh=0.3~\dex$, we find that the maximum GC metallicity in a moving redshift window of width $\Delta z\sim0.5$ is a suitable tracer of the host galaxy's in-situ metal enrichment history.\footnote{The appropriate magnitude of the metallicity shift depends on the metallicity gradient of the galaxy, which is set by the subgrid physics included in the simulations. It may therefore differ in other simulations and may require modification for observational applications. However, in the absence of further evidence, we maintain the metallicity shift advocated here.}

Finally, \autoref{fig:ageztrees} makes an important modification to the commonly-posited picture that metal-poor GCs formed ex-situ, in accreted satellites \citep[e.g.][]{renaud17}, even if this is restricted to GCs that do not show disc-like kinematics and chemistries \citep[e.g.][]{leaman13}. Defining `metal-poor' as $\feh<-1$, we indeed see that metal-poor GCs having formed at $z<2$ are typically associated with the satellite branch. This fits the general idea that, at a given age, the more metal-poor stars formed in lower-mass galaxies due to the galaxy mass-metallicity relation. However, about 50~per~cent of the metal-poor GCs that formed at $z>2$ are associated with the main progenitor (see the colour coding of the data points in \autoref{fig:ageztrees}), forming in-situ at times when the main progenitor has not yet undergone significant metal enrichment. This is a natural phase in the formation of the host galaxy and shows that it is an oversimplification (or even incorrect) to equate `metal-poor' to `accreted' when describing the origins of GCs, as is often done in the literature. More broadly, it is questionable how useful the distinction between `in-situ' and `ex-situ' is when discussing the many low-mass progenitors at high redshift $z\ga2$ that merge to form the central spheroid of the host galaxy. We therefore recommend reserving the term `ex-situ' for accretion events taking place at redshift for which the satellite branch in the GC age-metallicity distribution has clearly separated from the main branch (i.e.~$z\la2.5$), after the peak of the cosmic star and GC formation history \citep{reinacampos19}, when the central spheroid has formed and accretion events mainly contribute to the spatially extended (`halo') GC population.

The above discussion establishes the framework for reconstructing galaxy merger trees using the GC age-metallicity distribution. While promising, three caveats are in order when considering potential applications of this method.
\begin{enumerate}
\item
GCs on a satellite branch only provide an upper limit to the time of accretion. Their existence indicates the presence of a lower-mass satellite at the time of their formation, but does not reveal when that satellite accreted onto the main branch.
\item
The central galaxies and their satellites do not only grow by merging, but mostly by gas accretion from the circumgalactic medium. This process typically lowers the metallicity at which stars are being formed in the accreting galaxy. A good example is provided by MW14 (top-right panel in \autoref{fig:ageztrees}), which undergoes several (major) mergers between $z=0.5$ and $z=0$, with each of the galaxies bringing in their own satellite population. The intense merger activity is accompanied by an episode of enhanced gas accretion, leading to the formation of metal-poor GCs with metallicities up to $\Delta\feh=1.4~\dex$ below the main branch without an associated satellite progenitor. About half of these formed within 8~kpc of the centre of the main progenitor, indicating in-situ star formation. This illustrates that the presence of low-metallicity GCs at low redshift ($z<1$) without an associated satellite branch extending to high redshifts ($z>2$) should be attributed to enhanced gas accretion, which may be merger-induced.
\item
When using the most metal-rich GCs to infer the enrichment history of the main branch, isolated high-metallicity outliers should be ignored. MW15 (middle-left panel of \autoref{fig:ageztrees} provides a good example -- it hosts one GC at $\{\tau,\feh\}=\{11~\gyr,-0.4\}$ that is offset from the main branch by $\Delta\feh=1~\dex$, which formed during a central starburst in a galaxy that was later accreted onto the main branch.
\end{enumerate}
Keeping these caveats in mind, a qualitative interpretation of the GC age-metallicity distribution in terms of the host galaxy's merger tree is clearly supported by the presented simulations. We now proceed to quantify this further by placing the GC age-metallicity distribution in the context of the evolution of the \eagle Recal-L025N0752 galaxies in the age-metallicity plane.

\subsection{Galaxy evolution in age-metallicity space} \label{sec:galrecon}
\begin{figure*}
\includegraphics[width=0.935\hsize]{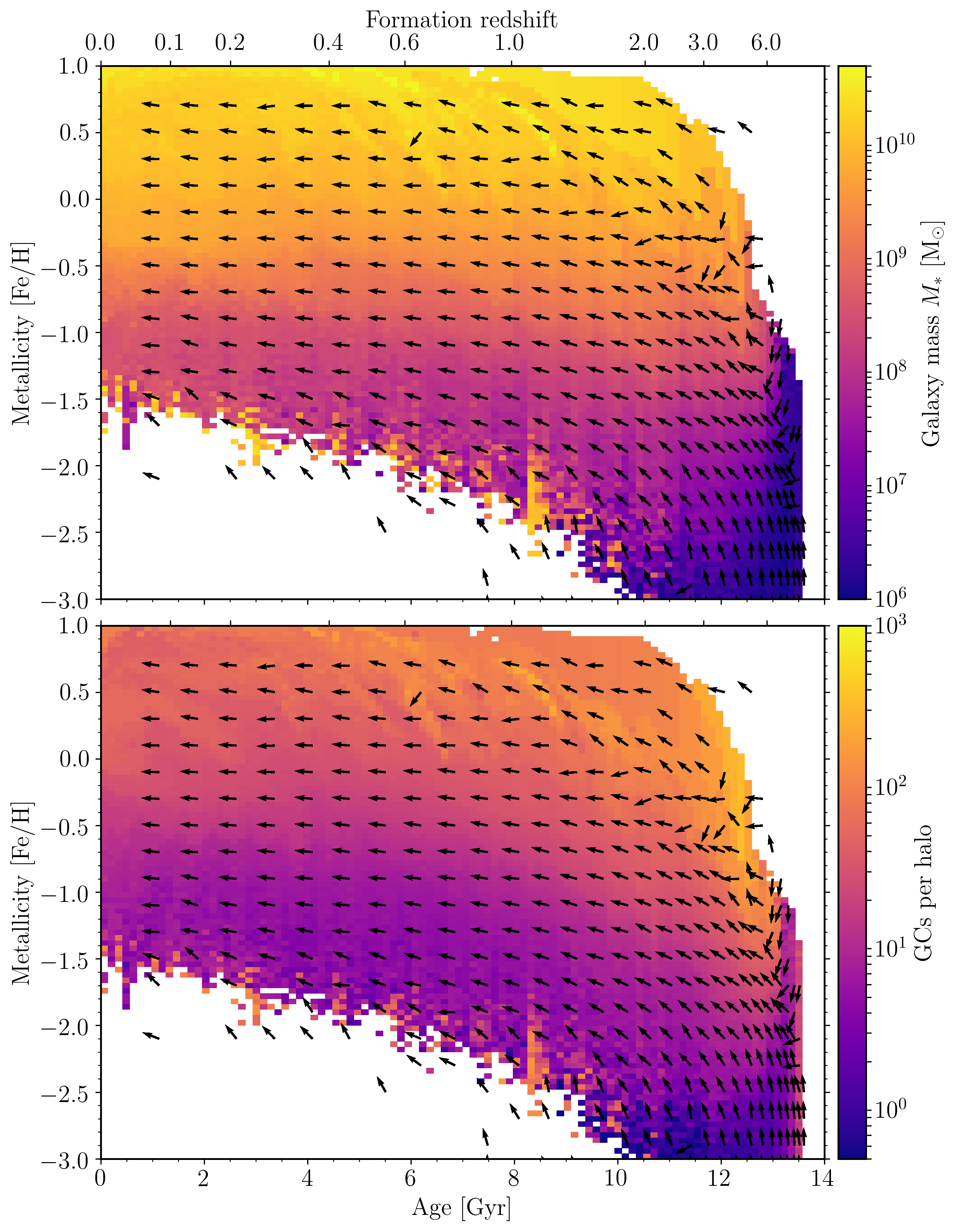}%
\caption{
\label{fig:agezgalaxy}
Age-metallicity evolution of star formation for all galaxies with halo masses $\mvir<3\times10^{12}~\msun$ within the \eagle Recal-L025N0752 simulation. Top panel: at each age-metallicity coordinate, the colour indicates the median host galaxy stellar mass of all stellar particles formed since the previous snapshot. Bottom panel: the colour indicates the median number of GCs brought in by a galaxy forming stars at that age-metallicity coordinate, based on its projected $z=0$ halo mass and the observed relation between the GC system mass and the dark matter halo mass at $z=0$ \citep[$M_{\rm GCs}/\mvir=3\times10^{-5}$,][]{durrell14,harris17}. In both panels, vectors indicate the evolution of the host's median metallicity of newly-formed stars towards the next snapshot. As shown in Section~\ref{sec:hist}, GCs trace the star formation activity and enrichment history of the host galaxy, implying that this figure can be used to reconstruct the merger tree of a galaxy from its GC population.
}
\end{figure*}
It is possible to infer a galaxy's merger tree using its GC age-metallicity distribution and to estimate the properties of its progenitors. We do this by considering the median properties of the parent galaxies of newly-formed stars at each age-metallicity coordinate. For this purpose, we use all galaxies with halo masses $\mvir<3\times10^{12}~\msun$ within the entire \eagle Recal-L025N0752 volume. Specifically, we consider the galaxy stellar mass and the expected number of GCs hosted by galaxies of that mass, which will allow us to determine how many progenitors are needed to assemble the entire GC population (see the discussion in Section~\ref{sec:ageztrees}). The median host stellar mass at the time of formation is straightforward to determine using the simulations, and the result is shown across age-metallicity space by the colour coding in the top panel of \autoref{fig:agezgalaxy}.

It is less straightforward to determine the number of GCs contributed by each progenitor galaxy, because the \eagle Recal-L025N0752 volume has no model for GC formation and evolution as in \emosaics. We therefore use the observed relation at $z=0$ between the dark matter halo mass and GC system mass from \citet{durrell14} and \citet{harris17} to predict the number of GCs per halo.\footnote{Note that the \emosaics galaxies are consistent with the observed relation between the GC system mass and the halo mass at $z=0$. We will demonstrate this in a future paper. For the present work, this motivates the extension of the observed relation to the entire \eagle Recal-L025N0752 volume.} At each age-metallicity coordinate, we first determine the instantaneous parent subhalo mass of the new-born stars. These are then converted to the projected halo mass at $z=0$ by performing power law fits between the $z=0$ central halo mass and the main progenitor halo mass at each redshift, thus connecting each age-metallicity coordinate to a (median) halo mass at $z=0$. The observed relation of $M_{\rm GCs}/\mvir=3\times10^{-5}$ from \citet{durrell14} and \citet{harris17} then provides the total GC population mass $M_{\rm GCs}$. Finally, we convert this to a total number of GCs by assuming a mean GC mass of $\overline{M}=4.7\times10^5~\msun$.\footnote{This is the mean mass obtained from the best-fitting evolved Schechter function to the observed GC mass function from \citet{jordan07} by integration over the interval $M>10^5~\msun$. Due to the slight overproduction of GCs in \emosaics at the low end of this mass range (see Section~\ref{sec:phys}), the mean mass in \emosaics is a factor of 1.4 lower at $\overline{M}=3.4\times10^5~\msun$. While the bottom panel of \autoref{fig:agezgalaxy} thus technically underestimates the expected number of GCs contributed by the galaxies in the \emosaics simulations by a factor of 1.4, we choose to adopt the higher mean mass to maintain consistency with future observational applications of this figure.} The result is shown across age-metallicity space by the colour coding in the bottom panel of \autoref{fig:agezgalaxy}. This approach for estimating the number of GCs contributed by a galaxy assumes that it survives past the peak of the GC formation history. If it is accreted and disrupted prior to that time, GC formation is cut short and the number of GCs per halo shown in \autoref{fig:agezgalaxy} provides an upper limit.

In addition to the colouring by the expected instantaneous stellar mass of the host galaxy and total expected number of GCs per halo, both panels of \autoref{fig:agezgalaxy} also show a vector field indicating the population-averaged direction in which a galaxy forming stars at each age-metallicity coordinate evolves in the \eagle Recal-L025N0752 volume. On average, the branches of merger trees projected into age-metallicity space are thus expected to align with the displayed vector fields. This is helpful when trying to identify groups of GCs that likely belong to the same progenitor.

Overall, the median stellar masses and numbers of GCs per halo in \autoref{fig:agezgalaxy} exhibit qualitatively similar behaviour. Both quantities increase with metallicity, showing that metal-rich galaxies are more massive and contribute a larger number of GCs. At fixed metallicity, they also increase with cosmic time, indicating that galaxies grow their mass and number of GCs with time, and also evolve to higher metallicities due to star formation-driven metal enrichment. This latter property of the galaxies is mirrored in the fact that the vectors typically point towards increasing metallicities. Somewhat counterintuitively, both panels also show that some of the most metal-poor stars forming at low redshifts ($z<2$ and $\feh<-1.5$) are born in massive haloes. These orange and yellow features represent tidal stripping from accreted satellites. This highlights the same ambiguity between accreted satellites and gas accretion as in the discussion of MW14 and \autoref{fig:ageztrees} in Section~\ref{sec:ageztrees}, which in a small number of cases obstructs the unambiguous identification of the progenitor galaxies using the GC population. It may be possible to avoid such features by using a more sophisticated way of identifying the host galaxy haloes of the formed stars and GCs, but this would not yield significant additional insight -- after all, very few GCs reside in this part of age-metallicity space \citep[e.g.][]{forbes10,dotter10,dotter11,vandenberg13}. Given the small expected number of cases where the ambiguity between low or high halo masses may arise, it is advisable to apply a simple `visual interpolation' of the blue and purple background around the orange and yellow features.

Finally, we point out a handful of bright yellow features at high ($\feh>-0.5$) metallicities (e.g.~at $z=0.6$, $z=1$, and $z=5$). These represent episodes of rapid galaxy growth, which is sometimes driven by major mergers, during which the characteristic metallicity rapidly increases. Due to the high metallicities, this part of age-metallicity space is dominated by a small number of massive galaxies, which allows their individual evolutionary tracks to show up when taking the median over the entire galaxy population. As with the contamination by massive galaxies at very low metallicities and low redshifts, very few GCs are associated with these age-metallicity coordinates, implying that a simple visual interpolation of the colour field around these features suffices to deal with such individual cases.

\subsection{GCs ages and metallicities as tracers of the host galaxy mass growth} \label{sec:gcrecon}
\begin{figure*}
\includegraphics[width=\hsize]{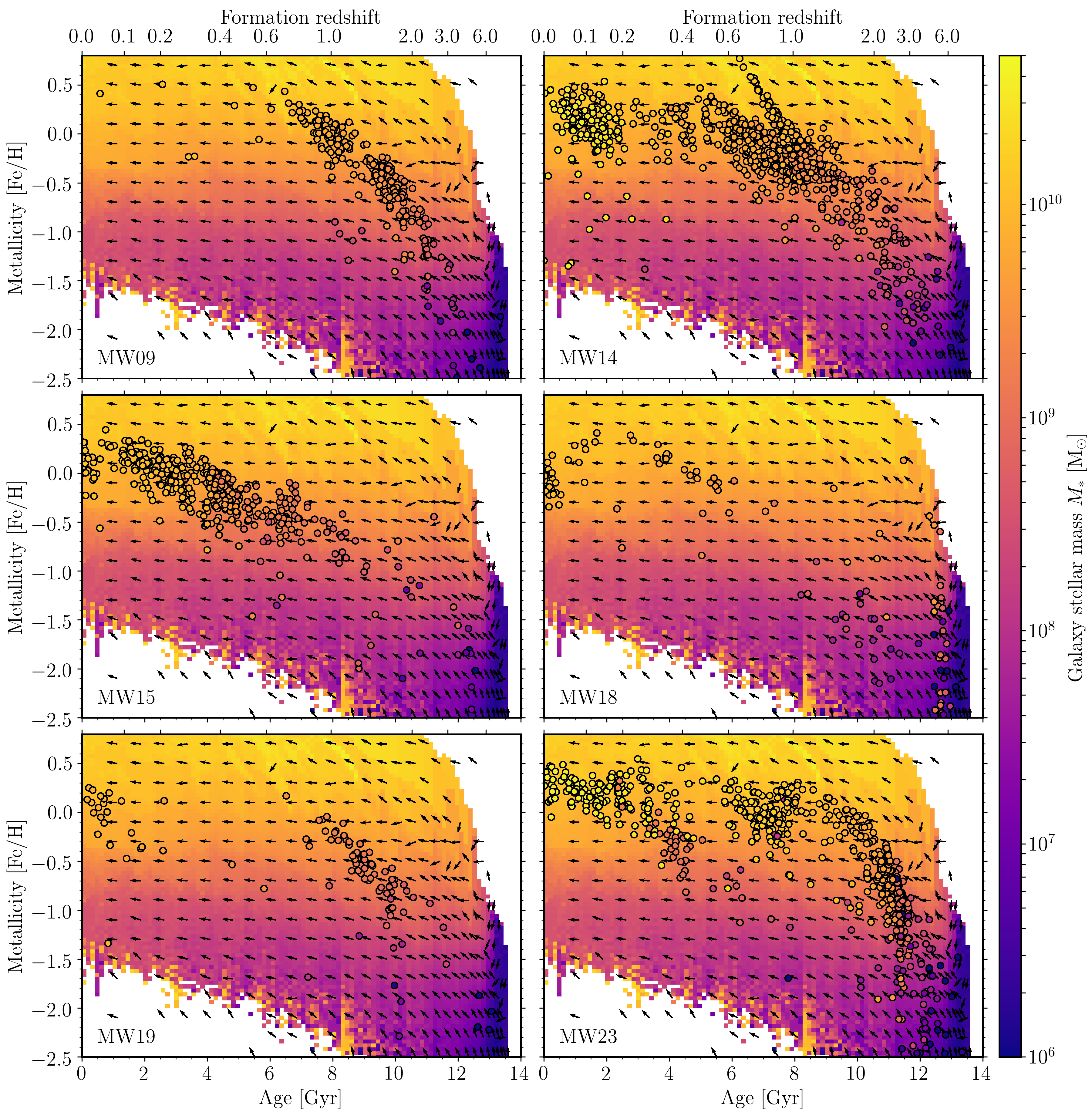}%
\caption{
\label{fig:agezmatch}
Age-metallicity distribution of GCs for six of our Milky Way-mass galaxies at $z=0$ from \autoref{fig:agez} (MW09, MW14, MW15, MW18, MW19, MW23; indicated in the bottom-left corner of each panel), with the symbol colour indicating the host galaxy stellar mass at the time of GC formation. The background colour shows the median host galaxy stellar mass of stellar particles born at each age-metallicity coordinate from \autoref{fig:agezgalaxy}, with vectors indicating the host's median metallicity evolution. The colour similarity between the symbols and the background shows that the distribution of GCs in age-metallicity space is consistent with the star formation activity of the host galaxy. In conjunction with the identification of satellite branches as in \autoref{fig:ageztrees} (which broadly follow the vector field in this figure), this enables reconstructing the merger tree of the host galaxy.}
\end{figure*}

We now evaluate the accuracy of our method for reconstructing the host galaxy merger tree using GCs. \autoref{fig:agezmatch} shows the GC age-metallicity distributions for the six example galaxies used throughout this paper, coloured by the host galaxy stellar mass at the time of their formation. The background shows the expected host stellar mass from the \eagle Recal-L025N0752 volume as in the top panel of \autoref{fig:agezgalaxy}, with its evolution represented by the vector field. The similarity between the symbol colours and the local background colour shows that GCs have metallicities corresponding to the metallicity at which the host galaxy is forming stars at that time. As a result, \autoref{fig:agezgalaxy} can indeed be used to infer the host galaxy mass at the time of GC formation. In addition, the sequences of GCs tracing the branches of the merger trees in \autoref{fig:ageztrees} align with the background vector field, implying that GC formation also traces the enrichment history of the host galaxy. As a result, the vectors can be used to connect `streams' of GCs in age-metallicity space with a common progenitor host galaxy. Together, this shows that the GC age-metallicity distribution can be used to broadly reconstruct the merger tree of a galaxy from its GC population.

\begin{figure*}
\includegraphics[width=\hsize]{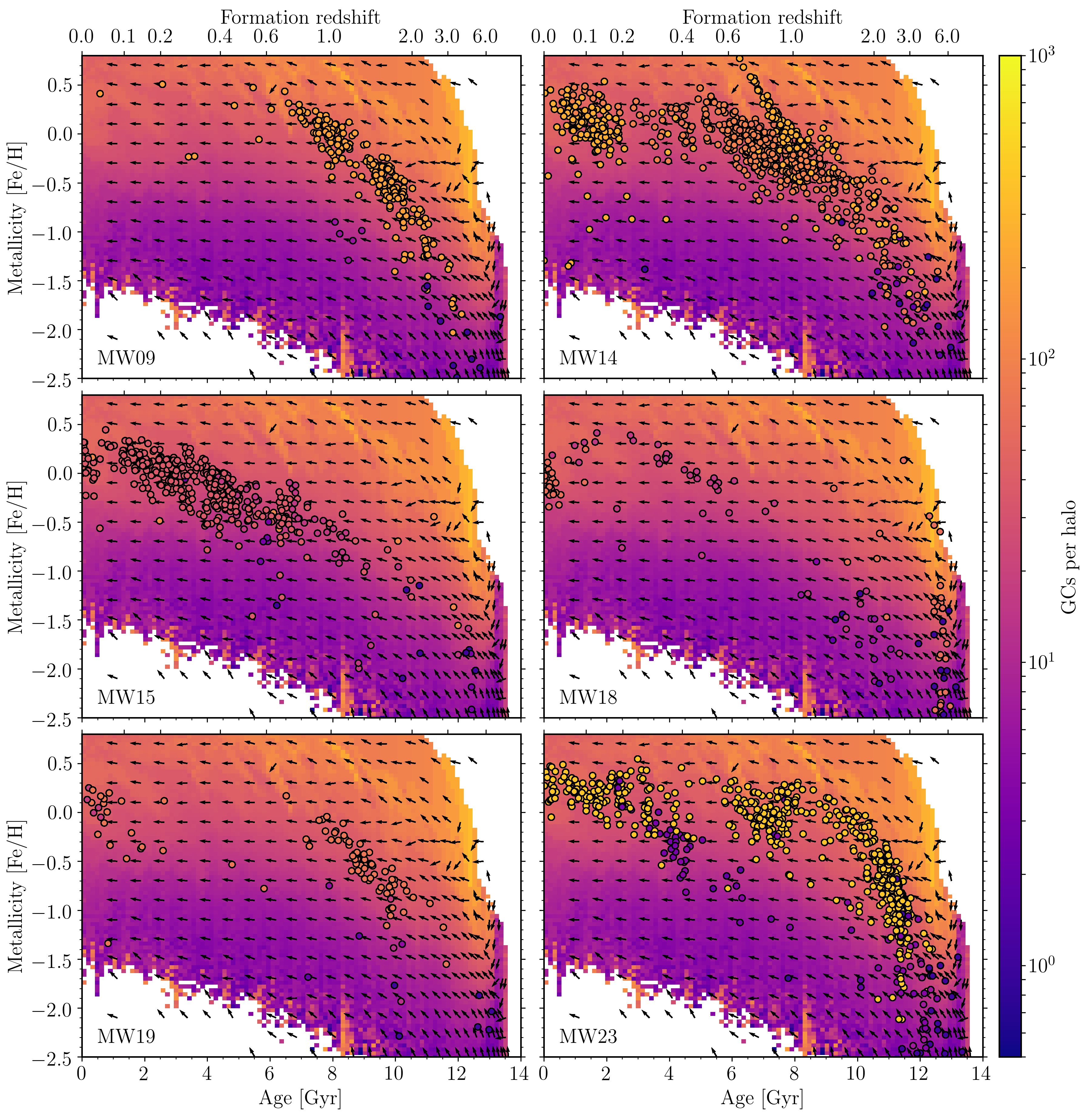}%
\caption{
\label{fig:agezn}
Age-metallicity distribution of GCs for six of our Milky Way-mass galaxies at $z=0$ from \autoref{fig:agez} (MW09, MW14, MW15, MW18, MW19, MW23; indicated in the bottom-left corner of each panel), with the symbol colour indicating the number of GCs with formation redshifts $z\geq1$ contributed by their original host galaxy. The background colour shows the median number of GCs expected to be hosted at $z=0$ by galaxies forming stars at each age-metallicity coordinate from \autoref{fig:agezgalaxy}, with vectors indicating the host's median metallicity evolution. By comparing the number of GCs to the number of GCs contributed per progenitor galaxy, it is possible to disentangle how many progenitor galaxies are needed to populate a branch in GC age-metallicity space. In combination with \autoref{fig:agezmatch}, this enables the reconstruction of the merger tree of a galaxy from its GC population.}
\end{figure*}

Of course, more than one progenitor galaxy may occupy the same part of age-metallicity space. To help identify such cases, \autoref{fig:agezn} replicates \autoref{fig:agezmatch}, this time with a background colour showing the expected number of GCs per halo from the bottom panel of \autoref{fig:agezgalaxy} and the symbol colour showing the actual number of GCs with formation redshift $z\geq1$ brought in by each GC's original host. By using this figure to infer the number of GCs expected to be hosted by a single progenitor galaxy at the age-metallicity coordinate of a (satellite) branch, it is possible to estimate how many individual progenitors may have populated that branch. This is an important ingredient for accurately reconstructing the main galaxy's merger tree. The similarity in colours between symbols in background shows that this approach works in principle, even if there exist stronger deviations here than in \autoref{fig:agezmatch}. This is caused by the scatter of the relation between halo mass and GC population mass used to calculate the background colours. Using the bottom panel of \autoref{fig:agezgalaxy} to estimate the number of progenitors thus provides a statistical average that is susceptible to (Poisson) noise.

We explore these ideas further with a number of examples, focusing on three galaxies that provide a good illustration of the diagnostic power of the method. We consider two galaxies with pronounced satellite branches (MW09 and MW18) and one with a single, wide main branch in the GC age-metallicity distribution (MW14).

\begin{enumerate}
\item[\textbf{MW09:}]
The main branch of MW09 (top-left panels of \autoref{fig:agezmatch} and \autoref{fig:agezn}) suggests stellar mass growth and metal enrichment histories of the central galaxy reaching $M_*\sim\{10^8,10^9,10^{10}\}~\msun$ and $\feh\sim\{-1.8,-1.1,-0.3\}$ at $z=\{3,2,1\}$, respectively, including the downward metallicity correction of $\Delta\feh=0.3~\dex$ for the main progenitor proposed in Section~\ref{sec:ageztrees}. As shown by \autoref{fig:ageztrees}, the true masses and metallicities at these redshifts are $M_*\sim\{4\times10^8,2\times10^9,7\times10^9\}~\msun$ and $\feh\sim\{-1.8,-1.0,-0.2\}$, which agrees well with the estimated growth and enrichment histories.

The satellite branch of MW09 suggests the existence of an intermediate-mass satellite, that grows from a stellar mass of $M_*\sim4\times10^7~\msun$ at $z=2.5$ to $M_*\sim4\times10^8~\msun$ at $z=1$ and merges into the main branch at $z<1$. Only one such satellite is expected, because the satellite branch contains 10--15 GCs and follows a track of $\sim10$ GCs per halo in \autoref{fig:agezn}. Comparison to the merger tree in \autoref{fig:trees} and \autoref{fig:ageztrees} shows that this characterisation is accurate -- the galaxy experienced one major accretion event. The satellite experienced a true mass evolution from $M_*\sim2\times10^7~\msun$ at $z=2.5$ to $M_*\sim5\times10^8~\msun$ at $z=1$, with the merger taking place just prior to $z=0$. This late merger confirms the discussion in Section~\ref{sec:ageztrees}, which highlighted that satellite branches in GC age-metallicity space only provide an upper limit to the lookback time of satellite accretion.

\item[\textbf{MW14:}]
The GC population of MW14 (top-right panels of \autoref{fig:agezmatch} and \autoref{fig:agezn}) has no unambiguous satellite branch, but stands out relative to the other galaxies through its dominant main branch. The main branch spans up to $\Delta\feh\sim1~\dex$ and hosts 240 `old' GCs (here assumed to correspond to formation redshifts $z>1$), whereas only 60--80 GCs per halo are expected based on \autoref{fig:agezn}. This implies the presence of about three similar-mass progenitors and thus two late ($z<1$) major mergers. Indeed, \autoref{fig:trees} and \autoref{fig:ageztrees} show that this galaxy formed through the coalescence of three similar-mass systems. Based on \autoref{fig:agezmatch}, these main progenitor galaxies experienced stellar mass growth and metal enrichment histories reaching $M_*\sim\{10^8,2\times10^9,6\times10^9\}~\msun$ and $\feh\sim\{-1.8,-1.0,-0.5\}$ at $z=\{3,2,1\}$, respectively, including the downward metallicity correction of $\Delta\feh=0.3~\dex$ (see Section~\ref{sec:ageztrees}). The (mean) true evolutionary histories of these galaxies from \autoref{fig:ageztrees} follow $M_*\sim\{3\times10^8,1.5\times10^9,5\times10^9\}~\msun$ and $\feh\sim\{-1.4,-1.0,-0.3\}$ at $z=\{3,2,1\}$, with factor-of-two variation around these numbers. As before, this is in satisfactory agreement with the estimated numbers. However, due to the width of the main branch, it is not possible to identify any progenitors other than the three most massive ones.

\item[\textbf{MW18:}]
The main branch of MW18 (middle-right panels of \autoref{fig:agezmatch} and \autoref{fig:agezn}) suggests stellar mass growth and metal enrichment histories of the central galaxy reaching $M_*\sim\{5\times10^9,10^{10},5\times10^9\}~\msun$ and $\feh\sim\{-0.2,-0.4,-0.5\}$ at $z=\{3,2,1\}$, respectively, again including the downward metallicity correction for the main progenitor. As shown by \autoref{fig:ageztrees}, the true masses and metallicities at these redshifts are $M_*\sim\{4\times10^9,5\times10^9,6\times10^9\}~\msun$ and $\feh\sim\{-0.6,-0.7,-0.2\}$, which again shows reasonable agreement with the estimated growth and enrichment histories. The rapid initial mass growth and enrichment at $z>3$ signals efficient in-situ star formation.

The satellite branch of MW18 has a width of $\Delta\feh\sim1~\dex$, which implies the existence of multiple satellite progenitors. At the bottom of the metallicity range, the satellites grow from a stellar mass of $M_*\sim3\times10^7~\msun$ at $z=2.5$ to $M_*\sim2\times10^8~\msun$ at $z=1$. The top of the metallicity range represents higher satellite masses, with $M_*\sim7\times10^7~\msun$ at $z=2.5$ and $M_*\sim5\times10^8~\msun$ at $z=1$. The entire satellite branch persists to $z<1$ and may merge into the main branch any time thereafter. In total, about 30 GCs are associated with the satellite branch, whereas \autoref{fig:agezn} predicts 5--20 GC per halo (the range reflects the width of the satellite branch). About half of the GCs are associated with the lower-metallicity envelope, suggesting roughly two low-mass ($\sim5$ GCs each) and one intermediate-mass ($\sim20$ GCs) progenitor. If we compare to the merger tree in \autoref{fig:trees} and \autoref{fig:ageztrees}, we see that this characterisation again provides a reasonable description of the merger tree -- the galaxy experienced one major accretion event at $z\approx0$ and two low-mass satellites merged before merging into the intermediate-mass satellite and eventually accreting onto the central galaxy. The intermediate-mass satellite experienced a true mass evolution from $M_*\sim8\times10^7~\msun$ at $z=2.5$ to $M_*\sim10^9~\msun$ at $z=1$, whereas the low-mass satellites evolved from $M_*\sim\{4,2\}\times10^7~\msun$ at $z=2.5$ to $M_*\sim\{3,1\}\times10^8~\msun$ at $z=1$, respectively. Again, these satellite masses provide an excellent match to the range inferred above from the GC age-metallicity distribution.
\end{enumerate}
The above three examples show that the GC age-metallicity distribution can be used to reconstruct part of the host galaxy's merger tree. In the above discussion, we infer a total of 15 galaxy masses and 9 metallicities, which exhibit a scatter around the 1:1 relation of $\sigma[\log{(M_*/M_{\rm true})}]=0.26$ (corresponding to a factor of 1.8 uncertainty) and $\sigma(\feh-\feh_{\rm true})=0.25$, respectively. These numbers lack major biases, as evidenced by the small mean offsets relative to the true values of $\overline{\log{(M_*/M_{\rm true})}}=-0.06$ (a factor of 0.87) and $\overline{\feh-\feh_{\rm true}}=-0.04$, respectively. The above precision and accuracy are sufficiently small to place meaningful constraints on the merger trees of observed galaxies and their GC populations.

In spite of this success, our method is not suitable for retrieving the {\it entire} merger tree. It is mostly sensitive to satellites that have been accreted since $z\sim2.5$ -- any earlier accretion events correspond to times at which the age-metallicity distribution is so steep that it can be hard to distinguish individual (satellite) branches. This is not a major concern, because the physical differentiation of central and satellite galaxies is arguably not meaningful at these early epochs. In addition, the lowest-redshift GCs on a satellite branch only provide an upper limit to the time of accretion. Given that most ex-situ GC formation has ceased at $z\la1$, the method therefore effectively allows individual progenitors at $z=1$--$2.5$ to be identified, where the upper end of the redshift range depends on when the satellite branch separates from the main branch. The effective minimum mass is $M_*\sim7\times10^7~\msun$ at $z=2$ and $M_*\sim10^8~\msun$ at $z=1$, because lower-mass galaxies rarely form GCs across that redshift range. Keeping these considerations in mind, we see that the partial reconstruction of merger trees from the GC age-metallicity distribution provides a promising complement to the correlations presented in Section~\ref{sec:hist} for inferring the assembly histories of galaxies of which GC ages and metallicities are known. We carry out such a reconstruction using the GC population of the Milky Way in \citet{kruijssen18c}.

\section{Discussion of caveats and future observational applications} \label{sec:disc}
In this paper, we have demonstrated how the age-metallicity distribution of GCs can be interpreted in terms of the formation and assembly history of the host galaxy. Broadly speaking, we have taken two complementary approaches.
\begin{enumerate}
\item
In Section~\ref{sec:hist}, we correlate quantities that parameterise the GC age-metallicity distribution to those describing the host galaxy formation and assembly history, identifying which correlations are statistically significant and thus can be used to trace galaxy formation and assembly.
\item
In Section~\ref{sec:recon}, we combine the GC age-metallicity distributions with the median stellar masses of the galaxies in the \eagle Recal-L025N0752 volume that form stars at each age-metallicity coordinate, allowing us to identify the masses and metallicities of the progenitor galaxies that contributed these GCs and perform a (partial) reconstruction of the host galaxy's merger tree.
\end{enumerate}
Both methods are potentially powerful means of inferring the formation and assembly histories of real-Universe galaxies for which the GC age-metallicity distribution has been determined. We now discuss some of the limitations to (or caveats of) this approach and explore its potential for future observational applications of our framework. We also include a comparison to previous models in Appendix~\ref{sec:appprevious}.

\subsection{Caveats of the presented methods}
The quantitative details of our results depend on the adopted selection functions of GCs and their host galaxies. This work focuses on a volume-limited sample of Milky Way-mass galaxies with halo masses $\log{\mvir/\msun}=11.85$--$12.48$. Galaxies of different masses have formation and assembly histories characterised by different collapse redshifts and satellite accretion rates \citep[e.g.][]{neistein08,fakhouri10,correa15,qu17}. As a result, the quantitative relations between the GC age-metallicity distributions and the galaxy-related quantities do not hold independently of the host galaxy mass, but require modification when applied to a different range of galaxy masses. This mass dependence is not a shortcoming of the presented framework, but a feature of galaxy formation and evolution that may be exploited in future work to further constrain the link between GC populations and galaxy formation across the entire galaxy mass range.

In this context, it is relevant to point out that the GC age-metallicity relation seems to follow a form of galaxy `downsizing', i.e.~the observation that the galaxy mass scale at which star formation proceeds most efficiently decreases with cosmic time \citep{cowie96}. Indeed, we find that GCs having formed in low-mass galaxies or satellites are often younger on average than those having formed in the main progenitor, even more so when considering GCs at fixed metallicity (but see \citealt{hughes19} for important details). Such a relation between formation redshift, metal enrichment rate, and galaxy mass implies that the derived relations must indeed change with galaxy mass. However, the GC population also exhibits a departure from downsizing, because the conditions favourable to GC formation (i.e.~a high CFE and maximum cluster mass) depend on redshift, due to the decrease with cosmic time of the gas accretion rates and ISM pressures. This redshift dependence weakens the effect of downsizing on the GC population relative to a scenario in which the GCs had formed in direct proportion to the host galaxy's stellar mass. As a result, we predict that GC ages depend less strongly on host stellar mass than the ages of field stars.

The presented correlations between GC-related and galaxy-related quantities also depend on the adopted GC mass and metallicity intervals. We have experimented with different mass or metallicity ranges and find that this leads to quantitative changes to the relation between the GC age-metallicity distribution and galaxy formation. This is not surprising -- after all, the metallicity of forming GCs depends on the metal enrichment history of the host galaxy and thus on its mass, formation, and assembly history. Likewise, changing the GC mass range leads to a formation redshift bias, because the environmental dependence of the maximum cluster mass induces variation of the upper GC mass function with redshift. Again, these are not fundamental shortcomings of our results, but represent intrinsic properties of the connection between GC populations and their host galaxies that should be kept in mind when applying our framework to observed systems. For this reason, we have chosen the metallicity interval ($-2.5<\feh<-0.5$) to match the range across which most of the age measurements of Galactic GCs have been carried out. The adopted GC mass range ($M>10^5~\msun$) covers the most massive GCs, for which \emosaics best matches observed GC populations (e.g.~\citetalias{pfeffer18}).

The above caveats are mainly relevant to the correlation between GC properties and host galaxy formation and assembly (Section~\ref{sec:hist}). Neither of them affect the method for reconstructing the host galaxy merger trees from the GC population presented in Section~\ref{sec:recon}, because that method places the selected GCs and galaxies in the broader context of the evolution and metal enrichment history of the galaxy population, covering a much wider galaxy mass range ($\mvir<3\times10^{12}~\msun$) and all of GC age-metallicity space. As a result, selection effects may only bias the inferred merger tree if the GC sample is incomplete, because an underestimated number of GCs in the bottom panel of \autoref{fig:agezgalaxy} would also lead to the underestimation of the number of progenitor galaxies required for assembling the GC population.

Finally, the \emosaics simulations themselves have shortcomings that affect our ability for identifying how GCs trace galaxy formation and assembly. Most prominently, the simulations currently lack an explicitly-modelled cold ISM, which leads to an underestimation of the cluster disruption rate by tidal perturbations from ISM substructure \citep{gieles06,elmegreen10b,kruijssen11}. Therefore, the simulations overpredict the number of GCs, especially those at young ages ($\tau\la6~\gyr$) and high metallicities ($\feh>-0.5$), as the lack of disruption allows disc clusters that are too young to have migrated into the halo to survive until $z=0$. We expect that the improved modelling of the ISM, star formation, and feedback in galaxy formation simulations coupled to the \mosaics cluster model will alleviate these problems. In this paper, we have minimised their impact by only considering GCs with $M>10^5~\msun$ and $-2.5<\feh<-0.5$, which provides a better description of observed GC populations than when including the entire simulated GC population. In addition, we have corrected the total number of GCs ($\ngc$) for the lack of disruption in Section~\ref{sec:hist}, thus ensuring that the relations between the corrected number of GCs ($\ngcp$) and galaxy-related quantities can reliably be applied to observations.

\subsection{Potential for observational applications}

Accounting for the considerations discussed above, the presented framework for reconstructing galaxy merger trees from Section~\ref{sec:recon} is suitable for application to any galaxy of which ages and metallicities have been measured for most of the GC population. For the correlations from Section~\ref{sec:hist}, the galaxy mass should fall within (or close to) the range $\log{\mvir/\msun}=11.85$--$12.48$ covered by the 25 \emosaics simulations. All of these conditions are satisfied by the Milky Way, which hosts 78 GCs with masses $M>10^5~\msun$ and metallicities $-2.5<\feh<-0.5$, of which the ages have been measured for 61 GCs (corresponding to 78~per~cent). The precision of these measurements is of the order $1$--$2~\gyr$, making the Milky Way the most obvious target for the observational application of the framework developed in this work. We carry out such an application in \citet{kruijssen18c}, where we constrain the Milky Way's mass growth, metal enrichment, and merger history, culminating in the partial reconstruction of its merger tree.

Another galaxy to which our method may be applied is M31, of which some GCs have measured ages and future observational facilities such as the upcoming thirty metre class telescopes may expand the sample. The part of our framework describing how to reconstruct the host galaxy merger tree from GCs has no restrictions on the host galaxy mass and therefore is also suitable for interpreting the GC populations of lower-mass galaxies like the Large Magellanic Cloud \citep[e.g.][]{wagnerkaiser17b}. In galaxies beyond the Local Group, the uncertainties on the GC age measurements are so large that they prohibit obtaining meaningful constraints on the host galaxy's formation and assembly history \citep[e.g.~$\sigma(\tau)\sim4~\gyr$ in NGC~4449 at $D=3.8~\mpc$, see][]{annibali18}. In view of these current limitations, we encourage future efforts for improving the precision of GC age measurements, both from spatially resolved and unresolved observations, be it through photometry or (integral field) spectroscopy. The diagnostic power of accurate GC ages for reconstructing galaxy formation and assembly is so large that future work should fully exploit this potential.

Perhaps most importantly, the demonstrated use of the GC age-metallicity distribution for tracing galaxy formation and assembly only represents a first example of the diagnostic power accessed by self-consistently modelling the co-formation and co-evolution of GCs and galaxies in a cosmological context. It is plausible that such correlations extend to other observables. The six correlations of galaxy-related quantities with the number of GCs with masses $M>10^5~\msun$ and metallicities $-2.5<\feh<-0.5$ listed in \autoref{tab:fit} are encouraging in this context, by demonstrating that relevant constraints on the host galaxy's formation and assembly history (i.e.~the maximum circular velocity, the dark matter halo concentration, the number of mergers, the number of progenitors, the number of tiny mergers, and the number of minor mergers) can be obtained without making use of the GC age information. Other properties of the GC population may carry a similar potential, possibly enabling the tracing of aspects of galaxy formation and evolution other than those identified in this paper. For instance, the spatial structure and kinematics of the GC population may encode information on the satellite accretion history of the host galaxy, especially in the outer halo, which in turn may enable the derivation of the fraction of stellar mass that formed ex-situ. Similar information may be drawn from (the metallicity dependence of) the specific frequency, or chemical abundances. In future work, we aim to explore how these GC-related quantities provide insight in the formation and assembly histories of the host galaxy.

\section{Summary and conclusions} \label{sec:concl}
We present the extension of the \emosaics project (\citetalias{pfeffer18}) to a volume-limited sample of 25 cosmological, hydrodynamical zoom-in simulations of Milky Way-mass galaxies that couple the semi-analytic \mosaics model for stellar cluster formation and evolution to the \eagle galaxy formation model. This extends the suite of 10 such simulations that we presented in \citetalias{pfeffer18} and includes all galaxies contained in the \eagle Recal-L025N0752 volume within the halo mass range $\log{\mvir/\msun}=11.85$--$12.48$. While the first paper focused on describing how our detailed, environmentally-dependent physical model for cluster formation and disruption naturally gives rise to GC populations in a cosmological context, the present work demonstrates how the resulting dependence of these GC populations' properties on the galactic environment can be used to constrain the formation and assembly histories of their host galaxies. We show that the integration of cluster formation and evolution in a cosmological context produces GC populations in good agreement with observations and fulfils the decades-old promise that GCs can be used for tracing galaxy formation and evolution. In this first application of the \emosaics simulations to this problem, we have specifically explored the diagnostic power of the distribution of GCs in age-metallicity space for reconstructing the formation and assembly history of the host galaxy. The main findings of this work are as follows.
\begin{enumerate}
\item
The \emosaics simulations reproduce several of the key demographics of GC populations at $z=0$, such as their specific frequencies, upper ($M>10^5~\msun$) mass functions, lower and intermediate ($-2.5<\feh<-1.0$) metallicity distributions, and spatial density profiles. The simulations do not perform well at lower GC masses and higher GC metallicities, because we do not model the cold ISM, which leads to an underestimation of the cluster disruption rate by tidal perturbations from ISM substructure. This is alleviated by restricting our analysis to $M>10^5~\msun$ and $-2.5<\feh<-0.5$ (also see Appendix~\ref{sec:appfeh}). A fundamental prediction of \emosaics is a considerable variety of specific frequencies and metallicity distributions between different galaxies, which reflect differences in galaxy assembly history. (Section~\ref{sec:valid})
\item
The GC age-metallicity distribution exhibits an enormous diversity across our 25 simulations. We find a close correspondence between concentrations of GCs in age-metallicity space and the peaks in the field star age-metallicity distribution, showing that the GC population probes the host galaxy's formation history. Most GC age-metallicity distributions are characterised by a main branch, which rises steeply towards high metallicities at old ages. This branch is found to correspond to the assembly of the main progenitor galaxy. In addition, many galaxies have one or more pronounced `satellite branches' in age metallicity space, which extend to younger ages at lower metallicities and contain the GCs that formed in accreted satellite galaxies and generally reside at galactocentric radii $R_{\rm gc}>10~\kpc$. The above results illustrate that the variety of GC age-metallicity distributions traces differences in the formation and assembly histories of the host galaxies. It also underlines that a single simulation of a GC population holds little diagnostic power -- reproducing the age-metallicity distribution of Galactic GCs is a matter of picking a simulated galaxy with the `right' formation and assembly history. (Section~\ref{sec:gcvariety})
\item
We characterise the GC age-metallicity distribution through 13 quantitative metrics that plausibly carry the imprint of specific events or processes driving the formation and assembly of the host galaxy. These are the median, interquartile range, skewness, and kurtosis of both the GC ages and GC metallicities, as well as the product and ratio of the interquartile ranges, the slope and zero point of a function fitted to the GC age-metallicity distribution, and the number of GCs in the considered mass ($M>10^5~\msun$) and metallicity ($-2.5<\feh<-0.5$) range. These GC-related quantities are later correlated with the formation and assembly histories of the host galaxies. (Section~\ref{sec:gcmetrics})
\item
The 25 simulated galaxies exhibit a great variety of formation and assembly histories, specifically in terms of their star formation histories, metal enrichment histories, and merger trees. Qualitatively, the differences between these galaxies are reflected by the GC age-metallicity distributions. For instance, the presence of broad satellite branches in age-metallicity space indicates a large number of minor mergers, whereas a wide and rich main branch hints at the occurrence of major mergers. Steep GC age-metallicity distributions are found in galaxies undergoing rapid growth and metal enrichment. (Section~\ref{sec:galform})
\item
The galaxy formation and assembly histories are characterised through 30 quantitative metrics. Among others, these galaxy-related quantities describe instantaneous galaxy properties (e.g.~halo mass, concentration), mass growth histories (e.g.~lookback times at which the main progenitor attains certain fractions of the halo mass or stellar mass), and merger tree properties (e.g.~lookback times of different merger types, number of various merger types, also as a function of redshift, number of progenitors, ex-situ fractions). The expansion of the \emosaics suite to 25 simulations increases the dynamic range of these quantities to such a degree that meaningful correlations with GC-related quantities can be identified. (Section~\ref{sec:galmetrics})
\item
We carry out a correlation analysis between all 13 GC-related quantities and all 30 galaxy-related quantities, resulting in a total of 390 possible correlations. Using a conservative statistical selection criterion, we retain 20 statistically (highly) significant correlations. The GC-related quantities that provide these correlations are the median GC age, $\widetilde{\tau}$, the GC age interquartile range, $\iqr(\tau)$, the interquartile range aspect ratio, $\riqr\equiv\iqr(\feh)/\iqr(\tau)$, the slope of the GC age-metallicity distribution, $\dfehdt$, and the number of GCs with masses $M>10^5~\msun$ and metallicities $-2.5<\feh<-0.5$, $\ngcp$. Together, these are shown to trace 12 key diagnostics describing galaxy formation and assembly, which are listed in \autoref{tab:fit}. These correlations are well-defined, with relative uncertainties on the galaxy-related quantities of 9--70~per~cent, and reveal several important relations between GC populations and the host galaxy's assembly history. We find that GCs trace the early ($z>1$) mass growth and merger history of the host, before half of the halo mass and stellar mass have been assembled. Late ($z<1$) mass growth and merging does not affect the properties of the GC population. Perhaps surprisingly, the GC age-metallicity distribution is almost insensitive to the major merger history of the host, but is strongly correlated with its minor merger statistics. We provide an interpretation of these results in the context of galaxy formation in the $\Lambda$CDM cosmogony. (Section~\ref{sec:correlations})
\item
By projecting the galaxy merger trees into the age-metallicity plane, we demonstrate that the detailed structure of individual GC age-metallicity distributions traces the topology of the host galaxy's merger tree and the enrichment histories of the progenitor galaxies. Contrary to the interpretation commonly found in the literature that metal-poor GCs formed ex-situ in accreted satellites, we find that about 50~per~cent of the metal-poor ($\feh<-1$) GCs at $z>2$ formed in-situ, during the early assembly and enrichment of the main progenitor. Due to the multitude of low-mass progenitors and rapid merging during this early phase of galaxy growth, we recommend to reserve the term `ex-situ GCs' for accretion events taking place at $z<2$, when the central spheroid has formed and accretion events unambiguously contribute to the spatially extended (`halo') GC population. (Section~\ref{sec:ageztrees})
\item
We use all galaxies with $\mvir<3\times10^{12}~\msun$ in the \eagle Recal-L025N0752 volume to quantify the evolution of galaxies in age-metallicity space. For each age-metallicity coordinate, we determine the median of the instantaneous stellar mass and the eventual number of GCs contributed by a galaxy forming stars at that coordinate based on its $z=0$ halo mass and the GC mass-halo mass relation of \citet{durrell14} and \citet{harris17}.\footnote{This approach for estimating the number of GCs contributed by a galaxy assumes that it survives past the peak of the GC formation history. If it is accreted and disrupted prior to that time, GC formation is cut short and the estimated number of GCs per halo provides an upper limit. However, the reasonable agreement in \autoref{fig:agezn} between the actual number of GCs contributed by each host (symbol colours) and the estimated number (background colour) shows that this potential bias is not a major concern.} In addition, we generate vector fields indicating the direction in which the star formation activity of galaxies evolves in the age-metallicity plane. This provides a method for interpreting the GC age-metallicity distributions in terms of the host galaxy's merger tree -- progenitor galaxies should generally evolve along the vector fields, with stellar masses and numbers of GCs per progenitor given by the median properties of the \eagle galaxies. (Section~\ref{sec:galrecon})
\item
We show that the GC host galaxy stellar masses and number of GCs contributed by the host that we infer from the their location in age-metallicity space agree well with their true values. This validates the idea that the GC age-metallicity distribution can be used to reconstruct the merger tree of the host galaxy. We demonstrate the process by reconstructing the merger trees of three example galaxies from our suite of 25 simulations, chosen to have rich, yet different merger histories. The method successfully recovers the metal enrichment histories of the main progenitor, identifies the satellite progenitors with masses $M_*>10^8~\msun$ at $z=1$, and retrieves the stellar mass growth histories of both the main progenitor and the accreted satellites from $z=2.5$ to $z=1$. The accuracy with which these quantities are retrieved is approximately $0.25~\dex$, corresponding to a factor of 1.8 uncertainty, which provides a physically meaningful reconstruction of the progenitor population. Due to the crowding of the progenitor galaxies in age-metallicity space at $z>2.5$, it is challenging to uniquely identify progenitors at the highest redshifts. This method therefore generally allows individual progenitors at $z=1$--$2.5$ to be identified, where the upper end of this range depends on the redshift at which the satellite branch separates from the main branch, with minimum progenitor masses of $M_*\sim7\times10^7~\msun$ at $z=2$ and $M_*\sim10^8~\msun$ at $z=1$. Lower-mass galaxies are missed due to their low numbers of GCs. It is not possible to unambiguously determine the times at which these galaxies merge with the main progenitor -- the age of the youngest GC on each branch provides an upper limit to the lookback time of the accretion event. (Section~\ref{sec:gcrecon})
\end{enumerate}

The above findings quantitatively link the GC age-metallicity distribution to the formation and assembly history of the host galaxy and demonstrate the diagnostic power gained by self-consistently modelling the co-formation and co-evolution of GCs and galaxies in a cosmological context. By quantifying this connection, the \emosaics simulations explicitly realise the largely unfulfilled potential of tracing galaxy formation and assembly using the properties of (observed) GC populations. In a follow-up paper \citep{kruijssen18c}, we apply the resulting insights to the Galactic GC population and reconstruct the formation and assembly history of the Milky Way.

While this paper largely focuses on the GC age-metallicity distribution because of its direct connection to the host galaxy's formation and assembly history, we also expect other properties of the GC population to trace specific aspects of galaxy formation. For instance, future work in the \emosaics project will investigate the relation between galaxy formation and assembly and the GC formation history, the spatial and kinematic structure of GC populations, the GC mass-metallicity distribution, and many other GC-related diagnostics that are observationally accessible. Thanks to greatly improved numerical models and the computational power of modern high-performance computing facilities, the multi-scale and multi-physics problem of GC formation during galaxy formation and assembly has become tractable. In combination with the arrival of new observational facilities that will observe GC populations across cosmic time, such as the James Webb Space Telescope and the thirty metre class ground-based observatories, these developments herald a golden age for studies of GC populations and their connection to galaxy formation and assembly.

\section*{Acknowledgements}
JMDK gratefully acknowledges funding from the German Research Foundation (DFG) in the form of an Emmy Noether Research Group (grant number KR4801/1-1, PI Kruijssen), from the European Research Council (ERC) under the European Union's Horizon 2020 research and innovation programme via the ERC Starting Grant MUSTANG (grant agreement number 714907, PI Kruijssen), and from Sonderforschungsbereich SFB 881 ``The Milky Way System'' (subproject B2) of the DFG. JP and NB gratefully acknowledge funding from the ERC under the European Union's Horizon 2020 research and innovation programme via the ERC Consolidator Grant Multi-Pop (grant agreement number 646928, PI Bastian). NB and RAC are Royal Society University Research Fellows. This work made use of high performance computing facilities at Liverpool John Moores University, partly funded by the Royal Society and LJMU's Faculty of Engineering and Technology. This work used the DiRAC Data Centric system at Durham University, operated by the Institute for Computational Cosmology on behalf of the STFC DiRAC HPC Facility (\url{www.dirac.ac.uk}). This equipment was funded by BIS National E-infrastructure capital grant ST/K00042X/1, STFC capital grants ST/H008519/1 and ST/K00087X/1, STFC DiRAC Operations grant ST/K003267/1 and Durham University. DiRAC is part of the National E-Infrastructure. This work has made use of Numpy \citep{vanderwalt11}, Scipy \citep{jones01}, and Astropy \citep{astropy13}. All figures in this paper were produced using the Python library Matplotlib \citep{hunter07}. The authors thank Marta Reina-Campos, Benjamin Keller, and M\'{e}lanie Chevance for advice and helpful discussions during the development of this work. We thank an anonymous referee for comments that improved the presentation of this paper.

\bibliographystyle{mnras}
\bibliography{mybib}

\appendix

\section{Metrics used in this paper} \label{sec:appmetrics}

\subsection{Metrics used for describing GC age-metallicity distributions} \label{sec:appmetrics_gc}

\begin{table*}
  \caption{Quantities describing the GC age-metallicity distributions in the 25 Milky Way-mass, $L^*$ galaxies at $z=0$ in the cosmological zoom-in simulations considered in this work. From left to right, the columns show: simulation ID; median GC age $\widetilde{\tau}$ in Gyr; GC age interquartile range $\iqr(\tau)$ in Gyr; GC age skewness $S(\tau)$; GC age excess kurtosis $K(\tau)$; median GC metallicity $\widetilde{\feh}$; GC metallicity interquartile range $\iqr(\feh)$; GC metallicity skewness $S(\feh)$; GC metallicity excess kurtosis $K(\feh)$; combined interquartile range $\iqr^2\equiv\iqr(\tau)\times\iqr(\feh)$ in Gyr; interquartile range aspect ratio $\riqr\equiv\iqr(\feh)/\iqr(\tau)$ in Gyr$^{-1}$; best-fitting slope $\dfehdt$ of the function from equation~(\ref{eq:fit}) indicating the rapidity of metal enrichment in the progenitor galaxies as traced by GCs; best-fitting intercept $\feh_0$ of the function from equation~(\ref{eq:fit}) indicating the typical `initial' GC metallicity at $1~\gyr$ after the Big Bang; number of GCs (defined as clusters with present-day masses of $M>10^5~\msun$) in the metallicity range $-2.5<\feh<-0.5$ that is considered throughout this work. To give an indication of the typical values and dynamic ranges, the final three rows list the median, the interquartile range, and the total range (i.e.~$\max-\min$) of each column.} 
\label{tab:gcmetrics}
  \begin{tabular}{l c c c c c c c c c c c c c}
   \hline
   Name & $\widetilde{\tau}$ & $\iqr(\tau)$ & $S(\tau)$ & $K(\tau)$ & $\widetilde{\feh}$ & $\iqr(\feh)$ & $S(\feh)$ & $K(\feh)$ & $\iqr^2$ & $\riqr$ & $\frac{{\rm d[Fe/H]}}{{\rm d}\log t}$ & $\feh_0$ & $N_{\rm GC}$ \\ 
   \hline
   MW00 & $10.92$ & $1.39$ & $0.48$ & $-0.76$ & $-1.18$ & $0.88$ & $-0.35$ & $-1.07$ & $1.22$ & $0.63$ & $4.40$ & $-3.00$ & $94$ \\ 
   MW01 & $10.88$ & $1.30$ & $-0.44$ & $0.33$ & $-0.98$ & $0.69$ & $-1.03$ & $0.45$ & $0.90$ & $0.53$ & $4.18$ & $-2.99$ & $104$ \\ 
   MW02 & $11.52$ & $1.33$ & $-1.14$ & $2.33$ & $-0.91$ & $0.83$ & $-0.77$ & $-0.54$ & $1.10$ & $0.62$ & $6.58$ & $-3.31$ & $313$ \\ 
   MW03 & $11.10$ & $1.40$ & $-0.20$ & $0.49$ & $-0.89$ & $0.78$ & $-0.80$ & $-0.44$ & $1.09$ & $0.56$ & $3.78$ & $-2.59$ & $160$ \\ 
   MW04 & $11.43$ & $0.83$ & $-0.65$ & $4.19$ & $-0.87$ & $0.65$ & $-1.15$ & $0.32$ & $0.54$ & $0.78$ & $5.30$ & $-2.79$ & $152$ \\ 
   MW05 & $11.91$ & $0.64$ & $0.14$ & $0.58$ & $-0.76$ & $0.59$ & $-1.37$ & $1.04$ & $0.38$ & $0.92$ & $6.29$ & $-2.62$ & $306$ \\ 
   MW06 & $10.30$ & $0.71$ & $-3.15$ & $28.09$ & $-0.71$ & $0.33$ & $-1.67$ & $2.18$ & $0.23$ & $0.46$ & $3.10$ & $-2.47$ & $199$ \\ 
   MW07 & $10.65$ & $2.39$ & $-0.40$ & $-0.57$ & $-0.85$ & $0.78$ & $-0.95$ & $-0.32$ & $1.86$ & $0.33$ & $2.48$ & $-2.20$ & $73$ \\ 
   MW08 & $11.97$ & $0.84$ & $-2.11$ & $6.56$ & $-1.01$ & $0.36$ & $-1.24$ & $2.09$ & $0.30$ & $0.43$ & $2.77$ & $-1.69$ & $50$ \\ 
   MW09 & $10.28$ & $1.08$ & $0.27$ & $0.63$ & $-0.88$ & $0.61$ & $-1.18$ & $0.69$ & $0.66$ & $0.56$ & $4.04$ & $-3.06$ & $100$ \\ 
   MW10 & $10.13$ & $1.09$ & $-0.03$ & $0.04$ & $-0.81$ & $0.62$ & $-1.15$ & $0.31$ & $0.68$ & $0.57$ & $3.29$ & $-2.73$ & $297$ \\ 
   MW11 & $10.73$ & $2.15$ & $-1.66$ & $2.84$ & $-0.99$ & $0.86$ & $-0.54$ & $-1.07$ & $1.85$ & $0.40$ & $3.29$ & $-2.69$ & $72$ \\ 
   MW12 & $10.49$ & $1.32$ & $-1.72$ & $9.79$ & $-0.98$ & $0.80$ & $-0.74$ & $-0.46$ & $1.06$ & $0.61$ & $5.58$ & $-3.90$ & $253$ \\ 
   MW13 & $11.10$ & $2.56$ & $-0.33$ & $-1.16$ & $-1.32$ & $0.84$ & $-0.36$ & $-0.97$ & $2.15$ & $0.33$ & $3.11$ & $-2.47$ & $99$ \\ 
   MW14 & $10.43$ & $2.07$ & $-1.92$ & $3.54$ & $-0.97$ & $0.67$ & $-0.67$ & $-0.54$ & $1.39$ & $0.32$ & $2.57$ & $-2.50$ & $139$ \\ 
   MW15 & $7.90$ & $3.46$ & $0.05$ & $-0.38$ & $-0.75$ & $0.70$ & $-1.08$ & $-0.11$ & $2.42$ & $0.20$ & $2.44$ & $-2.73$ & $73$ \\ 
   MW16 & $11.06$ & $1.54$ & $-0.51$ & $-0.05$ & $-0.99$ & $0.85$ & $-0.71$ & $-0.70$ & $1.31$ & $0.55$ & $3.25$ & $-2.48$ & $209$ \\ 
   MW17 & $9.22$ & $1.74$ & $-0.88$ & $3.95$ & $-1.02$ & $0.61$ & $-0.90$ & $0.01$ & $1.06$ & $0.35$ & $3.60$ & $-3.31$ & $101$ \\ 
   MW18 & $12.45$ & $2.26$ & $-2.04$ & $5.69$ & $-1.59$ & $0.70$ & $0.18$ & $-0.60$ & $1.58$ & $0.31$ & $2.86$ & $-2.49$ & $60$ \\ 
   MW19 & $9.73$ & $0.95$ & $-1.93$ & $8.33$ & $-0.77$ & $0.56$ & $-1.34$ & $0.70$ & $0.53$ & $0.59$ & $3.52$ & $-3.06$ & $57$ \\ 
   MW20 & $10.19$ & $1.50$ & $0.71$ & $0.12$ & $-0.90$ & $0.66$ & $-0.93$ & $0.04$ & $0.99$ & $0.44$ & $2.33$ & $-2.21$ & $87$ \\ 
   MW21 & $11.80$ & $0.91$ & $-1.82$ & $3.55$ & $-1.15$ & $0.85$ & $-0.54$ & $-0.86$ & $0.77$ & $0.93$ & $3.50$ & $-2.37$ & $135$ \\ 
   MW22 & $10.46$ & $1.40$ & $-0.37$ & $0.28$ & $-0.91$ & $0.88$ & $-0.84$ & $-0.42$ & $1.23$ & $0.63$ & $3.60$ & $-2.92$ & $192$ \\ 
   MW23 & $11.29$ & $0.44$ & $-3.15$ & $10.94$ & $-0.90$ & $0.49$ & $-1.36$ & $1.25$ & $0.22$ & $1.11$ & $4.77$ & $-3.05$ & $327$ \\ 
   MW24 & $9.36$ & $2.63$ & $0.36$ & $-1.07$ & $-0.91$ & $0.58$ & $-1.09$ & $0.27$ & $1.53$ & $0.22$ & $2.33$ & $-2.38$ & $69$ \\ 
   \hline
   Median & $10.73$ & $1.39$ & $-0.51$ & $0.63$ & $-0.91$ & $0.69$ & $-0.93$ & $-0.11$ & $1.06$ & $0.55$ & $3.50$ & $-2.69$ & $104$ \\ 
   IQR & $1.01$ & $1.12$ & $1.79$ & $4.15$ & $0.12$ & $0.22$ & $0.44$ & $0.99$ & $0.73$ & $0.27$ & $1.32$ & $0.53$ & $126$ \\ 
   Range & $4.55$ & $3.02$ & $3.86$ & $29.25$ & $0.88$ & $0.55$ & $1.85$ & $3.25$ & $2.21$ & $0.91$ & $4.25$ & $2.21$ & $277$ \\ 
   \hline
  \end{tabular} 
\end{table*}
In \autoref{tab:gcmetrics}, we list the quantities describing the GC age-metallicity distributions of the 25 simulations of Milky Way-mass galaxies considered in this work. These quantities are listed in the table caption and discussed in Section~\ref{sec:gcmetrics}.

\subsection{Metrics used for describing galaxy properties, growth histories, and merger trees} \label{sec:appmetrics_gal}

\begin{table*}
  \caption{Quantities describing the properties and growth histories of the 25 Milky Way-mass, $L^*$ galaxies at $z=0$ in the cosmological zoom-in simulations considered in this work. The structure of the table follows that of \autoref{tab:gcmetrics}. See Section~\ref{sec:galmetrics} for a detailed discussion of the listed quantities.} 
\label{tab:galmetric1}
  \begin{tabular}{l c c c c c c c c c c c c c c}
   \hline
   Name & $\mvir$ & $\rvir$ & $\vmax$ & $\rvmax$ & $\cnfw$ & $\ttf$ & $\tfz$ & $\tsf$ & $\tmax$ & $\ta$ & $\za$ & $\tf$ & $\zf$ & $\delta_t$ \\ 
    & $[10^{12}~\msun]$ & $[\kpc]$ & $[\kms]$ & $[\kpc]$ &  & $[\gyr]$ & $[\gyr]$ & $[\gyr]$ & $[\gyr]$ & $[\gyr]$ &  & $[\gyr]$ &  &  \\ 
   \hline
   MW00 & $0.88$ & $202$ & $177.3$ & $28.7$ & $11.34$ & $11.57$ & $11.04$ & $10.39$ & $0.00$ & $9.42$ & $1.46$ & $9.55$ & $1.51$ & $0.013$ \\ 
   MW01 & $1.33$ & $232$ & $184.5$ & $17.2$ & $8.12$ & $10.78$ & $9.31$ & $5.50$ & $0.00$ & $8.04$ & $1.02$ & $8.12$ & $1.05$ & $0.010$ \\ 
   MW02 & $1.95$ & $263$ & $215.2$ & $45.3$ & $9.27$ & $11.55$ & $10.62$ & $6.56$ & $0.00$ & $9.27$ & $1.40$ & $9.38$ & $1.44$ & $0.011$ \\ 
   MW03 & $1.49$ & $241$ & $191.8$ & $28.1$ & $8.25$ & $11.13$ & $10.15$ & $6.97$ & $0.00$ & $9.07$ & $1.33$ & $9.15$ & $1.36$ & $0.009$ \\ 
   MW04 & $1.04$ & $214$ & $170.2$ & $37.4$ & $8.28$ & $11.48$ & $9.48$ & $6.10$ & $0.00$ & $9.39$ & $1.44$ & $9.86$ & $1.64$ & $0.048$ \\ 
   MW05 & $1.18$ & $223$ & $188.7$ & $27.2$ & $10.91$ & $12.42$ & $10.41$ & $8.11$ & $0.00$ & $11.60$ & $2.92$ & $11.63$ & $2.96$ & $0.002$ \\ 
   MW06 & $0.91$ & $204$ & $172.1$ & $25.5$ & $9.84$ & $11.01$ & $10.37$ & $6.51$ & $0.00$ & $7.47$ & $0.89$ & $7.50$ & $0.90$ & $0.004$ \\ 
   MW07 & $0.72$ & $189$ & $150.8$ & $35.6$ & $7.68$ & $10.40$ & $8.96$ & $7.11$ & $0.00$ & $5.77$ & $0.58$ & $6.40$ & $0.68$ & $0.099$ \\ 
   MW08 & $0.74$ & $190$ & $156.2$ & $8.5$ & $8.21$ & $11.62$ & $8.38$ & $6.14$ & $4.12$ & $7.27$ & $0.85$ & $7.27$ & $0.85$ & $0.000$ \\ 
   MW09 & $0.74$ & $191$ & $159.0$ & $20.0$ & $9.87$ & $11.58$ & $8.74$ & $6.15$ & $0.00$ & $7.43$ & $0.88$ & $7.82$ & $0.97$ & $0.050$ \\ 
   MW10 & $2.30$ & $278$ & $211.5$ & $66.7$ & $7.96$ & $10.69$ & $7.92$ & $7.27$ & $1.35$ & $7.12$ & $0.82$ & $8.74$ & $1.22$ & $0.185$ \\ 
   MW11 & $1.40$ & $236$ & $154.3$ & $154.8$ & $4.84$ & $9.38$ & $6.37$ & $1.77$ & $0.00$ & $8.39$ & $1.12$ & $8.46$ & $1.14$ & $0.008$ \\ 
   MW12 & $2.19$ & $274$ & $197.7$ & $64.2$ & $7.60$ & $10.59$ & $9.63$ & $3.08$ & $0.00$ & $8.44$ & $1.13$ & $8.90$ & $1.27$ & $0.051$ \\ 
   MW13 & $2.41$ & $282$ & $208.2$ & $52.0$ & $7.79$ & $11.01$ & $8.88$ & $4.46$ & $0.00$ & $9.34$ & $1.42$ & $9.41$ & $1.45$ & $0.008$ \\ 
   MW14 & $2.21$ & $275$ & $203.3$ & $6.2$ & $8.44$ & $8.74$ & $5.97$ & $5.26$ & $3.24$ & $3.60$ & $0.31$ & $7.35$ & $0.87$ & $0.510$ \\ 
   MW15 & $1.46$ & $239$ & $176.5$ & $110.4$ & $3.35$ & $7.44$ & $6.06$ & $0.79$ & $0.00$ & $3.60$ & $0.31$ & $3.77$ & $0.33$ & $0.045$ \\ 
   MW16 & $2.08$ & $269$ & $230.4$ & $2.2$ & $7.81$ & $9.91$ & $9.35$ & $9.09$ & $7.96$ & $7.95$ & $1.00$ & $9.24$ & $1.39$ & $0.139$ \\ 
   MW17 & $1.93$ & $263$ & $216.7$ & $1.1$ & $5.51$ & $9.97$ & $9.26$ & $6.95$ & $0.00$ & $7.24$ & $0.84$ & $7.61$ & $0.92$ & $0.048$ \\ 
   MW18 & $1.77$ & $255$ & $185.4$ & $104.1$ & $5.11$ & $11.03$ & $3.88$ & $2.26$ & $0.00$ & $5.48$ & $0.54$ & $9.20$ & $1.37$ & $0.404$ \\ 
   MW19 & $1.58$ & $245$ & $151.1$ & $215.1$ & $4.35$ & $4.96$ & $3.76$ & $2.25$ & $1.35$ & $2.88$ & $0.24$ & $3.18$ & $0.26$ & $0.091$ \\ 
   MW20 & $0.94$ & $206$ & $150.8$ & $51.9$ & $6.90$ & $10.61$ & $6.56$ & $4.90$ & $2.32$ & $6.57$ & $0.71$ & $7.17$ & $0.83$ & $0.084$ \\ 
   MW21 & $1.32$ & $231$ & $177.3$ & $43.8$ & $8.77$ & $10.03$ & $8.30$ & $6.61$ & $4.12$ & $4.99$ & $0.47$ & $11.19$ & $2.49$ & $0.554$ \\ 
   MW22 & $1.40$ & $236$ & $193.8$ & $36.9$ & $9.43$ & $11.05$ & $8.54$ & $7.88$ & $6.69$ & $7.34$ & $0.86$ & $8.37$ & $1.11$ & $0.123$ \\ 
   MW23 & $1.53$ & $243$ & $207.0$ & $39.4$ & $11.20$ & $11.55$ & $8.06$ & $5.04$ & $0.00$ & $6.25$ & $0.66$ & $6.83$ & $0.76$ & $0.085$ \\ 
   MW24 & $1.15$ & $221$ & $169.9$ & $48.6$ & $6.77$ & $9.32$ & $8.33$ & $7.67$ & $3.24$ & $7.58$ & $0.92$ & $7.83$ & $0.97$ & $0.032$ \\ 
   \hline
   Median & $1.40$ & $236$ & $184.5$ & $37.4$ & $8.12$ & $10.78$ & $8.74$ & $6.15$ & $0.00$ & $7.43$ & $0.88$ & $8.37$ & $1.11$ & $0.048$ \\ 
   IQR & $0.89$ & $49$ & $33.4$ & $26.5$ & $2.38$ & $1.51$ & $1.56$ & $2.21$ & $2.32$ & $2.19$ & $0.47$ & $1.89$ & $0.52$ & $0.089$ \\ 
   Range & $1.69$ & $94$ & $79.6$ & $214.0$ & $8.00$ & $7.46$ & $7.28$ & $9.61$ & $7.96$ & $8.72$ & $2.69$ & $8.45$ & $2.69$ & $0.554$ \\ 
   \hline
  \end{tabular} 
\end{table*}
In \autoref{tab:galmetric1}, we list the quantities describing the properties and growth histories of the 25 simulations of Milky Way-mass galaxies considered in this work. Likewise, \autoref{tab:galmetric2}, lists the quantities describing the merger trees of the simulated galaxies. All quantities listed in both tables are discussed in Section~\ref{sec:galmetrics}.

\begin{table*}
  \caption{Quantities describing the merger trees of the 25 Milky Way-mass, $L^*$ galaxies at $z=0$ in the cosmological zoom-in simulations considered in this work. The structure of the table follows that of \autoref{tab:gcmetrics}. The two listed time-scales ($\tmm$ and $\tam$) have units of $\gyr$. When calculating the median, interquartile range, or total range of the lookback time ($\tmm$) and redshift ($\zmm$) of the last major merger, we assign values of $\tmm=13.64~\gyr$ and $\zmm=20$ to those simulations that do not have any major mergers. See Section~\ref{sec:galmetrics} for a detailed discussion of the listed quantities.} 
\label{tab:galmetric2}
  \begin{tabular}{l c c c c c c c c c c c c c c c c}
   \hline
   Name & $\tmm$ & $\zmm$ & $\tam$ & $\zam$ & $r_t$ & $\nbrz$ & $\nbr$ & $\rz$ & $\nleaf$ & $\rbl$ & $N_{<1:100}$ & $N_{1:100-1:4}$ & $N_{>1:4}$ & $r_{\rm mm}$ & $\fexs$ & $\fexgc$ \\ 
   \hline
   MW00 & $9.50$ & $1.49$ & $2.32$ & $0.18$ & $0.38$ & $10$ & $25$ & $0.40$ & $34$ & $0.74$ & $14$ & $6$ & $5$ & $0.25$ & $0.10$ & $0.26$ \\ 
   MW01 & -- & -- & $1.35$ & $0.10$ & $0.00$ & $7$ & $16$ & $0.44$ & $21$ & $0.76$ & $6$ & $10$ & $0$ & $0.00$ & $0.08$ & $0.26$ \\ 
   MW02 & $12.66$ & $5.04$ & $0.00$ & $0.00$ & $0.08$ & $11$ & $25$ & $0.44$ & $38$ & $0.66$ & $17$ & $5$ & $3$ & $0.14$ & $0.10$ & $0.26$ \\ 
   MW03 & $9.50$ & $1.49$ & $0.00$ & $0.00$ & $0.31$ & $6$ & $14$ & $0.43$ & $24$ & $0.58$ & $6$ & $4$ & $4$ & $0.40$ & $0.12$ & $0.29$ \\ 
   MW04 & $10.87$ & $2.24$ & $6.69$ & $0.74$ & $0.41$ & $6$ & $9$ & $0.67$ & $18$ & $0.50$ & $3$ & $4$ & $2$ & $0.29$ & $0.18$ & $0.34$ \\ 
   MW05 & $12.77$ & $5.49$ & $0.00$ & $0.00$ & $0.08$ & $13$ & $24$ & $0.54$ & $31$ & $0.77$ & $17$ & $6$ & $1$ & $0.04$ & $0.08$ & $0.15$ \\ 
   MW06 & -- & -- & $1.35$ & $0.10$ & $0.00$ & $4$ & $14$ & $0.29$ & $16$ & $0.88$ & $9$ & $5$ & $0$ & $0.00$ & $0.04$ & $0.12$ \\ 
   MW07 & $11.17$ & $2.48$ & $2.32$ & $0.18$ & $0.23$ & $4$ & $8$ & $0.50$ & $12$ & $0.67$ & $3$ & $2$ & $3$ & $0.60$ & $0.10$ & $0.18$ \\ 
   MW08 & -- & -- & $8.86$ & $1.26$ & $0.00$ & $2$ & $3$ & $0.67$ & $4$ & $0.75$ & $2$ & $1$ & $0$ & $0.00$ & $0.03$ & $0.09$ \\ 
   MW09 & $10.87$ & $2.24$ & $0.00$ & $0.00$ & $0.21$ & $6$ & $14$ & $0.43$ & $23$ & $0.61$ & $5$ & $7$ & $2$ & $0.17$ & $0.11$ & $0.21$ \\ 
   MW10 & $4.12$ & $0.37$ & $2.32$ & $0.18$ & $0.84$ & $3$ & $22$ & $0.14$ & $42$ & $0.52$ & $14$ & $3$ & $5$ & $0.29$ & $0.34$ & $0.54$ \\ 
   MW11 & $11.17$ & $2.48$ & $0.00$ & $0.00$ & $0.19$ & $4$ & $10$ & $0.40$ & $15$ & $0.67$ & $4$ & $3$ & $3$ & $0.43$ & $0.16$ & $0.35$ \\ 
   MW12 & -- & -- & $2.32$ & $0.18$ & $0.00$ & $7$ & $27$ & $0.26$ & $41$ & $0.66$ & $14$ & $13$ & $0$ & $0.00$ & $0.22$ & $0.38$ \\ 
   MW13 & $12.66$ & $5.04$ & $0.00$ & $0.00$ & $0.08$ & $9$ & $17$ & $0.53$ & $27$ & $0.63$ & $9$ & $7$ & $1$ & $0.06$ & $0.08$ & $0.20$ \\ 
   MW14 & $0.00$ & $0.00$ & $0.00$ & $0.00$ & $1.00$ & $4$ & $14$ & $0.29$ & $41$ & $0.34$ & $6$ & $5$ & $3$ & $0.27$ & $0.46$ & $0.56$ \\ 
   MW15 & $4.12$ & $0.37$ & $2.32$ & $0.18$ & $0.84$ & $0$ & $4$ & $0.00$ & $8$ & $0.50$ & $1$ & $2$ & $1$ & $0.33$ & $0.29$ & $0.48$ \\ 
   MW16 & $7.96$ & $1.00$ & $0.00$ & $0.00$ & $0.42$ & $8$ & $21$ & $0.38$ & $38$ & $0.55$ & $10$ & $8$ & $3$ & $0.17$ & $0.48$ & $0.60$ \\ 
   MW17 & $7.35$ & $0.86$ & $1.35$ & $0.10$ & $0.52$ & $2$ & $12$ & $0.17$ & $30$ & $0.40$ & $5$ & $2$ & $5$ & $0.71$ & $0.28$ & $0.54$ \\ 
   MW18 & $0.00$ & $0.00$ & $0.00$ & $0.00$ & $1.00$ & $11$ & $17$ & $0.65$ & $29$ & $0.59$ & $10$ & $5$ & $2$ & $0.13$ & $0.38$ & $0.49$ \\ 
   MW19 & -- & -- & $0.00$ & $0.00$ & $0.00$ & $2$ & $6$ & $0.33$ & $9$ & $0.67$ & $3$ & $3$ & $0$ & $0.00$ & $0.20$ & $0.33$ \\ 
   MW20 & -- & -- & $0.00$ & $0.00$ & $0.00$ & $3$ & $8$ & $0.38$ & $14$ & $0.57$ & $3$ & $5$ & $0$ & $0.00$ & $0.17$ & $0.28$ \\ 
   MW21 & $0.00$ & $0.00$ & $0.00$ & $0.00$ & $1.00$ & $11$ & $19$ & $0.58$ & $40$ & $0.48$ & $9$ & $5$ & $5$ & $0.36$ & $0.62$ & $0.57$ \\ 
   MW22 & $5.98$ & $0.62$ & $0.00$ & $0.00$ & $0.57$ & $4$ & $19$ & $0.21$ & $33$ & $0.58$ & $9$ & $7$ & $3$ & $0.19$ & $0.30$ & $0.48$ \\ 
   MW23 & $11.17$ & $2.48$ & $0.00$ & $0.00$ & $0.19$ & $7$ & $29$ & $0.24$ & $39$ & $0.74$ & $20$ & $8$ & $1$ & $0.04$ & $0.14$ & $0.34$ \\ 
   MW24 & $7.35$ & $0.86$ & $4.12$ & $0.37$ & $0.67$ & $2$ & $6$ & $0.33$ & $17$ & $0.35$ & $2$ & $1$ & $3$ & $1.00$ & $0.26$ & $0.55$ \\ 
   \hline
   Median & $10.87$ & $2.24$ & $0.00$ & $0.00$ & $0.23$ & $6$ & $14$ & $0.40$ & $27$ & $0.61$ & $6$ & $5$ & $2$ & $0.17$ & $0.17$ & $0.34$ \\ 
   IQR & $5.42$ & $4.62$ & $2.32$ & $0.18$ & $0.49$ & $5$ & $12$ & $0.21$ & $22$ & $0.14$ & $7$ & $4$ & $2$ & $0.30$ & $0.20$ & $0.23$ \\ 
   Range & $13.64$ & $20.00$ & $8.86$ & $1.26$ & $1.00$ & $13$ & $26$ & $0.67$ & $38$ & $0.53$ & $19$ & $12$ & $5$ & $1.00$ & $0.59$ & $0.51$ \\ 
   \hline
  \end{tabular} 
\end{table*}

\section{Spearman correlation coefficients and $p$-values} \label{sec:appspear}
In \autoref{tab:correlations_sr}, we list the Spearman rank order correlation coefficients between the GC-related quantities from \autoref{tab:gcmetrics} and the galaxy formation-related quantities from \autoref{tab:galmetric1} and \autoref{tab:galmetric2}. The associated $p$-values that these correlations arise due to random chance are listed in \autoref{tab:correlations_sp}. For the discussion and physical interpretation of these results, see Section~\ref{sec:correlations}.

\begin{table*}
  \caption{Spearman correlation coefficients between the GC-related quantities from \autoref{tab:gcmetrics} (columns) and the galaxy formation-related quantities from \autoref{tab:galmetric1} and \autoref{tab:galmetric2} (rows). Because the Spearman correlation coefficient is rank-ordered, we have removed the redshifts, which exhibit correlations that are identical to those of the lookback times. Correlations that are statistically significant according to their $p$-values in \autoref{tab:correlations_sp} (see the text for details) are marked in red.} 
\label{tab:correlations_sr}
  \begin{tabular}{l c c c c c c c c c c c c c}
   \hline
   Quantity & $\color{red}\widetilde{\tau}$ & $\color{red}{\rm IQR}(\tau)$ & $S(\tau)$ & $K(\tau)$ & $\widetilde{\rm [Fe/H]}$ & ${\rm IQR}({\rm [Fe/H]})$ & $S({\rm [Fe/H]})$ & $K({\rm [Fe/H]})$ & ${\rm IQR}^2$ & $\color{red}\riqr$ & $\color{red}\frac{{\rm d[Fe/H]}}{{\rm d}\log t}$ & ${\rm [Fe/H]}_0$ & $\color{red}N_{\rm GC}$ \\ 
   \hline
   $\mvir$ & $-0.06$ & $0.25$ & $-0.22$ & $0.04$ & $-0.24$ & $0.22$ & $0.38$ & $-0.38$ & $0.27$ & $-0.07$ & $0.12$ & $-0.33$ & $0.35$ \\ 
   $\rvir$ & $-0.06$ & $0.25$ & $-0.22$ & $0.04$ & $-0.24$ & $0.22$ & $0.38$ & $-0.38$ & $0.27$ & $-0.07$ & $0.12$ & $-0.33$ & $0.35$ \\ 
   $\color{red}\vmax$ & $0.16$ & $0.02$ & $-0.12$ & $-0.02$ & $-0.29$ & $0.23$ & $0.29$ & $-0.28$ & $0.09$ & $0.19$ & $0.37$ & $-0.35$ & $\color{red}0.68$ \\ 
   $\rvmax$ & $-0.13$ & $0.23$ & $0.03$ & $-0.04$ & $0.10$ & $0.17$ & $0.15$ & $-0.22$ & $0.26$ & $-0.06$ & $-0.13$ & $-0.07$ & $-0.25$ \\ 
   $\color{red}\cnfw$ & $0.41$ & $\color{red}-0.63$ & $0.01$ & $0.11$ & $0.09$ & $-0.03$ & $-0.19$ & $0.19$ & $-0.49$ & $\color{red}0.67$ & $0.53$ & $-0.12$ & $\color{red}0.60$ \\ 
   $\color{red}\tau_{25}$ & $\color{red}0.61$ & $\color{red}-0.55$ & $0.09$ & $0.12$ & $-0.02$ & $-0.11$ & $-0.20$ & $0.26$ & $-0.46$ & $0.51$ & $\color{red}0.54$ & $-0.11$ & $0.32$ \\ 
   $\color{red}\tau_{50}$ & $0.30$ & $-0.30$ & $0.20$ & $-0.09$ & $0.02$ & $0.16$ & $-0.03$ & $-0.00$ & $-0.19$ & $0.41$ & $\color{red}0.59$ & $-0.21$ & $0.51$ \\ 
   $\tau_{75}$ & $0.09$ & $-0.14$ & $0.39$ & $-0.40$ & $0.02$ & $0.12$ & $-0.05$ & $-0.01$ & $-0.09$ & $0.31$ & $0.18$ & $0.11$ & $0.36$ \\ 
   $\tau_{\rm max}$ & $-0.10$ & $0.02$ & $-0.03$ & $-0.12$ & $-0.14$ & $0.03$ & $0.04$ & $-0.03$ & $-0.03$ & $-0.00$ & $-0.42$ & $0.41$ & $-0.10$ \\ 
   $\color{red}\tau_{\rm a}$ & $0.37$ & $-0.16$ & $0.33$ & $-0.21$ & $-0.13$ & $0.25$ & $0.10$ & $-0.15$ & $-0.04$ & $0.33$ & $\color{red}0.57$ & $-0.19$ & $0.34$ \\ 
   $\color{red}\tau_{\rm f}$ & $\color{red}0.60$ & $-0.14$ & $0.18$ & $-0.15$ & $-0.38$ & $0.49$ & $0.44$ & $-0.47$ & $0.05$ & $0.44$ & $0.50$ & $-0.05$ & $0.38$ \\ 
   $\delta_t$ & $-0.16$ & $0.13$ & $-0.13$ & $0.03$ & $-0.09$ & $0.20$ & $0.27$ & $-0.25$ & $0.14$ & $0.07$ & $-0.18$ & $-0.02$ & $0.08$ \\ 
   $\tau_{\rm mm}$ & $0.31$ & $-0.24$ & $0.16$ & $-0.14$ & $0.21$ & $-0.06$ & $-0.27$ & $0.15$ & $-0.17$ & $0.31$ & $0.50$ & $-0.19$ & $0.26$ \\ 
   $\tau_{\rm am}$ & $-0.20$ & $0.00$ & $0.10$ & $-0.09$ & $0.12$ & $-0.25$ & $-0.27$ & $0.28$ & $-0.05$ & $-0.19$ & $-0.14$ & $0.07$ & $-0.22$ \\ 
   $r_{t}$ & $-0.08$ & $0.42$ & $0.12$ & $-0.27$ & $-0.19$ & $0.29$ & $0.40$ & $-0.43$ & $0.46$ & $-0.19$ & $-0.26$ & $0.11$ & $-0.02$ \\ 
   $\color{red}\nbrz$ & $\color{red}0.75$ & $-0.24$ & $-0.05$ & $0.02$ & $-0.36$ & $0.42$ & $0.40$ & $-0.38$ & $-0.03$ & $0.49$ & $\color{red}0.56$ & $-0.11$ & $0.45$ \\ 
   $\color{red}\nbr$ & $0.40$ & $-0.32$ & $-0.06$ & $0.06$ & $-0.23$ & $0.35$ & $0.27$ & $-0.27$ & $-0.15$ & $\color{red}0.62$ & $\color{red}0.64$ & $-0.41$ & $\color{red}0.76$ \\ 
   $\color{red}\rz$ & $\color{red}0.78$ & $-0.17$ & $-0.08$ & $0.03$ & $-0.30$ & $0.10$ & $0.16$ & $-0.10$ & $-0.07$ & $0.11$ & $0.13$ & $0.38$ & $-0.23$ \\ 
   $\color{red}N_{\rm leaf}$ & $0.21$ & $-0.17$ & $-0.08$ & $0.06$ & $-0.29$ & $0.32$ & $0.36$ & $-0.35$ & $-0.06$ & $0.46$ & $0.45$ & $-0.34$ & $\color{red}0.72$ \\ 
   $r_{\rm bl}$ & $0.35$ & $-0.47$ & $-0.23$ & $0.26$ & $0.13$ & $-0.17$ & $-0.32$ & $0.34$ & $-0.40$ & $0.28$ & $0.34$ & $-0.08$ & $0.06$ \\ 
   $\color{red}N_{<1:100}$ & $0.43$ & $-0.35$ & $-0.15$ & $0.14$ & $-0.18$ & $0.24$ & $0.20$ & $-0.20$ & $-0.20$ & $\color{red}0.57$ & $\color{red}0.59$ & $-0.34$ & $\color{red}0.75$ \\ 
   $\color{red}N_{1:100-1:4}$ & $0.29$ & $-0.28$ & $-0.01$ & $0.06$ & $-0.21$ & $0.31$ & $0.20$ & $-0.15$ & $-0.15$ & $0.48$ & $0.51$ & $-0.33$ & $\color{red}0.55$ \\ 
   $N_{>1:4}$ & $-0.02$ & $0.28$ & $0.24$ & $-0.36$ & $-0.27$ & $0.46$ & $0.46$ & $-0.54$ & $0.35$ & $0.07$ & $0.06$ & $-0.04$ & $0.14$ \\ 
   $r_{\rm mm}$ & $-0.23$ & $0.50$ & $0.31$ & $-0.41$ & $-0.05$ & $0.30$ & $0.29$ & $-0.40$ & $0.52$ & $-0.26$ & $-0.20$ & $0.10$ & $-0.13$ \\ 
   $f_{\rm ex,*}$ & $-0.25$ & $0.31$ & $-0.06$ & $-0.01$ & $-0.14$ & $0.25$ & $0.34$ & $-0.36$ & $0.30$ & $-0.08$ & $-0.23$ & $-0.04$ & $0.04$ \\ 
   $f_{\rm ex,GC}$ & $-0.26$ & $0.34$ & $-0.06$ & $-0.04$ & $-0.27$ & $0.27$ & $0.40$ & $-0.40$ & $0.32$ & $-0.11$ & $-0.18$ & $-0.08$ & $0.07$ \\ 
   \hline
  \end{tabular} 
\end{table*}

\begin{table*}
  \caption{Logarithm of the $p$-values of Spearman correlation coefficients between the GC-related quantities from \autoref{tab:gcmetrics} (columns) and the galaxy formation-related quantities from \autoref{tab:galmetric1} and \autoref{tab:galmetric2} (rows). Because the Spearman correlation coefficient is rank-ordered, we have removed the redshifts, which exhibit correlations that are identical to those of the lookback times. Statistically significant correlations (see the text) are marked in red.} 
\label{tab:correlations_sp}
  \begin{tabular}{l c c c c c c c c c c c c c}
   \hline
   Quantity & $\color{red}\widetilde{\tau}$ & $\color{red}{\rm IQR}(\tau)$ & $S(\tau)$ & $K(\tau)$ & $\widetilde{\rm [Fe/H]}$ & ${\rm IQR}({\rm [Fe/H]})$ & $S({\rm [Fe/H]})$ & $K({\rm [Fe/H]})$ & ${\rm IQR}^2$ & $\color{red}\riqr$ & $\color{red}\frac{{\rm d[Fe/H]}}{{\rm d}\log t}$ & ${\rm [Fe/H]}_0$ & $\color{red}N_{\rm GC}$ \\ 
   \hline
   $\mvir$ & $-0.11$ & $-0.64$ & $-0.54$ & $-0.07$ & $-0.59$ & $-0.55$ & $-1.22$ & $-1.21$ & $-0.71$ & $-0.13$ & $-0.24$ & $-0.98$ & $-1.07$ \\ 
   $\rvir$ & $-0.11$ & $-0.64$ & $-0.54$ & $-0.07$ & $-0.59$ & $-0.55$ & $-1.22$ & $-1.21$ & $-0.71$ & $-0.13$ & $-0.24$ & $-0.98$ & $-1.07$ \\ 
   $\color{red}\vmax$ & $-0.35$ & $-0.03$ & $-0.25$ & $-0.03$ & $-0.78$ & $-0.58$ & $-0.78$ & $-0.74$ & $-0.17$ & $-0.44$ & $-1.17$ & $-1.09$ & $\color{red}-3.73$ \\ 
   $\rvmax$ & $-0.26$ & $-0.57$ & $-0.05$ & $-0.07$ & $-0.20$ & $-0.38$ & $-0.31$ & $-0.53$ & $-0.66$ & $-0.10$ & $-0.27$ & $-0.14$ & $-0.64$ \\ 
   $\color{red}\cnfw$ & $-1.37$ & $\color{red}-3.15$ & $-0.02$ & $-0.22$ & $-0.18$ & $-0.05$ & $-0.45$ & $-0.43$ & $-1.91$ & $\color{red}-3.57$ & $-2.17$ & $-0.25$ & $\color{red}-2.78$ \\ 
   $\color{red}\tau_{25}$ & $\color{red}-2.96$ & $\color{red}-2.32$ & $-0.17$ & $-0.24$ & $-0.04$ & $-0.21$ & $-0.49$ & $-0.69$ & $-1.66$ & $-2.02$ & $\color{red}-2.26$ & $-0.22$ & $-0.93$ \\ 
   $\color{red}\tau_{50}$ & $-0.82$ & $-0.82$ & $-0.48$ & $-0.18$ & $-0.04$ & $-0.35$ & $-0.06$ & $-0.00$ & $-0.44$ & $-1.36$ & $\color{red}-2.69$ & $-0.49$ & $-2.00$ \\ 
   $\tau_{75}$ & $-0.17$ & $-0.29$ & $-1.28$ & $-1.30$ & $-0.04$ & $-0.25$ & $-0.10$ & $-0.02$ & $-0.18$ & $-0.90$ & $-0.42$ & $-0.23$ & $-1.10$ \\ 
   $\tau_{\rm max}$ & $-0.20$ & $-0.04$ & $-0.05$ & $-0.24$ & $-0.31$ & $-0.06$ & $-0.07$ & $-0.06$ & $-0.05$ & $-0.00$ & $-1.43$ & $-1.40$ & $-0.21$ \\ 
   $\color{red}\tau_{\rm a}$ & $-1.14$ & $-0.34$ & $-0.98$ & $-0.52$ & $-0.26$ & $-0.65$ & $-0.20$ & $-0.31$ & $-0.07$ & $-0.96$ & $\color{red}-2.49$ & $-0.43$ & $-1.03$ \\ 
   $\color{red}\tau_{\rm f}$ & $\color{red}-2.80$ & $-0.30$ & $-0.41$ & $-0.33$ & $-1.22$ & $-1.86$ & $-1.58$ & $-1.73$ & $-0.08$ & $-1.56$ & $-1.98$ & $-0.09$ & $-1.20$ \\ 
   $\delta_t$ & $-0.34$ & $-0.28$ & $-0.27$ & $-0.06$ & $-0.17$ & $-0.48$ & $-0.72$ & $-0.64$ & $-0.30$ & $-0.12$ & $-0.42$ & $-0.03$ & $-0.14$ \\ 
   $\tau_{\rm mm}$ & $-0.71$ & $-0.49$ & $-0.29$ & $-0.25$ & $-0.41$ & $-0.09$ & $-0.59$ & $-0.27$ & $-0.31$ & $-0.71$ & $-1.55$ & $-0.37$ & $-0.54$ \\ 
   $\tau_{\rm am}$ & $-0.48$ & $-0.00$ & $-0.20$ & $-0.16$ & $-0.24$ & $-0.65$ & $-0.73$ & $-0.75$ & $-0.10$ & $-0.43$ & $-0.30$ & $-0.13$ & $-0.53$ \\ 
   $r_{t}$ & $-0.16$ & $-1.41$ & $-0.24$ & $-0.71$ & $-0.45$ & $-0.81$ & $-1.31$ & $-1.48$ & $-1.67$ & $-0.45$ & $-0.67$ & $-0.22$ & $-0.04$ \\ 
   $\color{red}\nbrz$ & $\color{red}-4.74$ & $-0.62$ & $-0.10$ & $-0.04$ & $-1.10$ & $-1.46$ & $-1.33$ & $-1.23$ & $-0.06$ & $-1.92$ & $\color{red}-2.47$ & $-0.21$ & $-1.64$ \\ 
   $\color{red}\nbr$ & $-1.31$ & $-0.92$ & $-0.11$ & $-0.11$ & $-0.58$ & $-1.08$ & $-0.71$ & $-0.73$ & $-0.32$ & $\color{red}-3.05$ & $\color{red}-3.30$ & $-1.36$ & $\color{red}-4.92$ \\ 
   $\color{red}\rz$ & $\color{red}-5.43$ & $-0.39$ & $-0.14$ & $-0.05$ & $-0.84$ & $-0.20$ & $-0.36$ & $-0.21$ & $-0.12$ & $-0.22$ & $-0.27$ & $-1.21$ & $-0.57$ \\ 
   $\color{red}\nleaf$ & $-0.50$ & $-0.39$ & $-0.16$ & $-0.11$ & $-0.80$ & $-0.90$ & $-1.10$ & $-1.08$ & $-0.10$ & $-1.72$ & $-1.64$ & $-1.02$ & $\color{red}-4.24$ \\ 
   $\rbl$ & $-1.08$ & $-1.76$ & $-0.58$ & $-0.68$ & $-0.27$ & $-0.39$ & $-0.93$ & $-1.01$ & $-1.32$ & $-0.75$ & $-1.01$ & $-0.16$ & $-0.11$ \\ 
   $\color{red}N_{<1:100}$ & $-1.52$ & $-1.09$ & $-0.32$ & $-0.29$ & $-0.40$ & $-0.60$ & $-0.47$ & $-0.48$ & $-0.47$ & $\color{red}-2.57$ & $\color{red}-2.77$ & $-1.03$ & $\color{red}-4.73$ \\ 
   $\color{red}N_{1:100-1:4}$ & $-0.78$ & $-0.76$ & $-0.02$ & $-0.10$ & $-0.50$ & $-0.87$ & $-0.46$ & $-0.31$ & $-0.33$ & $-1.80$ & $-2.04$ & $-0.99$ & $\color{red}-2.36$ \\ 
   $N_{>1:4}$ & $-0.03$ & $-0.76$ & $-0.62$ & $-1.09$ & $-0.70$ & $-1.67$ & $-1.67$ & $-2.28$ & $-1.07$ & $-0.13$ & $-0.11$ & $-0.06$ & $-0.29$ \\ 
   $r_{\rm mm}$ & $-0.57$ & $-1.99$ & $-0.87$ & $-1.40$ & $-0.10$ & $-0.85$ & $-0.78$ & $-1.34$ & $-2.10$ & $-0.66$ & $-0.46$ & $-0.19$ & $-0.27$ \\ 
   $f_{\rm ex,*}$ & $-0.65$ & $-0.87$ & $-0.10$ & $-0.01$ & $-0.29$ & $-0.65$ & $-1.04$ & $-1.12$ & $-0.82$ & $-0.15$ & $-0.57$ & $-0.07$ & $-0.07$ \\ 
   $f_{\rm ex,GC}$ & $-0.66$ & $-1.03$ & $-0.11$ & $-0.08$ & $-0.73$ & $-0.70$ & $-1.30$ & $-1.34$ & $-0.94$ & $-0.21$ & $-0.41$ & $-0.15$ & $-0.12$ \\ 
   \hline
  \end{tabular} 
\end{table*}

Here we define which correlations in \autoref{tab:correlations_sr} and \autoref{tab:correlations_sp} are considered to have a high statistical significance. This definition should account for the fact that each correlation has a non-zero probability of arising due to random chance. For an individual correlation, we consider it to be significant if the probability $p$ that it arises from random chance is $p<\pref$ with $\pref=0.05$. However, when correlating a large number of data sets with each other, some fraction of these correlations may appear significant even if they are not. For instance, among 100 entirely random correlations, five are expected to be significant with a $p$-value of $p<0.05$. To remedy the problem of detecting spurious correlations, we adjust the threshold $p$-value $\pref$ by applying the Holm-Bonferroni method \citep{holm79}. This is an update to the original Bonferroni correction, which divides the desired $p$-value by the number of correlations evaluated $\ncorr$, for an effective $p$-value of $\peff=p/\ncorr$. Doing so represents an extreme measure, because it makes the strong assumption that all considered variable pairs are independent, which is rarely satisfied and certainly does not apply to the quantities in \autoref{tab:gcmetrics}--\autoref{tab:galmetric2}. This shortcoming of the Bonferroni correction is remedied to some extent by the Holm-Bonferroni method, which technically does not assume the variables are independent and in practice makes a weaker assumption of independence. Given a set of correlations, this method orders them by increasing $p$-value. For a rank order $i\geq1$, the effective maximum $p$-value needed for reaching the desired confidence level $\pref$ is then defined as
\be
\label{eq:peff}
\peff = \frac{\pref}{\ncorr+1-i} .
\ee
Because $\peff$ increases with $i$, the threshold for statistical significance relaxes as one goes further up the list of sorted correlations. As such, this method evaluates the likelihood of spurious correlations {\it for the remainder} of the list rather than the entire list. This ensures that the probability of a false positive is always $p<\pref$.

Even if the Holm-Bonferroni method is more appropriate for variable sets that are not independent, it still overcorrects for spurious correlations if variables are near-duplicates of each other. This is easily verified using equation~(\ref{eq:peff}) -- if we duplicate each quantity, the ranked list of correlations is twice as long and the same $p$-value is only reached after trawling through twice the number of correlations. Doubling $\ncorr$ and $i$ in equation~(\ref{eq:peff}), we see that this results in $\peff$ being halved (provided that $\ncorr\gg1$ and $i\gg1$). This shows that even the Holm-Bonferroni method is sensitive to including variables that are known to not be independent, or variables that are known in advance to not correlate.

Acknowledging these considerations regarding the independence of the variables, we need to set the number of correlations $\ncorr$ in equation~(\ref{eq:peff}). This depends on the question at hand -- if we ask which of all possible variable pairs between \autoref{tab:gcmetrics} and the combination of \autoref{tab:galmetric1} and \autoref{tab:galmetric2} show a significant correlation, then we should set $\ncorr=13\times30=390$. However, this again assumes that all variables are independent, whereas we have prior knowledge that the considered quantities are not. For instance, visual inspection of \autoref{fig:agez} shows that $\iqr(\tau)$ effectively sets $\iqr^2$ and $\riqr$, because $\iqr(\feh)$ is very similar for all galaxies (also see \autoref{tab:gcmetrics}). In addition, most of the galaxy-related quantities in \autoref{tab:galmetric1} and \autoref{tab:galmetric2} are known to correlate at some level. Specifically, the virial masses and radii, as well as the maximum velocities are not independent, and neither are the lookback times and redshifts. Condensing these points into a single number, we find 24 `independent' galaxy-related quantities, although even in this remaining set important dependences may remain.

In view of these concerns, we proceed by asking for each galaxy-related quantity independently whether it correlates with any of the GC-related quantities. This implies that a total number of 13 correlations are evaluated per galaxy-related quantity. Accounting for the relation between the various renditions of the interquartile range, 11 of these are not trivially dependent on each other. We therefore set $\ncorr=11$ and disregard the correlations with the two highest $p$-values found for each galaxy-related quantity when stepping through the rank-ordered list during the evaluation of equation~(\ref{eq:peff}). In combination with $\pref=0.05$, equation~(\ref{eq:peff}) thus shows that $\peff$ varies from $4.5\times10^{-3}$ to $5\times10^{-2}$, such that $\log{\peff}$ ranges from $-2.34$ to $-1.30$, depending on the rank $i$. Given that this guarantees significant correlations at an equivalent $p$-value of $0.05$ for each individual galaxy-related quantity, we may expect only one out of the 24 `independent' galaxy-related quantities to show a spurious (and relatively low-significance) correlation with the GC-related quantities. This represents a worst-case scenario, because even when using an effective number of 24 galaxy-related quantities, their independence is still overestimated. Therefore, we deem this a satisfactorily conservative setup.

\section{Pearson correlation coefficients and $p$-values} \label{sec:appcorr}
In \autoref{tab:correlations_pr}--\ref{tab:correlations_pplog}, we evaluate the correlations between the GC-related quantities from \autoref{tab:gcmetrics} and the galaxy formation-related quantities from \autoref{tab:galmetric1} and \autoref{tab:galmetric2} by listing the Pearson correlation coefficients and the logarithms of the $p$-values that these correlations arise due to random chance. We consider both the original quantities (\autoref{tab:correlations_pr} and \autoref{tab:correlations_pp}) and their logarithms (\autoref{tab:correlations_prlog} and \autoref{tab:correlations_pplog}). The Pearson correlation coefficient indicates how well a linear correlation fits the data, implying that the inclusion of logarithms in \autoref{tab:correlations_prlog} and \autoref{tab:correlations_pplog} evaluates the same for a power law relation. Prior to taking the logarithm, we have applied appropriate transformations to avoid any cases in which the logarithm of zero is taken. Across all tables, statistically significant correlations according to their Spearman rank $p$-values from \autoref{tab:correlations_sp} are shown in red. The quantities in the first row and first column are coloured according to their highest correlation coefficients, thus reflecting how well-constrained the quantities in the first column are and how useful the quantities in the first row are for constraining them. Because the correlation coefficients and $p$-values are highly similar for the lookback times and equivalent redshifts, we omit the latter from these tables.

\begin{table*}
  \caption{Pearson correlation coefficients between the GC-related quantities from \autoref{tab:gcmetrics} (columns) and the galaxy formation-related quantities from \autoref{tab:galmetric1} and \autoref{tab:galmetric2} (rows). Correlations that are statistically significant according to their $p$-values in \autoref{tab:correlations_sp} (see the text for details) are marked in red.}
\label{tab:correlations_pr}
  \begin{tabular}{l c p{0.65cm} c c c c c c c c p{0.65cm} p{0.85cm} c}
   \hline
   Quantity & $\color{red}\widetilde{\tau}$ & $\color{red}{\rm IQR}(\tau)$ & $S(\tau)$ & $K(\tau)$ & $\widetilde{\rm [Fe/H]}$ & ${\rm IQR}({\rm [Fe/H]})$ & $S({\rm [Fe/H]})$ & $K({\rm [Fe/H]})$ & ${\rm IQR}^2$ & $\color{red}\riqr$ & $\color{red}\frac{{\rm d[Fe/H]}}{{\rm d}\log t}$ & ${\rm [Fe/H]}_0$ & $\color{red}N_{\rm GC}$ \\ 
   \hline
   $\mvir$ & $-0.05$ & $0.21$ & $-0.15$ & $-0.10$ & $-0.27$ & $0.33$ & $0.38$ & $-0.43$ & $0.28$ & $-0.09$ & $0.10$ & $-0.38$ & $0.36$ \\ 
   $\rvir$ & $-0.06$ & $0.21$ & $-0.16$ & $-0.09$ & $-0.26$ & $0.34$ & $0.37$ & $-0.44$ & $0.27$ & $-0.06$ & $0.12$ & $-0.40$ & $0.37$ \\ 
   $\color{red}\vmax$ & $0.10$ & $-0.03$ & $-0.05$ & $-0.08$ & $-0.18$ & $0.26$ & $0.22$ & $-0.27$ & $0.05$ & $0.17$ & $0.32$ & $-0.37$ & $\color{red}0.62$ \\ 
   $\rvmax$ & $-0.24$ & $0.22$ & $-0.17$ & $0.07$ & $0.01$ & $0.08$ & $0.09$ & $-0.14$ & $0.23$ & $-0.13$ & $-0.12$ & $-0.15$ & $-0.26$ \\ 
   $\color{red}\cnfw$ & $0.49$ & $\color{red}-0.63$ & $-0.01$ & $0.13$ & $0.10$ & $-0.07$ & $-0.21$ & $0.23$ & $-0.54$ & $\color{red}0.64$ & $0.50$ & $-0.05$ & $\color{red}0.57$ \\ 
   $\color{red}\ttf$ & $\color{red}0.61$ & $\color{red}-0.38$ & $0.12$ & $-0.01$ & $-0.21$ & $0.00$ & $0.07$ & $0.13$ & $-0.30$ & $0.36$ & $\color{red}0.43$ & $0.05$ & $0.42$ \\ 
   $\color{red}\tfz$ & $0.20$ & $-0.34$ & $0.24$ & $0.01$ & $0.21$ & $0.10$ & $-0.21$ & $0.08$ & $-0.25$ & $0.33$ & $\color{red}0.53$ & $-0.19$ & $0.45$ \\ 
   $\tsf$ & $0.23$ & $-0.36$ & $0.35$ & $-0.19$ & $0.10$ & $0.07$ & $-0.12$ & $0.04$ & $-0.31$ & $0.29$ & $0.19$ & $0.13$ & $0.30$ \\ 
   $\tmax$ & $0.05$ & $-0.05$ & $0.03$ & $-0.17$ & $-0.02$ & $0.14$ & $0.06$ & $-0.09$ & $-0.02$ & $0.01$ & $-0.33$ & $0.36$ & $-0.01$ \\ 
   $\color{red}\ta$ & $0.41$ & $-0.28$ & $0.33$ & $-0.14$ & $-0.03$ & $0.17$ & $0.01$ & $-0.02$ & $-0.17$ & $0.28$ & $\color{red}0.60$ & $-0.15$ & $0.39$ \\ 
%   $\color{red}\za$ & $0.40$ & $-0.29$ & $0.34$ & $-0.17$ & $0.07$ & $0.11$ & $-0.07$ & $0.04$ & $-0.19$ & $0.36$ & $\color{red}0.67$ & $-0.12$ & $0.42$ \\ 
   $\color{red}\tf$ & $\color{red}0.68$ & $-0.30$ & $0.15$ & $-0.17$ & $-0.38$ & $0.36$ & $0.37$ & $-0.28$ & $-0.16$ & $0.41$ & $0.47$ & $0.00$ & $0.39$ \\ 
%   $\color{red}\zf$ & $\color{red}0.62$ & $-0.33$ & $0.15$ & $-0.18$ & $-0.24$ & $0.30$ & $0.23$ & $-0.20$ & $-0.20$ & $0.53$ & $0.53$ & $0.03$ & $0.39$ \\ 
   $\delta_t$ & $0.22$ & $0.06$ & $-0.29$ & $-0.02$ & $-0.40$ & $0.18$ & $0.44$ & $-0.32$ & $0.06$ & $0.06$ & $-0.29$ & $0.20$ & $-0.06$ \\ 
   $\tmm$ & $0.12$ & $-0.23$ & $0.28$ & $-0.15$ & $0.33$ & $-0.00$ & $-0.38$ & $0.25$ & $-0.14$ & $0.23$ & $0.55$ & $-0.30$ & $0.25$ \\ 
%   $\zmm$ & $0.35$ & $-0.21$ & $0.13$ & $-0.10$ & $0.14$ & $0.03$ & $-0.21$ & $0.18$ & $-0.11$ & $0.27$ & $0.66$ & $-0.19$ & $0.38$ \\ 
   $\tam$ & $0.01$ & $-0.06$ & $0.01$ & $0.06$ & $0.11$ & $-0.41$ & $-0.28$ & $0.41$ & $-0.18$ & $-0.14$ & $-0.08$ & $0.29$ & $-0.23$ \\ 
%   $\zam$ & $0.13$ & $-0.13$ & $-0.07$ & $0.07$ & $0.06$ & $-0.44$ & $-0.26$ & $0.45$ & $-0.24$ & $-0.10$ & $-0.08$ & $0.37$ & $-0.24$ \\ 
   $r_{t}$ & $-0.13$ & $0.39$ & $0.05$ & $-0.26$ & $-0.31$ & $0.22$ & $0.41$ & $-0.39$ & $0.34$ & $-0.18$ & $-0.37$ & $0.15$ & $-0.11$ \\ 
   $\color{red}\nbrz$ & $\color{red}0.75$ & $-0.30$ & $-0.02$ & $-0.09$ & $-0.48$ & $0.39$ & $0.43$ & $-0.30$ & $-0.12$ & $0.53$ & $\color{red}0.61$ & $-0.16$ & $0.40$ \\ 
   $\color{red}\nbr$ & $0.42$ & $-0.41$ & $-0.12$ & $0.06$ & $-0.19$ & $0.32$ & $0.21$ & $-0.21$ & $-0.24$ & $\color{red}0.62$ & $\color{red}0.65$ & $-0.52$ & $\color{red}0.78$ \\ 
   $\color{red}\rz$ & $\color{red}0.81$ & $-0.24$ & $-0.07$ & $-0.08$ & $-0.45$ & $0.04$ & $0.28$ & $0.00$ & $-0.18$ & $0.18$ & $0.17$ & $0.42$ & $-0.22$ \\ 
   $\color{red}\nleaf$ & $0.29$ & $-0.25$ & $-0.08$ & $-0.08$ & $-0.27$ & $0.39$ & $0.35$ & $-0.39$ & $-0.11$ & $0.45$ & $0.43$ & $-0.44$ & $\color{red}0.67$ \\ 
   $\rbl$ & $0.38$ & $-0.47$ & $-0.28$ & $0.45$ & $0.12$ & $-0.24$ & $-0.28$ & $0.44$ & $-0.38$ & $0.33$ & $0.36$ & $-0.04$ & $0.22$ \\ 
   $\color{red}N_{<1:100}$ & $0.44$ & $-0.45$ & $-0.21$ & $0.16$ & $-0.12$ & $0.13$ & $0.08$ & $-0.04$ & $-0.32$ & $\color{red}0.64$ & $\color{red}0.66$ & $-0.42$ & $\color{red}0.84$ \\ 
   $\color{red}N_{1:100-1:4}$ & $0.27$ & $-0.30$ & $-0.09$ & $0.11$ & $-0.17$ & $0.30$ & $0.15$ & $-0.15$ & $-0.15$ & $0.41$ & $0.48$ & $-0.54$ & $\color{red}0.44$ \\ 
   $N_{>1:4}$ & $-0.00$ & $0.12$ & $0.26$ & $-0.41$ & $-0.19$ & $0.47$ & $0.39$ & $-0.54$ & $0.20$ & $0.03$ & $-0.04$ & $-0.04$ & $0.06$ \\ 
   $r_{\rm mm}$ & $-0.39$ & $0.50$ & $0.31$ & $-0.35$ & $0.03$ & $0.17$ & $0.11$ & $-0.30$ & $0.41$ & $-0.38$ & $-0.33$ & $0.11$ & $-0.30$ \\ 
   $f_{\rm ex,*}$ & $-0.07$ & $0.19$ & $-0.08$ & $-0.17$ & $-0.26$ & $0.32$ & $0.39$ & $-0.45$ & $0.20$ & $0.00$ & $-0.29$ & $0.04$ & $-0.00$ \\ 
   $f_{\rm ex,GC}$ & $-0.30$ & $0.32$ & $0.04$ & $-0.25$ & $-0.19$ & $0.32$ & $0.34$ & $-0.47$ & $0.30$ & $-0.12$ & $-0.27$ & $-0.15$ & $0.02$ \\ 
   \hline
  \end{tabular} 
\end{table*}

\begin{table*}
  \caption{Logarithm of the $p$-values of Pearson correlation coefficients between the GC-related quantities from \autoref{tab:gcmetrics} (columns) and the galaxy formation-related quantities from \autoref{tab:galmetric1} and \autoref{tab:galmetric2} (rows). Correlations that are statistically significant according to their $p$-values in \autoref{tab:correlations_sp} (see the text for details) are marked in red.}
\label{tab:correlations_pp}
  \begin{tabular}{l c p{0.65cm} c c c c c c c c p{0.65cm} p{0.85cm} c}
   \hline
   Quantity & $\color{red}\widetilde{\tau}$ & $\color{red}{\rm IQR}(\tau)$ & $S(\tau)$ & $K(\tau)$ & $\widetilde{\rm [Fe/H]}$ & ${\rm IQR}({\rm [Fe/H]})$ & $S({\rm [Fe/H]})$ & $K({\rm [Fe/H]})$ & ${\rm IQR}^2$ & $\color{red}\riqr$ & $\color{red}\frac{{\rm d[Fe/H]}}{{\rm d}\log t}$ & ${\rm [Fe/H]}_0$ & $\color{red}N_{\rm GC}$ \\ 
   \hline
   $\mvir$ & $-0.10$ & $-0.52$ & $-0.31$ & $-0.19$ & $-0.73$ & $-0.95$ & $-1.19$ & $-1.52$ & $-0.75$ & $-0.18$ & $-0.20$ & $-1.21$ & $-1.12$ \\ 
   $\rvir$ & $-0.11$ & $-0.50$ & $-0.35$ & $-0.18$ & $-0.68$ & $-1.00$ & $-1.18$ & $-1.56$ & $-0.73$ & $-0.10$ & $-0.25$ & $-1.33$ & $-1.16$ \\ 
   $\color{red}\vmax$ & $-0.19$ & $-0.05$ & $-0.10$ & $-0.16$ & $-0.42$ & $-0.66$ & $-0.54$ & $-0.70$ & $-0.09$ & $-0.39$ & $-0.91$ & $-1.15$ & $\color{red}-3.05$ \\ 
   $\rvmax$ & $-0.59$ & $-0.54$ & $-0.38$ & $-0.12$ & $-0.02$ & $-0.15$ & $-0.17$ & $-0.30$ & $-0.55$ & $-0.28$ & $-0.25$ & $-0.32$ & $-0.69$ \\ 
   $\color{red}\cnfw$ & $-1.86$ & $\color{red}-3.17$ & $-0.02$ & $-0.27$ & $-0.20$ & $-0.13$ & $-0.51$ & $-0.58$ & $-2.25$ & $\color{red}-3.27$ & $-1.95$ & $-0.09$ & $\color{red}-2.52$ \\ 
   $\color{red}\ttf$ & $\color{red}-2.96$ & $\color{red}-1.24$ & $-0.24$ & $-0.02$ & $-0.50$ & $-0.01$ & $-0.13$ & $-0.26$ & $-0.84$ & $-1.12$ & $\color{red}-1.51$ & $-0.10$ & $-1.41$ \\ 
   $\color{red}\tfz$ & $-0.48$ & $-1.04$ & $-0.60$ & $-0.02$ & $-0.49$ & $-0.21$ & $-0.49$ & $-0.15$ & $-0.64$ & $-0.96$ & $\color{red}-2.17$ & $-0.43$ & $-1.61$ \\ 
   $\tsf$ & $-0.58$ & $-1.14$ & $-1.09$ & $-0.43$ & $-0.20$ & $-0.13$ & $-0.24$ & $-0.07$ & $-0.90$ & $-0.79$ & $-0.44$ & $-0.27$ & $-0.82$ \\ 
   $\tmax$ & $-0.09$ & $-0.08$ & $-0.06$ & $-0.39$ & $-0.03$ & $-0.30$ & $-0.10$ & $-0.18$ & $-0.04$ & $-0.02$ & $-0.98$ & $-1.12$ & $-0.02$ \\ 
   $\color{red}\ta$ & $-1.37$ & $-0.77$ & $-0.99$ & $-0.30$ & $-0.04$ & $-0.38$ & $-0.01$ & $-0.03$ & $-0.37$ & $-0.75$ & $\color{red}-2.83$ & $-0.33$ & $-1.25$ \\ 
%   $\color{red}\za$ & $-1.34$ & $-0.80$ & $-1.00$ & $-0.37$ & $-0.13$ & $-0.22$ & $-0.13$ & $-0.07$ & $-0.45$ & $-1.10$ & $\color{red}-3.56$ & $-0.24$ & $-1.46$ \\ 
   $\color{red}\tf$ & $\color{red}-3.72$ & $-0.86$ & $-0.31$ & $-0.38$ & $-1.22$ & $-1.11$ & $-1.16$ & $-0.77$ & $-0.34$ & $-1.40$ & $-1.78$ & $-0.00$ & $-1.28$ \\ 
%   $\color{red}\zf$ & $\color{red}-3.05$ & $-0.99$ & $-0.32$ & $-0.40$ & $-0.61$ & $-0.82$ & $-0.59$ & $-0.48$ & $-0.49$ & $-2.19$ & $-2.16$ & $-0.05$ & $-1.29$ \\ 
   $\delta_t$ & $-0.55$ & $-0.11$ & $-0.80$ & $-0.03$ & $-1.34$ & $-0.40$ & $-1.53$ & $-0.94$ & $-0.12$ & $-0.11$ & $-0.78$ & $-0.48$ & $-0.10$ \\ 
   $\tmm$ & $-0.20$ & $-0.46$ & $-0.62$ & $-0.26$ & $-0.77$ & $-0.00$ & $-0.96$ & $-0.52$ & $-0.24$ & $-0.47$ & $-1.84$ & $-0.69$ & $-0.52$ \\ 
%   $\zmm$ & $-0.84$ & $-0.41$ & $-0.22$ & $-0.16$ & $-0.25$ & $-0.04$ & $-0.41$ & $-0.34$ & $-0.19$ & $-0.57$ & $-2.68$ & $-0.35$ & $-0.95$ \\ 
   $\tam$ & $-0.02$ & $-0.11$ & $-0.01$ & $-0.11$ & $-0.22$ & $-1.35$ & $-0.76$ & $-1.36$ & $-0.42$ & $-0.31$ & $-0.14$ & $-0.80$ & $-0.59$ \\ 
%   $\zam$ & $-0.27$ & $-0.27$ & $-0.13$ & $-0.14$ & $-0.11$ & $-1.55$ & $-0.70$ & $-1.62$ & $-0.62$ & $-0.21$ & $-0.15$ & $-1.17$ & $-0.61$ \\ 
   $r_{t}$ & $-0.28$ & $-1.28$ & $-0.09$ & $-0.68$ & $-0.87$ & $-0.54$ & $-1.40$ & $-1.29$ & $-1.01$ & $-0.40$ & $-1.15$ & $-0.33$ & $-0.22$ \\ 
   $\color{red}\nbrz$ & $\color{red}-4.83$ & $-0.82$ & $-0.04$ & $-0.18$ & $-1.79$ & $-1.26$ & $-1.52$ & $-0.82$ & $-0.24$ & $-2.16$ & $\color{red}-2.93$ & $-0.34$ & $-1.33$ \\ 
   $\color{red}\nbr$ & $-1.43$ & $-1.38$ & $-0.24$ & $-0.11$ & $-0.45$ & $-0.92$ & $-0.50$ & $-0.51$ & $-0.59$ & $\color{red}-3.05$ & $\color{red}-3.33$ & $-2.11$ & $\color{red}-5.37$ \\ 
   $\color{red}\rz$ & $\color{red}-6.15$ & $-0.62$ & $-0.13$ & $-0.15$ & $-1.65$ & $-0.06$ & $-0.74$ & $-0.01$ & $-0.41$ & $-0.41$ & $-0.38$ & $-1.46$ & $-0.54$ \\ 
   $\color{red}\nleaf$ & $-0.81$ & $-0.63$ & $-0.16$ & $-0.15$ & $-0.70$ & $-1.30$ & $-1.04$ & $-1.28$ & $-0.22$ & $-1.64$ & $-1.50$ & $-1.57$ & $\color{red}-3.64$ \\ 
   $\rbl$ & $-1.21$ & $-1.75$ & $-0.77$ & $-1.59$ & $-0.24$ & $-0.61$ & $-0.76$ & $-1.55$ & $-1.23$ & $-0.98$ & $-1.09$ & $-0.07$ & $-0.54$ \\ 
   $\color{red}N_{<1:100}$ & $-1.56$ & $-1.62$ & $-0.49$ & $-0.35$ & $-0.24$ & $-0.28$ & $-0.16$ & $-0.07$ & $-0.91$ & $\color{red}-3.22$ & $\color{red}-3.49$ & $-1.46$ & $\color{red}-6.71$ \\ 
   $\color{red}N_{1:100-1:4}$ & $-0.72$ & $-0.85$ & $-0.17$ & $-0.22$ & $-0.38$ & $-0.82$ & $-0.33$ & $-0.33$ & $-0.32$ & $-1.40$ & $-1.83$ & $-2.29$ & $\color{red}-1.54$ \\ 
   $N_{>1:4}$ & $-0.01$ & $-0.25$ & $-0.70$ & $-1.35$ & $-0.45$ & $-1.76$ & $-1.29$ & $-2.30$ & $-0.48$ & $-0.05$ & $-0.08$ & $-0.07$ & $-0.11$ \\ 
   $r_{\rm mm}$ & $-1.29$ & $-1.97$ & $-0.89$ & $-1.09$ & $-0.05$ & $-0.37$ & $-0.23$ & $-0.84$ & $-1.40$ & $-1.22$ & $-0.96$ & $-0.22$ & $-0.83$ \\ 
   $f_{\rm ex,*}$ & $-0.13$ & $-0.44$ & $-0.16$ & $-0.39$ & $-0.69$ & $-0.93$ & $-1.27$ & $-1.59$ & $-0.47$ & $-0.00$ & $-0.80$ & $-0.08$ & $-0.00$ \\ 
   $f_{\rm ex,GC}$ & $-0.86$ & $-0.93$ & $-0.07$ & $-0.65$ & $-0.43$ & $-0.92$ & $-1.02$ & $-1.73$ & $-0.82$ & $-0.23$ & $-0.71$ & $-0.32$ & $-0.03$ \\ 
   \hline
  \end{tabular} 
\end{table*}

\begin{table*}
  \caption{Pearson correlation coefficients between the logarithm of the GC-related quantities from \autoref{tab:gcmetrics} (columns) and the logarithm of the galaxy formation-related quantities from \autoref{tab:galmetric1} and \autoref{tab:galmetric2} (rows). Correlations that are statistically significant according to their $p$-values in \autoref{tab:correlations_sp} (see the text for details) are marked in red.}
\label{tab:correlations_prlog}
  \begin{tabular}{p{1cm} p{0.65cm} p{0.65cm} p{0.65cm} p{0.85cm} p{0.75cm} p{1.2cm} p{1.1cm} p{1.6cm} p{0.65cm} p{0.65cm} p{0.65cm} p{0.85cm} p{0.65cm}}
   \hline
 \multicolumn{1}{c}{$\log$} & \multicolumn{13}{c}{$\log(\ldots)$}\\ \cline{2-14} \rule{-2pt}{3ex} 
   Quantity & $\color{red}\thub-\widetilde{\tau}$ & $\color{red}{\rm IQR}(\tau)$ & $-S(\tau)$ & $K(\tau)+3$ & $-\widetilde{\rm [Fe/H]}$ & ${\rm IQR}({\rm [Fe/H]})$ & $-S({\rm [Fe/H]})$ & $K({\rm [Fe/H]})+3$ & ${\rm IQR}^2$ & $\color{red}\riqr$ & $\color{red}\frac{{\rm d[Fe/H]}}{{\rm d}\log t}$ & $-{\rm [Fe/H]}_0$ & $\color{red}N_{\rm GC}$ \\ 
   \hline
   $\mvir$ & $0.05$ & $0.23$ & $-0.24$ & $0.02$ & $0.25$ & $0.37$ & $-0.31$ & $-0.41$ & $0.32$ & $-0.04$ & $0.14$ & $0.43$ & $0.38$ \\ 
   $\rvir$ & $0.05$ & $0.23$ & $-0.24$ & $0.02$ & $0.25$ & $0.37$ & $-0.31$ & $-0.41$ & $0.32$ & $-0.04$ & $0.14$ & $0.43$ & $0.38$ \\ 
   $\color{red}\vmax$ & $-0.09$ & $-0.01$ & $-0.30$ & $-0.05$ & $0.21$ & $0.26$ & $-0.23$ & $-0.25$ & $0.09$ & $0.17$ & $0.35$ & $0.39$ & $\color{red}0.66$ \\ 
   $\rvmax$ & $-0.01$ & $0.07$ & $0.02$ & $0.01$ & $-0.07$ & $0.14$ & $0.01$ & $-0.11$ & $0.11$ & $0.00$ & $0.01$ & $0.10$ & $-0.14$ \\ 
   $\color{red}\cnfw$ & $-0.40$ & $\color{red}-0.60$ & $-0.10$ & $0.02$ & $-0.02$ & $-0.09$ & $-0.04$ & $0.17$ & $-0.48$ & $\color{red}0.64$ & $0.48$ & $0.01$ & $\color{red}0.60$ \\ 
   $\color{red}\thub-\ttf$ & $\color{red}0.61$ & $\color{red}0.49$ & $0.13$ & $0.02$ & $-0.12$ & $0.11$ & $-0.05$ & $-0.21$ & $0.41$ & $-0.50$ & $\color{red}-0.55$ & $0.04$ & $-0.45$ \\ 
   $\color{red}\thub-\tfz$ & $0.14$ & $0.31$ & $0.22$ & $0.11$ & $0.13$ & $-0.05$ & $0.15$ & $-0.03$ & $0.21$ & $-0.39$ & $\color{red}-0.58$ & $-0.21$ & $-0.51$ \\ 
   $\thub-\tsf$ & $0.11$ & $0.21$ & $0.47$ & $0.36$ & $0.01$ & $-0.11$ & $0.19$ & $0.06$ & $0.12$ & $-0.31$ & $-0.20$ & $0.08$ & $-0.31$ \\ 
   $\thub-\tmax$ & $0.04$ & $-0.03$ & $0.08$ & $0.16$ & $-0.06$ & $-0.14$ & $0.10$ & $0.12$ & $-0.07$ & $-0.05$ & $0.28$ & $0.32$ & $-0.07$ \\ 
   $\color{red}\thub-\tau_{\rm a}$ & $0.35$ & $0.26$ & $0.33$ & $0.23$ & $0.02$ & $-0.13$ & $0.14$ & $0.02$ & $0.14$ & $-0.37$ & $\color{red}-0.64$ & $-0.14$ & $-0.43$ \\ 
%   $\color{red}1+z_{\rm a}$ & $-0.35$ & $-0.25$ & $-0.33$ & $-0.23$ & $-0.01$ & $0.13$ & $-0.14$ & $-0.02$ & $-0.14$ & $0.37$ & $\color{red}0.63$ & $0.14$ & $0.43$ \\ 
   $\color{red}\thub-\tau_{\rm f}$ & $\color{red}0.64$ & $0.27$ & $0.18$ & $0.14$ & $-0.32$ & $-0.32$ & $0.33$ & $0.28$ & $0.08$ & $-0.51$ & $-0.52$ & $-0.00$ & $-0.45$ \\ 
%   $\color{red}1+z_{\rm f}$ & $\color{red}-0.64$ & $-0.26$ & $-0.19$ & $-0.14$ & $0.33$ & $0.32$ & $-0.34$ & $-0.28$ & $-0.07$ & $0.50$ & $0.51$ & $0.00$ & $0.45$ \\ 
   $1+\delta_t$ & $-0.26$ & $0.09$ & $0.13$ & $0.13$ & $0.38$ & $0.20$ & $-0.21$ & $-0.31$ & $0.15$ & $0.01$ & $-0.28$ & $-0.17$ & $-0.01$ \\ 
   $\thub-\tmm$ & $0.21$ & $0.27$ & $0.01$ & $0.15$ & $0.21$ & $0.00$ & $-0.01$ & $-0.16$ & $0.23$ & $-0.29$ & $-0.61$ & $-0.23$ & $-0.28$ \\ 
%   $1+z_{\rm mm}$ & $-0.18$ & $-0.27$ & $-0.03$ & $-0.16$ & $-0.22$ & $-0.01$ & $0.02$ & $0.17$ & $-0.23$ & $0.29$ & $0.60$ & $0.24$ & $0.28$ \\ 
   $\thub-\tau_{\rm am}$ & $0.13$ & $0.14$ & $-0.04$ & $-0.11$ & $0.06$ & $0.45$ & $-0.25$ & $-0.41$ & $0.28$ & $0.11$ & $0.10$ & $0.43$ & $0.30$ \\ 
%   $1+z_{\rm am}$ & $-0.10$ & $-0.12$ & $0.03$ & $0.10$ & $-0.07$ & $-0.44$ & $0.25$ & $0.41$ & $-0.26$ & $-0.13$ & $-0.10$ & $-0.41$ & $-0.29$ \\ 
   $1+r_{t}$ & $0.05$ & $0.38$ & $-0.22$ & $-0.23$ & $0.28$ & $0.29$ & $-0.21$ & $-0.38$ & $0.39$ & $-0.27$ & $-0.34$ & $-0.08$ & $-0.05$ \\ 
   $\color{red}\nbrz$ & $\color{red}-0.65$ & $-0.14$ & $0.06$ & $-0.11$ & $0.38$ & $0.46$ & $-0.40$ & $-0.40$ & $0.09$ & $0.47$ & $\color{red}0.57$ & $0.24$ & $0.44$ \\ 
   $\color{red}\nbr$ & $-0.32$ & $-0.31$ & $-0.17$ & $0.04$ & $0.24$ & $0.38$ & $-0.31$ & $-0.31$ & $-0.08$ & $\color{red}0.59$ & $\color{red}0.62$ & $0.54$ & $\color{red}0.76$ \\ 
   $\color{red}1+\rz$ & $\color{red}-0.81$ & $-0.20$ & $0.22$ & $0.02$ & $0.45$ & $0.02$ & $-0.17$ & $-0.05$ & $-0.14$ & $0.25$ & $0.18$ & $-0.42$ & $-0.21$ \\ 
   $\color{red}\nleaf$ & $-0.18$ & $-0.12$ & $-0.22$ & $-0.07$ & $0.29$ & $0.47$ & $-0.38$ & $-0.43$ & $0.09$ & $0.42$ & $0.46$ & $0.53$ & $\color{red}0.70$ \\ 
   $1+\rbl$ & $-0.34$ & $-0.50$ & $0.23$ & $0.31$ & $-0.14$ & $-0.29$ & $0.18$ & $0.34$ & $-0.48$ & $0.41$ & $0.38$ & $0.02$ & $0.15$ \\ 
   $\color{red}N_{<1:100}$ & $-0.45$ & $-0.44$ & $-0.08$ & $0.18$ & $0.24$ & $0.19$ & $-0.23$ & $-0.17$ & $-0.26$ & $\color{red}0.63$ & $\color{red}0.62$ & $0.42$ & $\color{red}0.75$ \\ 
   $\color{red}N_{1:100-1:4}$ & $-0.24$ & $-0.29$ & $0.04$ & $0.11$ & $0.18$ & $0.35$ & $-0.25$ & $-0.24$ & $-0.08$ & $0.55$ & $0.54$ & $0.52$ & $\color{red}0.56$ \\ 
   $N_{>1:4}$ & $0.11$ & $0.09$ & $-0.32$ & $-0.05$ & $0.13$ & $0.36$ & $-0.30$ & $-0.40$ & $0.18$ & $0.03$ & $-0.13$ & $0.04$ & $-0.06$ \\ 
   $1+r_{\rm mm}$ & $0.35$ & $0.51$ & $-0.32$ & $-0.40$ & $0.02$ & $0.26$ & $-0.17$ & $-0.32$ & $0.48$ & $-0.44$ & $-0.33$ & $-0.07$ & $-0.24$ \\ 
   $1+\fexs$ & $0.05$ & $0.25$ & $-0.04$ & $-0.04$ & $0.26$ & $0.36$ & $-0.22$ & $-0.42$ & $0.32$ & $-0.07$ & $-0.27$ & $0.02$ & $0.06$ \\ 
   $1+\fexgc$ & $0.26$ & $0.35$ & $-0.17$ & $-0.14$ & $0.20$ & $0.39$ & $-0.19$ & $-0.43$ & $0.41$ & $-0.17$ & $-0.23$ & $0.22$ & $0.07$ \\ 
   \hline
  \end{tabular} 
\end{table*}

\begin{table*}
  \caption{Logarithm of the $p$-values of Pearson correlation coefficients between the logarithm of the GC-related quantities from \autoref{tab:gcmetrics} (columns) and the logarithm of the galaxy formation-related quantities from \autoref{tab:galmetric1} and \autoref{tab:galmetric2} (rows). Correlations that are statistically significant according to their $p$-values in \autoref{tab:correlations_sp} (see the text for details) are marked in red.}
\label{tab:correlations_pplog}
  \begin{tabular}{p{1cm} p{0.65cm} p{0.65cm} p{0.65cm} p{0.85cm} p{0.75cm} p{1.2cm} p{1.1cm} p{1.6cm} p{0.65cm} p{0.65cm} p{0.65cm} p{0.85cm} p{0.65cm}}
   \hline
 \multicolumn{1}{c}{$\log$} & \multicolumn{13}{c}{$\log(\ldots)$}\\ \cline{2-14} \rule{-2pt}{3ex} 
   Quantity & $\color{red}\thub-\widetilde{\tau}$ & $\color{red}{\rm IQR}(\tau)$ & $-S(\tau)$ & $K(\tau)+3$ & $-\widetilde{\rm [Fe/H]}$ & ${\rm IQR}({\rm [Fe/H]})$ & $-S({\rm [Fe/H]})$ & $K({\rm [Fe/H]})+3$ & ${\rm IQR}^2$ & $\color{red}\riqr$ & $\color{red}\frac{{\rm d[Fe/H]}}{{\rm d}\log t}$ & $-{\rm [Fe/H]}_0$ & $\color{red}N_{\rm GC}$ \\ 
   \hline
   $\mvir$ & $-0.10$ & $-0.57$ & $-0.48$ & $-0.03$ & $-0.66$ & $-1.17$ & $-0.85$ & $-1.40$ & $-0.90$ & $-0.08$ & $-0.31$ & $-1.48$ & $-1.21$ \\ 
   $\rvir$ & $-0.10$ & $-0.57$ & $-0.48$ & $-0.03$ & $-0.66$ & $-1.17$ & $-0.85$ & $-1.40$ & $-0.90$ & $-0.08$ & $-0.31$ & $-1.48$ & $-1.21$ \\ 
   $\color{red}\vmax$ & $-0.17$ & $-0.02$ & $-0.68$ & $-0.10$ & $-0.51$ & $-0.67$ & $-0.56$ & $-0.63$ & $-0.18$ & $-0.38$ & $-1.04$ & $-1.28$ & $\color{red}-3.54$ \\ 
   $\rvmax$ & $-0.01$ & $-0.14$ & $-0.03$ & $-0.02$ & $-0.13$ & $-0.30$ & $-0.02$ & $-0.22$ & $-0.22$ & $-0.00$ & $-0.01$ & $-0.19$ & $-0.30$ \\ 
   $\color{red}\cnfw$ & $-1.32$ & $\color{red}-2.79$ & $-0.16$ & $-0.04$ & $-0.04$ & $-0.18$ & $-0.07$ & $-0.38$ & $-1.83$ & $\color{red}-3.21$ & $-1.79$ & $-0.01$ & $\color{red}-2.78$ \\ 
   $\color{red}\thub-\tau_{25}$ & $\color{red}-2.92$ & $\color{red}-1.89$ & $-0.23$ & $-0.03$ & $-0.26$ & $-0.23$ & $-0.09$ & $-0.50$ & $-1.37$ & $-1.96$ & $\color{red}-2.34$ & $-0.07$ & $-1.61$ \\ 
   $\color{red}\thub-\tau_{50}$ & $-0.30$ & $-0.88$ & $-0.43$ & $-0.21$ & $-0.28$ & $-0.10$ & $-0.30$ & $-0.05$ & $-0.51$ & $-1.28$ & $\color{red}-2.65$ & $-0.51$ & $-2.05$ \\ 
   $\thub-\tau_{75}$ & $-0.22$ & $-0.50$ & $-1.36$ & $-1.13$ & $-0.02$ & $-0.22$ & $-0.42$ & $-0.11$ & $-0.23$ & $-0.88$ & $-0.46$ & $-0.15$ & $-0.87$ \\ 
   $\thub-\tau_{\rm max}$ & $-0.06$ & $-0.05$ & $-0.12$ & $-0.35$ & $-0.12$ & $-0.29$ & $-0.20$ & $-0.26$ & $-0.14$ & $-0.09$ & $-0.75$ & $-0.94$ & $-0.12$ \\ 
   $\color{red}\thub-\tau_{\rm a}$ & $-1.08$ & $-0.67$ & $-0.77$ & $-0.56$ & $-0.03$ & $-0.27$ & $-0.28$ & $-0.03$ & $-0.30$ & $-1.19$ & $\color{red}-3.24$ & $-0.30$ & $-1.47$ \\ 
%   $\color{red}1+z_{\rm a}$ & $-1.08$ & $-0.66$ & $-0.78$ & $-0.56$ & $-0.01$ & $-0.27$ & $-0.29$ & $-0.03$ & $-0.30$ & $-1.17$ & $\color{red}-3.19$ & $-0.30$ & $-1.47$ \\ 
   $\color{red}\thub-\tau_{\rm f}$ & $\color{red}-3.22$ & $-0.72$ & $-0.34$ & $-0.29$ & $-0.92$ & $-0.92$ & $-0.96$ & $-0.75$ & $-0.15$ & $-2.00$ & $-2.11$ & $-0.01$ & $-1.62$ \\ 
%   $\color{red}1+z_{\rm f}$ & $\color{red}-3.25$ & $-0.69$ & $-0.36$ & $-0.30$ & $-0.97$ & $-0.92$ & $-0.99$ & $-0.77$ & $-0.14$ & $-1.95$ & $-2.07$ & $-0.00$ & $-1.65$ \\ 
   $1+\delta_t$ & $-0.67$ & $-0.18$ & $-0.23$ & $-0.27$ & $-1.23$ & $-0.47$ & $-0.49$ & $-0.89$ & $-0.32$ & $-0.02$ & $-0.78$ & $-0.37$ & $-0.01$ \\ 
   $\thub-\tmm$ & $-0.41$ & $-0.59$ & $-0.01$ & $-0.26$ & $-0.40$ & $-0.01$ & $-0.01$ & $-0.30$ & $-0.48$ & $-0.66$ & $-2.25$ & $-0.46$ & $-0.62$ \\ 
%   $1+\zmm$ & $-0.34$ & $-0.58$ & $-0.04$ & $-0.30$ & $-0.45$ & $-0.01$ & $-0.03$ & $-0.32$ & $-0.47$ & $-0.65$ & $-2.21$ & $-0.49$ & $-0.61$ \\ 
   $\thub-\tam$ & $-0.27$ & $-0.29$ & $-0.06$ & $-0.23$ & $-0.10$ & $-1.64$ & $-0.63$ & $-1.39$ & $-0.74$ & $-0.23$ & $-0.19$ & $-1.47$ & $-0.83$ \\ 
%   $1+\zam$ & $-0.20$ & $-0.25$ & $-0.04$ & $-0.20$ & $-0.13$ & $-1.59$ & $-0.64$ & $-1.36$ & $-0.68$ & $-0.27$ & $-0.19$ & $-1.36$ & $-0.81$ \\ 
   $1+r_{t}$ & $-0.09$ & $-1.21$ & $-0.43$ & $-0.55$ & $-0.74$ & $-0.79$ & $-0.49$ & $-1.22$ & $-1.29$ & $-0.70$ & $-1.03$ & $-0.15$ & $-0.09$ \\ 
   $\color{red}\nbrz$ & $\color{red}-3.20$ & $-0.28$ & $-0.09$ & $-0.22$ & $-1.18$ & $-1.64$ & $-1.25$ & $-1.25$ & $-0.17$ & $-1.70$ & $\color{red}-2.46$ & $-0.60$ & $-1.48$ \\ 
   $\color{red}\nbr$ & $-0.93$ & $-0.88$ & $-0.32$ & $-0.08$ & $-0.62$ & $-1.21$ & $-0.87$ & $-0.90$ & $-0.16$ & $\color{red}-2.69$ & $\color{red}-3.00$ & $-2.28$ & $\color{red}-4.95$ \\ 
   $\color{red}1+\rz$ & $\color{red}-5.93$ & $-0.47$ & $-0.43$ & $-0.04$ & $-1.61$ & $-0.04$ & $-0.37$ & $-0.09$ & $-0.30$ & $-0.63$ & $-0.40$ & $-1.45$ & $-0.52$ \\ 
   $\color{red}\nleaf$ & $-0.42$ & $-0.24$ & $-0.45$ & $-0.12$ & $-0.79$ & $-1.75$ & $-1.15$ & $-1.51$ & $-0.18$ & $-1.44$ & $-1.70$ & $-2.18$ & $\color{red}-4.03$ \\ 
   $1+\rbl$ & $-1.02$ & $-1.96$ & $-0.47$ & $-0.88$ & $-0.29$ & $-0.78$ & $-0.39$ & $-1.02$ & $-1.84$ & $-1.36$ & $-1.23$ & $-0.03$ & $-0.31$ \\ 
   $\color{red}N_{<1:100}$ & $-1.63$ & $-1.57$ & $-0.14$ & $-0.41$ & $-0.60$ & $-0.45$ & $-0.55$ & $-0.38$ & $-0.66$ & $\color{red}-3.14$ & $\color{red}-3.08$ & $-1.43$ & $\color{red}-4.85$ \\ 
   $\color{red}N_{1:100-1:4}$ & $-0.62$ & $-0.80$ & $-0.06$ & $-0.22$ & $-0.42$ & $-1.08$ & $-0.63$ & $-0.59$ & $-0.15$ & $-2.36$ & $-2.25$ & $-2.11$ & $\color{red}-2.45$ \\ 
   $N_{>1:4}$ & $-0.19$ & $-0.15$ & $-0.58$ & $-0.07$ & $-0.23$ & $-0.88$ & $-0.66$ & $-1.04$ & $-0.34$ & $-0.04$ & $-0.22$ & $-0.05$ & $-0.10$ \\ 
   $1+r_{\rm mm}$ & $-1.07$ & $-2.06$ & $-0.75$ & $-1.34$ & $-0.03$ & $-0.68$ & $-0.36$ & $-0.92$ & $-1.84$ & $-1.56$ & $-0.97$ & $-0.13$ & $-0.61$ \\ 
   $1+\fexs$ & $-0.08$ & $-0.63$ & $-0.07$ & $-0.08$ & $-0.67$ & $-1.11$ & $-0.52$ & $-1.42$ & $-0.94$ & $-0.13$ & $-0.73$ & $-0.03$ & $-0.10$ \\ 
   $1+\fexgc$ & $-0.67$ & $-1.07$ & $-0.32$ & $-0.29$ & $-0.48$ & $-1.28$ & $-0.44$ & $-1.47$ & $-1.40$ & $-0.39$ & $-0.57$ & $-0.54$ & $-0.14$ \\ 
   \hline
  \end{tabular} 
\end{table*}

\section{Under-destruction of GCs} \label{sec:appfeh}

\subsection{Physical reason for an excess of metal-rich GCs}
\begin{figure}
\includegraphics[width=\hsize]{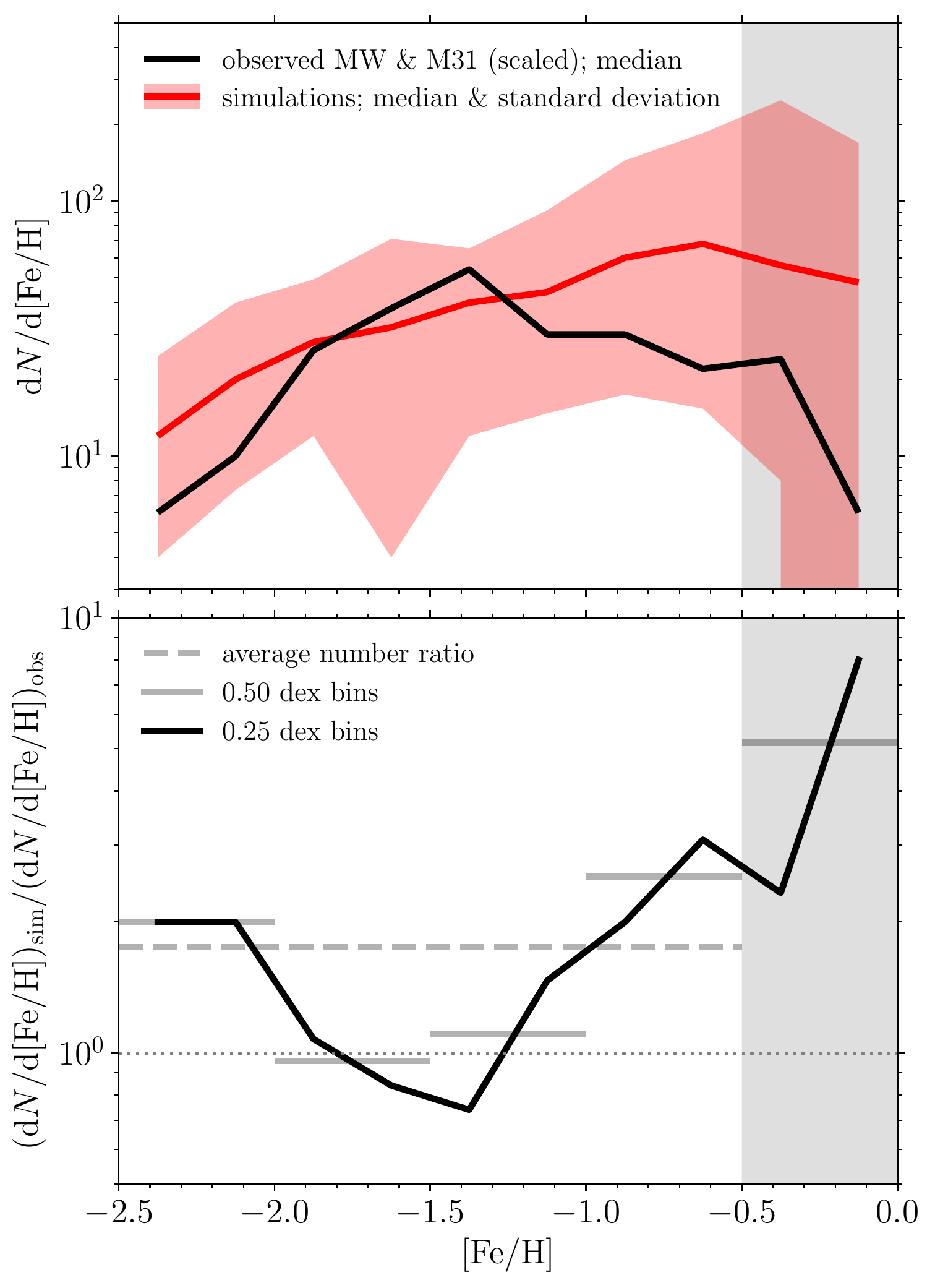}%
\caption{
\label{fig:feh_dis}
Comparison of simulated and observed GC metallicity distributions (cf.~\autoref{fig:compobs}) to quantify the under-disruption of GCs in \emosaics as a function of metallicity. Top panel: median $\feh$ distribution of GCs in the Milky Way and in M31 (scaled as in \autoref{fig:compobs}) in black, compared to the median of the simulations in red, with red shading indicating the $16^{\rm th}$ and $84^{\rm th}$ percentiles. Bottom panel: ratio between the simulated and observed median $\feh$ distributions from the top panel (black line). The grey solid and grey dashed lines show the mean ratio for bins of 0.5~dex in $\feh$ and for $-2.5<\feh<-0.5$, respectively. The \emosaics simulations overproduce GCs with $-1.0<\feh<-0.5$ by a factor of $\sim2.5$.
}
\end{figure}
As demonstrated in Section~\ref{sec:valid}, the \emosaics simulations have an excess of metal-rich GCs at the highest metallicities. To remedy this, we have restricted our analysis to $-2.5<\feh<-0.5$ throughout this work, which has the additional benefit that the sample of Galactic GCs with known ages covers roughly the same metallicity range. However, even at $-1.0<\feh<-0.5$, the simulations still contain too many GCs. This is quantified in \autoref{fig:feh_dis}, which compares the GC metallicity distribution obtained in the simulations to the median observed in M31 and the Milky Way. We choose to combine these two galaxies, because their GC metallicity distributions differ considerably (see \autoref{fig:compobs}) and together better represent the variety of distributions that may be expected from a larger sample (as is available with \emosaics).

The simulations overpredict the number of GCs by a factor of $\sim5$ for $-0.5<\feh<0.0$, justifying our choice to restrict the analysis of this paper to $-2.5<\feh<-0.5$. However, they still overpredict the number of GCs by a factor of $\sim2.5$ for $-1.0<\feh<-0.5$, which may influence the relations between the quantities characterising galaxy formation and assembly and those describing the GC age-metallicity distribution (see \autoref{tab:fit}). We demonstrate below that the diagnostic power of the inferred relations is negligibly affected by the excess of metal-rich GCs.

As discussed at length in \citet{pfeffer18} and Section~\ref{sec:valid}, the \emosaics simulations are not sufficiently effective at disrupting GCs due to the omission of a cold ISM. The simulations lack dense substructures like giant molecular clouds that otherwise would tidally disrupt stellar clusters. The total amount of GC mass loss that should have been applied had the simulations included a cold ISM increases with the ambient ISM pressure \citep[e.g.][]{elmegreen10b,kruijssen11,pfeffer18,li18} and the time spent in the natal environment \citep[e.g.][]{kruijssen15b}.

\begin{figure}
\includegraphics[width=\hsize]{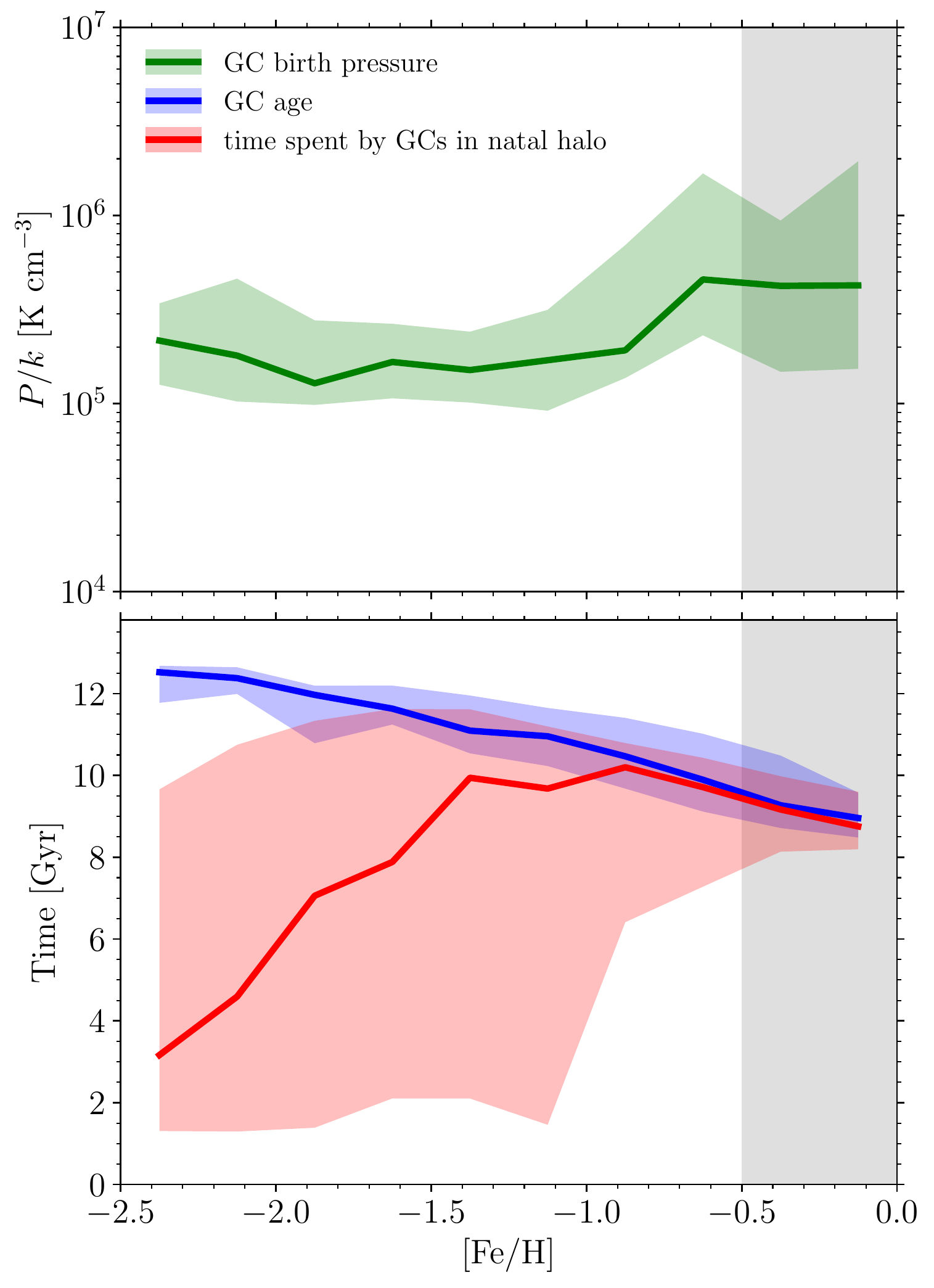}%
\caption{
\label{fig:age_pres}
Birth pressure (top panel), age, and time spent in the natal halo as a function of metallicity (bottom panel) for surviving GCs. In each metallicity bin, we determine the median value for the GCs in each of the 25 simulations. The median across the simulations is then shown as a solid line, with shading indicating the $16^{\rm th}$ and $84^{\rm th}$ percentiles. This figure shows that the GC birth pressure does not vary with metallicity, but metal-rich GCs spend their entire life in their natal environment, implying that the underestimation of tidal shock-driven disruption most strongly affects these GCs.
}
\end{figure}
\autoref{fig:age_pres} shows the GC birth pressure, age, and time spent in the natal halo as a function of the metallicity. It demonstrates that surviving GCs have similar birth pressures. This constancy results from two competing effects. First, there is a minimum gas surface density for the formation of GC progenitor clusters (with initial masses needed for long-term survival of $M\ga10^5~\msun$, cf.~\citealt{kruijssen15b}) in the adopted ICMF model \citep{reinacampos17}. This minimum gas surface density is $\Sigma_{\rm min}\sim50~\msun~\pc^2$, which translates to a minimum midplane ISM pressure as $P_{\rm min}\sim\pi G\Sigma_{\rm min}^2$, resulting in $P_{\rm min}/k\sim10^5~{\rm K}~\cmc$. We thus expect GCs to be born at pressures $P/k>10^5~{\rm K}~\cmc$, independently of metallicity. Secondly, the birth pressure probability distribution function steeply decreases with pressure \citep{pfeffer18}, such that clusters with low birth pressures are the most numerous. Because cluster disruption proceeds more rapidly at higher birth pressures, the surviving GCs typically have the lowest birth pressures needed for the formation of long-lived clusters, i.e.\ just above $P/k\sim10^5~{\rm K}~\cmc$. \autoref{fig:age_pres} shows that this is indeed the case.

While the disruption rate experienced by surviving GCs is roughly constant as a function of metallicity, the bottom panel of \autoref{fig:age_pres} shows that the time spent in the disruptive environment of the natal halo is a steep function of metallicity. At metallicities $\feh>-1.0$, there are almost no GCs that migrate out of the host halo in $<8~\gyr$, whereas lower-metallicity GCs generally spend a considerable fraction of their lives in a halo that they did not form in. This is a direct result of the galaxy mass-metallicity relation -- more metal-rich GCs formed in more massive haloes (and at later cosmic times), which implies a lower host galaxy merger rate and inefficient migration. Metal-poor GCs can be tidally stripped off their host relatively quickly by accretion onto a more massive galaxy, whereas metal-rich GCs spend more time in their disruptive natal environment. \autoref{fig:age_pres} thus demonstrates that the amount of GC mass loss that would have been applied had \emosaics included a cold ISM is expected to increase with metallicity, most strongly affecting metal-rich GCs with $\feh>-1.0$.

\subsection{Minor implications for the presented analysis}
We now quantify how the under-destruction of GCs at $-1.0<\feh<-0.5$ affects the metrics describing the GC age-metallicity distribution (\autoref{tab:gcmetrics}), the 20 statistically significant correlations between quantities characterising galaxy formation and assembly and those describing the GC age-metallicity distribution (\autoref{tab:fit}), and the constraints on the formation and assembly history of the Milky Way obtained by applying these correlations to the Galactic GC population (table~3 of \citealt{kruijssen18c}). This is done by artificially pruning the GC populations of the simulated galaxies by removing a randomly selected subset of 61~per~cent of the GCs with $-1.0<\feh<-0.5$ and repeating the correlation analysis of Sections~\ref{sec:agez} and~\ref{sec:hist}. This percentage is chosen, because \emosaics overpredicts the GCs with $-1.0<\feh<-0.5$ by a factor of $2.55$ (see \autoref{fig:feh_dis}).

\begin{table*}
  \caption{Repeat of \autoref{tab:gcmetrics} for the experiment described in Appendix~\ref{sec:appfeh}, in which the under-destruction of GCs is corrected for by removing a randomly selected subset of 61~per~cent of the GCs with $-1.0<\feh<-0.5$.}
\label{tab:gcmetrics_fehdis}
  \begin{tabular}{l c c c c c c c c c c c c c}
   \hline
   Name & $\widetilde{\tau}$ & $\iqr(\tau)$ & $S(\tau)$ & $K(\tau)$ & $\widetilde{\feh}$ & $\iqr(\feh)$ & $S(\feh)$ & $K(\feh)$ & $\iqr^2$ & $\iqr_{\rm div}$ & $\frac{{\rm d[Fe/H]}}{{\rm d}\log t}$ & $\feh_0$ & $N_{\rm GC}$ \\ 
   \hline
   MW00 & $11.08$ & $1.73$ & $0.27$ & $-1.15$ & $-1.44$ & $0.74$ & $0.00$ & $-0.99$ & $1.28$ & $0.43$ & $4.41$ & $-3.00$ & $72$ \\ 
   MW01 & $10.97$ & $1.31$ & $-0.43$ & $0.00$ & $-1.20$ & $0.59$ & $-0.70$ & $-0.02$ & $0.77$ & $0.45$ & $3.75$ & $-2.86$ & $72$ \\ 
   MW02 & $11.68$ & $1.34$ & $-1.30$ & $2.68$ & $-1.25$ & $0.90$ & $-0.19$ & $-0.99$ & $1.21$ & $0.67$ & $6.29$ & $-3.27$ & $208$ \\ 
   MW03 & $11.47$ & $1.57$ & $-0.08$ & $-0.85$ & $-1.23$ & $0.80$ & $-0.28$ & $-0.86$ & $1.26$ & $0.51$ & $3.89$ & $-2.62$ & $107$ \\ 
   MW04 & $11.75$ & $0.88$ & $0.20$ & $-0.40$ & $-1.23$ & $0.83$ & $-0.53$ & $-0.75$ & $0.73$ & $0.94$ & $5.75$ & $-2.89$ & $97$ \\ 
   MW05 & $11.96$ & $0.71$ & $-0.02$ & $0.22$ & $-1.07$ & $0.82$ & $-0.65$ & $-0.57$ & $0.58$ & $1.15$ & $6.40$ & $-2.65$ & $178$ \\ 
   MW06 & $10.47$ & $0.91$ & $-3.45$ & $24.61$ & $-0.88$ & $0.76$ & $-0.85$ & $-0.28$ & $0.69$ & $0.84$ & $3.00$ & $-2.46$ & $105$ \\ 
   MW07 & $11.48$ & $1.77$ & $-0.85$ & $-0.02$ & $-1.12$ & $1.04$ & $-0.36$ & $-1.07$ & $1.84$ & $0.59$ & $2.65$ & $-2.24$ & $47$ \\ 
   MW08 & $12.35$ & $0.96$ & $-2.07$ & $4.98$ & $-1.09$ & $0.28$ & $-1.04$ & $1.72$ & $0.27$ & $0.29$ & $2.49$ & $-1.66$ & $36$ \\ 
   MW09 & $10.90$ & $1.20$ & $-0.03$ & $0.31$ & $-1.14$ & $0.87$ & $-0.60$ & $-0.43$ & $1.04$ & $0.72$ & $4.05$ & $-3.06$ & $64$ \\ 
   MW10 & $10.28$ & $1.26$ & $-0.14$ & $-0.16$ & $-1.09$ & $0.85$ & $-0.49$ & $-0.83$ & $1.07$ & $0.67$ & $3.07$ & $-2.66$ & $180$ \\ 
   MW11 & $11.34$ & $1.76$ & $-1.67$ & $2.99$ & $-1.30$ & $1.00$ & $0.06$ & $-1.24$ & $1.76$ & $0.57$ & $3.29$ & $-2.72$ & $50$ \\ 
   MW12 & $10.69$ & $1.18$ & $-2.07$ & $10.75$ & $-1.31$ & $0.79$ & $-0.26$ & $-0.77$ & $0.93$ & $0.67$ & $5.61$ & $-3.93$ & $174$ \\ 
   MW13 & $12.09$ & $2.17$ & $-0.64$ & $-0.81$ & $-1.50$ & $0.76$ & $-0.08$ & $-0.85$ & $1.65$ & $0.35$ & $3.16$ & $-2.49$ & $78$ \\ 
   MW14 & $10.87$ & $2.25$ & $-1.93$ & $3.06$ & $-1.25$ & $0.71$ & $-0.16$ & $-0.87$ & $1.60$ & $0.32$ & $2.28$ & $-2.40$ & $95$ \\ 
   MW15 & $8.83$ & $4.40$ & $0.01$ & $-1.26$ & $-1.12$ & $1.11$ & $-0.39$ & $-1.17$ & $4.88$ & $0.25$ & $2.45$ & $-2.75$ & $44$ \\ 
   MW16 & $11.48$ & $1.64$ & $-0.94$ & $0.87$ & $-1.38$ & $0.90$ & $-0.25$ & $-1.08$ & $1.48$ & $0.55$ & $3.22$ & $-2.48$ & $143$ \\ 
   MW17 & $9.51$ & $1.45$ & $-0.94$ & $5.66$ & $-1.12$ & $0.65$ & $-0.57$ & $-0.46$ & $0.94$ & $0.45$ & $3.45$ & $-3.25$ & $73$ \\ 
   MW18 & $12.45$ & $2.17$ & $-0.69$ & $-0.84$ & $-1.65$ & $0.68$ & $0.03$ & $-0.55$ & $1.48$ & $0.31$ & $3.18$ & $-2.57$ & $55$ \\ 
   MW19 & $9.87$ & $1.25$ & $-1.57$ & $4.12$ & $-1.02$ & $0.95$ & $-0.63$ & $-0.94$ & $1.19$ & $0.76$ & $3.16$ & $-2.94$ & $33$ \\ 
   MW20 & $10.44$ & $1.87$ & $0.39$ & $-0.54$ & $-1.18$ & $0.80$ & $-0.46$ & $-0.55$ & $1.50$ & $0.43$ & $2.12$ & $-2.15$ & $58$ \\ 
   MW21 & $11.94$ & $0.77$ & $-2.73$ & $12.05$ & $-1.36$ & $0.78$ & $-0.19$ & $-0.93$ & $0.60$ & $1.01$ & $4.27$ & $-2.54$ & $102$ \\ 
   MW22 & $10.80$ & $1.59$ & $-0.57$ & $0.30$ & $-1.28$ & $0.93$ & $-0.26$ & $-0.97$ & $1.48$ & $0.58$ & $3.41$ & $-2.86$ & $125$ \\ 
   MW23 & $11.35$ & $0.60$ & $-3.01$ & $9.83$ & $-1.09$ & $0.68$ & $-0.83$ & $-0.11$ & $0.41$ & $1.13$ & $4.33$ & $-2.94$ & $205$ \\ 
   MW24 & $10.39$ & $3.51$ & $0.04$ & $-1.41$ & $-1.13$ & $0.68$ & $-0.54$ & $-0.75$ & $2.39$ & $0.19$ & $2.15$ & $-2.33$ & $44$ \\ 
   \hline
   Median & $11.08$ & $1.45$ & $-0.69$ & $0.30$ & $-1.20$ & $0.80$ & $-0.39$ & $-0.83$ & $1.21$ & $0.57$ & $3.29$ & $-2.66$ & $78$ \\ 
   IQR & $1.21$ & $0.59$ & $1.64$ & $4.66$ & $0.18$ & $0.19$ & $0.41$ & $0.42$ & $0.72$ & $0.30$ & $1.27$ & $0.46$ & $70$ \\ 
   Range & $3.62$ & $3.80$ & $3.84$ & $26.02$ & $0.77$ & $0.83$ & $1.10$ & $2.96$ & $4.62$ & $0.96$ & $4.28$ & $2.27$ & $175$ \\ 
   \hline
  \end{tabular} 
\end{table*}

\autoref{tab:gcmetrics_fehdis} shows how \autoref{tab:gcmetrics} changes after pruning the simulated GC populations from 61~per~cent of the GCs with $-1.0<\feh<-0.5$. The comparison of both tables shows that the metrics are not greatly affected. The median values of most columns agree to within the quoted interquartile ranges. The only exceptions are the median metallicity, $\widetilde{\feh}$, and its skewness, $S(\feh)$, but these metrics do not exhibit any statistically significant correlations with metrics describing galaxy formation and assembly. As a result, they are left out of the further analysis in this paper and thus their change does not affect the presented correlations. The reason that most metrics are only weakly affected is that they are defined by the entire distribution of GCs in age-metallicity space, which is dominated by the GCs with $-2.5<\feh<-1.0$.

\begin{table*}
\caption{Repeat of \autoref{tab:fit} for the experiment described in Appendix~\ref{sec:appfeh}, in which the under-destruction of GCs is corrected for by removing a randomly selected subset of 61~per~cent of the GCs with $-1.0<\feh<-0.5$. Contrary to \autoref{tab:fit}, these correlations now do not use the globally-corrected number of GCs ($\ngcp\equiv\ngc/\fcorr$), but the actual number of GCs $\ngc$.}
\label{tab:fit_feh}
\begin{tabular} {@{}ccccccccccc@{}}
  \hline
 Quantity ($y$) & [units] & Correlates with ($x$) & [units] & Spearman $r$ & $\log$~Spearman $p$ & Pearson $r$ & $\log$~Pearson $p$ & ${\rm d}y/{\rm d}x$ & $y_0$ & Scatter  \\ 
  \hline
  $\log{\vmax}$ & $[\kms]$ & $\log{\ngc}$ & [--] & $0.74$ & $-4.68$ & $0.74$ & $-4.55$ & $0.17$ & $1.94$ & $0.04$  \\
  $\cnfw$ & [--] & $\iqr(\tau)$ & $[\gyr]$ & $-0.54$ & $-2.29$ & $-0.58$ & $-2.63$ & $-1.42$ & $10.19$ & $1.66$ \\
  $\cnfw$ & [--] & $\riqr$ & $[\gyr^{-1}]$ & $0.48$ & $-1.83$ & $0.50$ & $-1.99$ & $3.92$ & $5.58$ & $1.76$ \\
  $\log{\cnfw}$ & [--] & $\log{\ngc}$ & [--] & $0.58$ & $-2.65$ & $0.61$ & $-2.91$ & $0.33$ & $0.24$ & $0.10$ \\
  $\ttf$ & $[\gyr]$ & $\widetilde{\tau}$ & $[\gyr]$ & $0.55$ & $-2.38$ & $0.60$ & $-2.82$ & $1.06$ & $-1.35$ & $1.22$ \\
  $\log{(\thub-\ttf)}$ & $[\gyr]$ & $\log{(\dfehdt)}$ & [--] & $-0.55$ & $-2.33$ & $-0.56$ & $-2.41$ & $-0.67$ & $0.87$ & $0.14$ \\
  $\log{(\thub-\tfz)}$ & $[\gyr]$ & $\log{(\dfehdt)}$ & [--] & $-0.60$ & $-2.83$ & $-0.58$ & $-2.62$ & $-0.61$ & $1.05$ & $0.12$ \\
  $\za$ & [--] & $\dfehdt$ & [--] & $0.58$ & $-2.61$ & $0.68$ & $-3.69$ & $0.30$ & $-0.13$ & $0.39$ \\
  $\tf$ & $[\gyr]$ & $\widetilde{\tau}$ & $[\gyr]$ & $0.57$ & $-2.51$ & $0.66$ & $-3.48$ & $1.42$ & $-7.49$ & $1.39$ \\
  $\nbrz$ & [--] & $\widetilde{\tau}$ & $[\gyr]$ & $0.70$ & $-3.97$ & $0.69$ & $-3.92$ & $2.75$ & $-24.60$ & $2.47$ \\
  $\nbrz$ & [--] & $\dfehdt$ & [--]  & $0.70$ & $-4.00$ & $0.68$ & $-3.80$ & $1.94$ & $-1.29$ & $2.50$ \\
  $\nbr$ & [--] & $\riqr$ & $[\gyr^{-1}]$ & $0.42$ & $-1.44$ & $0.47$ & $-1.75$ & $13.13$ & $7.52$ & $6.47$ \\
  $\nbr$ & [--] & $\dfehdt$ & [--] & $0.66$ & $-3.50$ & $0.63$ & $-3.08$ & $3.79$ & $1.39$ & $5.72$ \\
  $\nbr$ & [--] & $\ngc$ & [--] & $0.83$ & $-6.50$ & $0.84$ & $-6.71$ & $0.12$ & $4.08$ & $4.02$ \\
  $\rz$ & [--] & $\widetilde{\tau}$ & $[\gyr]$ & $0.87$ & $-7.72$ & $0.87$ & $-7.69$ & $0.17$ & $-1.45$ & $0.08$ \\
  $\log{\nleaf}$ & [--] & $\log{\ngc}$ & [--] & $0.79$ & $-5.59$ & $0.78$ & $-5.46$ & $0.86$ & $-0.30$ & $0.16$ \\
  $N_{<1:100}$ & [--] & $\riqr$ & $[\gyr^{-1}]$ & $0.44$ & $-1.53$ & $0.52$ & $-2.12$ & $10.48$ & $1.82$ & $4.49$ \\
  $N_{<1:100}$ & [--] & $\dfehdt$ & [--] & $0.60$ & $-2.84$ & $0.62$ & $-3.03$ & $2.70$ & $-1.89$ & $4.13$ \\
  $N_{<1:100}$ & [--] & $\ngc$ & [--] & $0.82$ & $-6.40$ & $0.87$ & $-7.63$ & $0.085$ & $-0.31$ & $2.64$ \\
  $\log{N_{1:100-1:4}}$ & [--] & $\log{\ngc}$ & [--] & $0.56$ & $-2.43$ & $0.61$ & $-2.95$ & $0.72$ & $-0.76$ & $0.22$ \\
  \hline
\end{tabular}
\end{table*}

Unsurprisingly, artificially pruning metal-rich GCs also has a weak effect on the inferred correlations between the quantities characterising galaxy formation and assembly and those describing the GC age-metallicity distribution. \autoref{tab:fit_feh} repeats \autoref{tab:fit} for the pruned GC samples and shows that most of the statistically significant correlations reported in \autoref{tab:fit} remain strong. The six correlations with the total number of GCs ($\ngc$) all become stronger after pruning metal-rich GCs, whereas the four correlations involving the age and metallicity interquartile ranges [$\iqr(\tau)$ and $\riqr$] become less significant. Correlations involving the median age ($\widetilde{\tau}$) or the slope of the age-metallicity distribution ($\dfehdt$) remain largely unchanged. For all 20 correlations, the coefficients describing the best-fitting linear relations and the scatter of the data exhibit little change relative to those in \autoref{tab:fit}.

\begin{table*}
\caption{Repeat of table~3 in \citet{kruijssen18c}, applying the 20 statistically significant correlations from \autoref{tab:fit} to the GC population of the Milky Way. For each correlation, the table lists the galaxy-related quantity predicted based on the properties of the Galactic GC population, including the $1\sigma$ uncertainties. For quantities constrained by multiple relations, the `Combined' columns show the weighted mean. Columns~4 and~5 list the original numbers from \citet{kruijssen18c}, whereas columns~6 and~7 list the same after correcting \emosaics for the under-destruction of GCs as in \autoref{tab:fit_feh}.}
\label{tab:fitmw}
\begin{tabular} {@{}cccclccl@{}}
  \hline
  &&&\multicolumn{2}{c}{Original \citep{kruijssen18c}} && \multicolumn{2}{c}{After pruning metal-rich GCs}\\
  \cline{4-5} \cline{7-8}
 Quantity & [units] & Obtained from & Value & Combined & & Value & Combined \\ 
  \hline
  $\vmax$ & $[\kms]$ & $\log{\ngcp}$ &           $180\pm17$   & & & $183\pm17$           \\
  $\cnfw$ & [--] & $\iqr(\tau)$ &                 $8.5\pm1.6$         & \hspace{-10pt}\rdelim\}{3}{10pt} \multirow{3}{*}{$8.0\pm1.0$} & & $8.5\pm1.7$         & \hspace{-10pt}\rdelim\}{3}{10pt} \multirow{3}{*}{$7.8\pm1.0$}      \\
  $\cnfw$ & [--] & $\riqr$ &                        $7.6\pm1.7$         &   & & $7.4\pm1.8$       \\
  $\cnfw$ & [--] & $\log{\ngcp}$ &                     $7.9\pm1.8$        &   & & $7.3\pm1.7$         \\
  $\ttf$ & $[\gyr]$ & $\widetilde{\tau}$ &   $11.8\pm1.2$        & \hspace{-10pt}\rdelim\}{2}{10pt} \multirow{2}{*}{$11.5\pm0.8$}  & & $11.6\pm1.2$         & \hspace{-10pt}\rdelim\}{2}{10pt} \multirow{2}{*}{$11.3\pm0.8$}                             \\
  $\ttf$ & $[\gyr]$ & $\log{\dfehdt}$ &                $11.2\pm1.0$        &   & & $11.2\pm1.0$                         \\
  $\tfz$ & $[\gyr]$ & $\log{\dfehdt}$ &               $9.4\pm1.4$        &   & & $9.4\pm1.4$                          \\
  $\za$ & [--] & $\dfehdt$ &                       $1.2\pm0.5$         &   & & $1.3\pm0.5$         \\
  $\tf$ & $[\gyr]$ & $\widetilde{\tau}$ &    $10.1\pm1.4$        &   & & $9.8\pm1.4$                            \\
  $\nbrz$ & [--] & $\widetilde{\tau}$ &       $9.9\pm2.3$        & \hspace{-10pt}\rdelim\}{2}{10pt} \multirow{2}{*}{$9.2\pm1.9$}  & & $9.0\pm2.5$         & \hspace{-10pt}\rdelim\}{2}{10pt} \multirow{2}{*}{$8.5\pm2.0$}                         \\
  $\nbrz$ & [--] & $\dfehdt$  &                  $7.4\pm3.4$         &    & & $7.7\pm3.4$                     \\
  $\nbr$ & [--] & $\riqr$ &                          $14.1\pm6.1$       & \hspace{-10pt}\rdelim\}{3}{10pt} \multirow{3}{*}{$15.1\pm3.3$} & & $13.8\pm6.6$         & \hspace{-10pt}\rdelim\}{3}{10pt} \multirow{3}{*}{$14.5\pm3.2$}         \\
  $\nbr$ & [--] & $\dfehdt$ &                     $18.9\pm7.4$       &   & & $18.9\pm7.3$            \\
  $\nbr$ & [--] & $\ngcp$ &                       $14.2\pm4.7$        &   & & $13.4\pm4.2$         \\
  $\rz$ & [--] & $\widetilde{\tau}$ &           $0.61\pm0.10$          &   & & $0.60\pm0.10$                    \\
  $\nleaf$ & [--] & $\log{\ngcp}$ &                     $24.1\pm10.2$      &   & & $21.2\pm8.1$             \\
  $N_{<1:100}$ & [--] & $\riqr$ &              $7.1\pm4.3$        & \hspace{-10pt}\rdelim\}{3}{10pt} \multirow{3}{*}{$7.9\pm2.2$}  & & $6.8\pm4.6$         & \hspace{-10pt}\rdelim\}{3}{10pt} \multirow{3}{*}{$7.1\pm2.2$}                  \\
  $N_{<1:100}$ & [--] & $\dfehdt$ &         $10.7\pm5.3$           &   & & $10.6\pm5.2$                      \\
  $N_{<1:100}$ & [--] & $\ngcp$ &            $7.4\pm3.0$         &    & & $6.3\pm2.7$                   \\
  $N_{1:100-1:4}$ & [--] & $\log{\ngcp}$ &       $4.4\pm2.3$         &    & & $4.0\pm2.1$                       \\
  \hline
\end{tabular}
\end{table*}

To conclude our investigation of how the under-destruction of GCs with $-1.0<\feh<-0.5$ affects the presented results, \autoref{tab:fitmw} shows the result of applying the correlations from \autoref{tab:fit_feh} to the GC population of the Milky Way (columns~6 and~7 in \autoref{tab:fitmw}). This is a partial repeat of table~3 of \citet{kruijssen18c}, which carries out the same application using the original correlations from \autoref{tab:fit} (columns~4 and~5 in \autoref{tab:fitmw}). Comparing the results for the original application and the artificial pruning experiment, we see that the under-destruction of metal-rich GCs has a statistically insignificant impact on the quantities describing the formation and assembly history of the Milky Way. In all cases, the inferred numbers change by considerably less than the quoted uncertainties. We conclude that the presented results are robust against the under-destruction of metal-rich GCs in \emosaics.

\section{Comparison to previous models} \label{sec:appprevious}

\emosaics is not the first project aimed at modelling the formation and/or evolution of GC populations in the cosmological context of galaxy formation and evolution. However, it differs in several critical aspects relative to previous models. We now briefly highlight the main differences and similarities to earlier studies.

Resolving the formation and evolution of entire GC populations in cosmological simulations down to $z=0$ will remain unachievable for the foreseeable future due to the enormous computational challenge implied by the large dynamic range spanned by the relevant time, mass, and spatial scales. For this reason, attempts at placing GC formation and evolution in the cosmological context of galaxy formation and evolution has always relied heavily on subgrid models for following the GCs (see section 3 of \citealt{kruijssen14c} and section 7 of \citealt{forbes18} for recent reviews on this topic). Previous models (including some of our own precursors) have generally made use of phenomenological or ad-hoc scalings to insert the GC population into a galactic setting. \emosaics retains several successes of previous work, but makes the critical step forward to achieve a physical, ab-initio subgrid model for GC formation and evolution, which thus enables predictive modelling.

The first class of models placing GCs in the context of galaxy formation generally insert GCs into dark matter-only simulations that have been post-processed to include baryons with a semi-analytic galaxy formation model \citep[e.g.][]{beasley02,moore06,muratov10,tonini13,choksi18}. This is accomplished by assuming either that the number of GCs is proportional to the stellar or gas mass of the host galaxy, or that it is set by the specific frequency as a function of galaxy mass observed at $z=0$. GC disruption is either accounted for partially in the form of evaporation driven by two-body relaxation \citep[which does not include disruption by tidal shocks]{muratov10,choksi18} or not at all \citep{beasley02,moore06,tonini13}. Several of these models \citep{beasley02,tonini13,choksi18} do not contain any spatial information on the GCs. While \emosaics shares the concept that GC formation is related to regular star formation in the host galaxy, it differs from these previous approaches in terms of the physical implementation of this idea. It traces the three-dimensional structure of the baryons in a self-consistent hydrodynamical simulation and models the formation and tidal disruption of clusters (not just GCs, which represent an emergent sub-population of all stellar clusters) in a way that accounts for the local, sub-galactic conditions of the ISM and the gravitational potential, on spatial scales comparable to the Jeans length of the warm photoionized ISM. This results in an environmental dependence of the CFE, the ICMF, and cluster disruption, as well as an explicit modelling of cluster migration from the dense, star-forming ISM into the thick disc or galaxy halo.

A step forward from the purely semi-analytic approach is made by hydrodynamical simulations using a recipe to identify sites of GC formation, either in isolated discs or mergers \citep[e.g.][]{bekki02,li04}, or in a cosmological context \citep[e.g.][]{kravtsov05,renaud17}. Despite including live baryons, these models have treated the formation and evolution of GCs largely phenomenologically. Following \citet{elmegreen97}, GC-forming gas is identified above a certain threshold in gas density \citep{li04,kravtsov05} or pressure \citep{bekki02}, where in the latter case GC formation proceeds entirely independently of field star formation. The formation of clusters with masses $M<10^5~\msun$ was not included, implying that these models form GC-like clusters by definition. In the model of \citet{renaud17}, GC formation is assumed to directly trace field star formation in space and time (through `particle tagging', implicitly assuming a constant CFE and a universal ICMF). None of these models include cluster disruption.

The first \emosaics precursor \citep[\mosaics,][]{kruijssen11} advanced from these studies by enabling the formation of (subgrid) clusters with masses down to $10^2~\msun$ and by including cluster disruption due to the local tidal field, generated by the total mass distribution including the substructured ISM. However, this original \mosaics model still assumed that clusters form in direct proportion to the SFR (implying a constant CFE) according to a universal power law ICMF with an exponential truncation at high masses. As shown in \citetalias{pfeffer18} \citep[also see][and \citealt{pfeffer19}]{reinacampos19}, the assumption of a constant CFE and ICMF as in \citet{kruijssen11} and \citet{renaud17} prohibits reproduction of several key properties of GC populations, such as their radial metallicity gradient, maximum mass scale, and median ages in excess of those of the field stars, as well as the properties of $z=0$ cluster populations. \emosaics now solves this problem by enabling the CFE and the high-mass truncation of the ICMF to vary continuously as a function of the local ISM properties (which is critical for reproducing observations, see below) according to physical models for these quantities, and by including GC destruction due to dynamical friction in post-processing.

An even more comprehensive approach will be to directly resolve the gas flows leading to GC formation in cosmological hydrodynamical simulations, either by modelling the GCs as sink particles \citep[e.g.][]{li17,li18} or by resolving them with a considerable number of star particles \citep[e.g.][]{kim18}. Attempts in this direction are highly promising, also because they are capable of completely resolving the ISM structures that generate the tidal shocks dominating GC disruption, but they face two important challenges. Firstly, the extreme computational cost of these models restricts them to high redshift, i.e.~$z>5$ \citep{kim18}, $z>3$ \citep{li17}, or $z>1.5$ \citep{li18}. It is clear that simulations resolving GCs will still take more than a decade before being able to follow the entire cluster population (i.e.~down to a few $10^2~\msun$) of a galaxy to $z=0$, whereas this goal is starting to get within reach for sink particle-based models. However, it is currently not computationally feasible for such models to probe the impact of galaxy formation and assembly on GC populations through suites of dozens of galaxy simulations like in \emosaics. Secondly, the spatially-resolved physics of star formation and feedback that govern star (cluster) formation are highly complex. Perhaps unsurprisingly, many of the high-resolution studies therefore struggle to reproduce some of the observed properties of cluster populations, such as their small radii, short formation time-scales, and the minimum and maximum mass scales of the ICMF. They also still require a subgrid model for modelling (mass loss by) collisional stellar dynamics, as these remain unresolved. These aspects remain important points of attention for future, improved high-resolution models of GC formation and evolution during galaxy formation.

Finally, there exists a family of models in which the formation of (metal-poor) GCs is assumed to be related to reionization \citep[e.g.][]{moore06,spitler12,griffen13,moran14,boylankolchin17}. Each of these models inserts GCs using a phenomenological recipe into dark matter-only environments, either using the output of a numerical simulation or (semi-)analytically. However, none of these studies demonstrate that GC formation is associated with reionization, nor do they include GC disruption. The only exception is the phenomenological approach by \citet{boylankolchin17}, who leaves mass loss as a free parameter and requires GCs to have been a factor of 5--10 times more massive at birth for them to have played a role in reionization. However, this is inconsistent with physical models of GC mass loss \citep[e.g.][]{webb15,kruijssen15b,reinacampos18}. It also assumes an ad hoc GC formation history to maximise the impact of GCs on reionization, with all metal-poor GCs having formed at $z>6$. This is inconsistent with observations showing that reionization preceded most of GC formation \citep[e.g.][also see the compilation of GC age measurements in \citealt{kruijssen18c}]{forbes15}, as well as with the predicted GC formation history from \emosaics \citep[see][]{reinacampos19}. Several of these reionization-related studies correlate the spatial and/or kinematic distribution of (metal-poor) GCs to that of dark matter haloes with a certain collapse redshift to argue that these GCs formed at that collapse redshift \citep{moore06,spitler12,moran14}, which effectively assumes that reionization truncated the formation of these GCs. However, the fundamental problem of this approach is that it ignores the environmental dependence of GC formation (through the CFE and the maximum mass limit) and GC disruption, which together can cause major changes to the GCs' spatial distribution and prohibit a naive interpretation in terms of a collapse redshift \citep[Reina-Campos et al.~in prep.]{kruijssen14c,forbes18}. This means that matching the radial profiles of GCs (within a certain metallicity range) to those of dark matter (within a certain collapse redshift range) provides only limited constraints on the formation redshifts of these GCs. Perhaps unsurprisingly in this context, models attributing a special role to reionization in relation to metal-poor GC formation are at odds with observed variety of metallicity distributions, including the existence of unambiguously unimodal ones \citep[e.g.][]{usher12}. For these reasons, \emosaics does not explicitly associate GC formation with reionization and treats the background radiation field by adopting the commonly-used model of a spatially-uniform, temporally-evolving radiation field comprising the cosmic microwave background and the metagalactic ultraviolet/X-ray background from \citet{haardt01}.

In summary, \emosaics models GC formation and evolution in a subgrid fashion, as has been done in previous work. However, it does not employ the phenomenological prescriptions used previously (e.g.~by assuming GC formation in proportion to the stellar mass or SFR of the host galaxy, equivalent to `particle tagging', or by identifying a threshold density or pressure above which GCs are allowed to form). Instead, it uses physical models that accurately describe cluster formation in the local Universe, in which the GC properties at formation vary continuously with the local environmental conditions in the ISM. As a result, the formation of stellar clusters and GCs represents a natural byproduct of star formation in our simulations. This is a critical difference relative to previous work, because it enables the formation of GC-like clusters (e.g.~$M>10^5~\msun$) even under the low-pressure conditions of local-Universe galaxies, albeit much more rarely than in the vigorously star-forming galaxies at high redshift. Given the existence of such clusters in the discs of nearby isolated galaxies \citep[e.g.][]{portegieszwart10,longmore14}, this unification of GC formation with regular star and cluster formation as a continuous function of the environmental conditions marks important progress.

In addition, \emosaics contains an environmentally-dependent model for cluster disruption due to the local tidal field, including ISM-independent mass loss mechanisms such as stellar evolution, tidal evaporation, disc shocking, and bulge shocking, but additionally (at least partially) accounting for mass loss due to tidal perturbations from the ISM. However, due to the limited resolution of \emosaics and its simplified treatment of the ISM, our simulations do not resolve all of the ISM structures generating the disruption. This is an important area of future improvement. None the less, it represents an important step forward relative to the earlier works discussed above, which either contain no GC disruption or only ISM-independent mass loss, thus omitting altogether the (often dominant) destructive power of tidal interactions with the ISM. Thanks to these physical ingredients, \emosaics enables the self-consistent modelling of GC formation and destruction during galaxy formation and assembly. As such, our simulations do not `insert' GCs into galaxy simulations based on some (semi-)empirical scaling relation, but `retrieve' them according to the evolving properties of the star-forming ISM.

\bsp

\label{lastpage}

\end{document}